\documentclass[aps,prc,showpacs,groupedaddress]{revtex4}

\usepackage{hyperref}

\usepackage{times}
\usepackage{graphicx}

\usepackage{multirow}
\usepackage{amssymb}
\usepackage{amsmath}
\usepackage{color}

\newcommand{\asy}{A}               
\newcommand{\ath}{A^{\rm th}}      

\newcommand{\q}{\mathrm}

\begin{document}

\title{Measurement of Parity-Violating Asymmetry in Electron-Deuteron
Inelastic Scattering}

\date{\today}

\collaboration{The PVDIS Collaboration}
\noaffiliation
\author{D.~Wang}
\affiliation{University of Virginia, Charlottesville, Virginia 22904, USA}

\author{K.~Pan}
\affiliation{Massachusetts Institute of Technology, Cambridge, Massachusetts 02139, USA}

\author{R.~Subedi} 
\thanks{now at Richland College, 
Dallas County Community College District, 
Dallas, Texas 75243, USA.}
\affiliation{University of Virginia, Charlottesville, Virginia 22904, USA}


\author{Z.~Ahmed}
\affiliation{Syracuse University, Syracuse, New York 13244, USA} 

\author{ K.~Allada}
\affiliation{University of Kentucky, Lexington, Kentucky 40506, USA}

\author{K.~A.~Aniol}
\affiliation{\mbox{California State University, Los Angeles}, Los Angeles, California 90032, USA }

\author{D.~S.~Armstrong}
\affiliation{College of William and Mary, Williamsburg, Virginia 23187, USA}

\author{J.~Arrington}
\affiliation{Physics Division, Argonne National Laboratory, Argonne, Illinois 60439, USA}

\author{ V.~Bellini}
\affiliation{Istituto Nazionale di Fisica Nucleare, Dipt.~di Fisica dell'Univ.~di Catania, I-95123 Catania, Italy}

\author{R.~Beminiwattha}
\affiliation{Ohio University, Athens, Ohio 45701, USA}

% ...

\begin{abstract}                
The parity-violating asymmetries
between a longitudinally-polarized electron beam and an unpolarized
deuterium target have been measured recently. 
The measurement covered two kinematic points in the deep inelastic scattering region 
and five in the nucleon resonance region.
We provide here details of the experimental setup, data analysis, 
and results on all asymmetry measurements including parity-violating
electron asymmetries and those of inclusive 
pion production and beam-normal asymmetries. 
The parity-violating deep-inelastic asymmetries were used to extract the electron-quark
weak effective couplings, and the resonance asymmetries provided the 
first evidence for quark-hadron duality in electroweak observables. 
These electron asymmetries and their interpretation were published
earlier, but are presented here in more detail. 
\end{abstract}


\maketitle



\section{Physics Motivation}
Parity symmetry implies that the physics laws behind a system remain the
same when the system undergoes a space-reversal (parity) transformation. 
A simplified
version of such transformation, in which only one dimension is reversed, 
mimics a mirror reflection, and thus parity 
symmetry is often called mirror symmetry. Among all known interactions
of nature, electromagnetic, strong, and gravitational forces respect
parity symmetry, but the weak force does not, as first postulated by 
Lee and Yang~\cite{Lee:1956qn}, and verified experimentally in nuclear
$\beta$-decay by Wu {\it et al.}~\cite{Wu:1957my}, in 1957. 

For spin-1/2 elementary particles (elementary fermions), 
the standard scheme to describe how they violate parity symmetry is 
to use their chirality, an abstract concept 
defined by the $\gamma^5$ Dirac matrix, 
the chiral operator in quantum electrodynamics. 
In the ultra-relativistic limit or for massless particles, chirality 
becomes the experimentally accessible helicity:
A particle is defined to be in a right(left)-handed helicity state, when
its spin as defined by the right-hand rule is 
in the same (opposite) direction as its linear momentum. 
Since parity transformation changes a right-handed chiral state to left-handed and
vice versa, parity violation implies that the fermion's weak charge must depend on 
the its chiral state.  This feature is 
different from the electric charge for the electromagnetic 
interaction, the color charge for the strong nuclear force, and the 
energy-momentum tensor for gravity.

In the decade that followed the first observation of parity violation, many 
theories were proposed to explain this phenomenon. Among them is 
the Glashow-Weinberg-Salam (GWS) 
theory~\cite{Glashow:1961tr,Weinberg:1967tq,Salam:1968rm} of 
electroweak unification. In this theory, the charged-weak force 
behind $\beta$-decays only acts on left-handed spin-1/2 elementary
particles (elementary fermions)  and right-handed anti-fermions, 
thus violates parity to the maximal 
degree.  The theory also predicted the existence of a new, neutral weak force 
carried by an electrically-neutral boson, the $Z^0$. Unlike the $W^\pm$ bosons
that carry the charged-weak force, the $Z^0$ does interact with both chiral 
states of all fermions and anti-fermions. For neutral-weak interactions, the
difference in the fermion's weak-interaction strengths
between its left- and right-handed chiral states is described by the weak
axial charge $g_A$, while the average of the 
two is called the weak vector charge $g_V$. 
In the GWS theory, $g_A$ 
equals the particle's weak isospin  $T_3$: $g_A=T_3=1/2$ for up, 
charm, top quarks and neutrinos, and $-1/2$ for down, strange and bottom quarks 
and electrons; and $g_V$ is related to the 
particle's $T_3$ and electric charge $Q$: $g_V=T_3 -2Q\sin^2\theta_W$, with
$\theta_W$ the weak mixing angle, a parameter that describes how the
electromagnetic interaction is unified with the weak force. 
Antiparticles have opposite weak isospin and electric charge, and thus opposite
$g_A$ and $g_V$ as their particle counterparts. 
The fact that $g_A=\pm 1/2$ for elementary fermions implies that they all have
a chirality preference in neutral-weak interactions. 

The $Z^0$ was soon
observed in the 1970's in both neutrino~\cite{Hasert:1973cr,Hasert:1974ju}
and electron scattering experiments~\cite{Prescott:1978tm,Prescott:1979dh}. 
In electron scattering, parity violation is observed by
a difference (an asymmetry) in the scattering cross sections between left-
and right-handed electrons from an unpolarized target:
\begin{eqnarray}
 A_{PV} &\equiv& \frac{\sigma_{R} - \sigma_{L}}{\sigma_{R} + \sigma_{L}}~.
\end{eqnarray}
In the most recent decades, parity-violating electron scattering (PVES) has
been used primarily in the elastic scattering region. In elastic kinematic
settings, the target nucleus remains whole during its interaction with the 
electron and the strong-interaction that binds quarks together to form the 
nucleon (or binds nucleons together
to form the nucleus) is not disturbed. Elastic PVES asymmetry
has been used to study the internal structure of the target that cannot be
revealed through electromagnetic interactions. For example, 
elastic scattering from the proton and light nuclei has been used to study
whether sea quarks contribute to the nucleon's structure, that is, whether
the strange and the anti-strange quarks are distributed differently after their creation. 
Such nucleon strange form factor experiments have been carried out at many
different facilities worldwide, such as the SAMPLE 
experiment~\cite{Mckeown:1989ir,Hasty:2001ep,Spayde:2003nr,Ito:2003mr,Beise:2004py} 
at MIT Bates, the A4 experiment at MAMI/Mainz~\cite{Maas:2004ta,Maas:2004dh,Baunack:2009gy}, 
the HAPPEX 
experiments~\cite{Aniol:2000at,Aniol:2004hp,Aniol:2005zf,Aniol:2005zg,Acha:2006my,Ahmed:2011vp} 
in JLab Hall A, and the $G0$ experiment~\cite{Beck:1989tg,Armstrong:2005hs,Androic:2009aa} 
in JLab Hall C.
In the recent PREX experiment~\cite{Abrahamyan:2012gp,Horowitz:2012tj}, 
elastic scattering from $^{208}$Pb  
has confirmed a difference in the spatial distributions between 
protons and neutrons inside this heavy nucleus.

On the other hand, of particular value to testing the Standard Model is the so-called 
deep inelastic scattering (DIS) regime, where the energy and momentum transferred from
the electron to the target are so high that the quarks are probed directly, and that 
the strong interaction among quarks 
become negligible due to the so-called ``asymptotic freedom'' phenomenon. 
The parity-violating deep inelastic scattering 
(PVDIS) asymmetry is determined by
the effective electron-quark couplings $C_{1q}$ and $C_{2q}$, weighted by
kinematic factors and the well-determined DIS structure functions. In the 
Standard Model tree-level diagram, the $C_{1q},C_{2q}$ couplings are the product of 
the electron and quark weak charges: $C_{1q}=2g_A^eg_V^q$ (the effective
electron-quark AV coupling), and $C_{2q}=2g_V^eg_A^q$ 
(the effective electron-quark VA coupling). 

The first PVES experiment~\cite{Prescott:1978tm,Prescott:1979dh}, E122 
at the Stanford Linear Accelerator Center (SLAC) by Prescott {\it et al.}, 
was performed in the DIS region and 
provided the first definitive measurement of the weak mixing angle 
$\sin^2\theta_W$. The E122 results were in good 
agreement with predictions from the GWS-theory, 
establishing it as a cornerstone of the now Standard Model of 
particle physics.
The thirty years that followed witnessed a vast amount of 
Standard-Model-test experiments.  Among those that determine the weak charges
of elementary particles, the most precise measurement of the
electron weak charges came from PVES on an electron 
target~\cite{Anthony:2005pm,Czarnecki:2005pe} that provided $C_{2e}=2g_V^eg_A^e$.
The best result on the effective electron-quark AV couplings $C_{1q}$ 
is from a combination~\cite{Androic:2013rhu} of elastic PVES~\cite{Aniol:2000at,Aniol:2004hp,Aniol:2005zf,Aniol:2005zg,Acha:2006my,Ahmed:2011vp,Beck:1989tg,Armstrong:2005hs,Androic:2009aa} 
and atomic parity violation 
experiments~\cite{Wood:1997zq,Bennett:1999pd,Ginges:2003qt,Dzuba:2012kx}. 

On the other hand, determination of the $C_{2q}$ couplings from PVES 
is difficult: For elastic scattering, the asymmetry component sensitive 
to the quark chirality (spin) is not directly determined by the $C_{2q}$, but by the
nucleon's axial form factor $G_A$. Extracting $C_{2q}$ from 
$G_A$~\cite{Hasty:2001ep,Spayde:2003nr,Ito:2003mr,Beise:2004py} 
depends on hadronic models and is subject to large uncertainties in the 
radiative corrections.  For DIS, the quark-chirality-dependent $C_{2q}$ 
contribution to the PVDIS asymmetry is kinematically suppressed because of 
angular momentum conservation, similar to the way in which the quark-spin-dependent
contribution to the unpolarized cross section is suppressed. 
The small value of 
$g_V^e$ further reduces the $C_{2q}$ contribution to the PVDIS asymmetry. 
Until the experiment reported here was carried out, the only direct data on 
$C_{2q}$ were from SLAC E122. 

In addition to DIS and elastic scattering, another kinematic region accessible in 
electron scattering is the nucleon resonance region. In this region, the nucleon 
is excited by the energy and momentum transferred 
from the electron, but the strong interaction among quarks is not negligible (unlike in 
DIS). The nucleon resonance region therefore provides a transition  
between the quark and gluon degrees of freedom of DIS to hadron degrees of freedom
of elastic scattering.  
Inclusive measurements in the nucleon resonance region have
demonstrated a remarkable feature called ``quark-hadron duality'',
first pointed out by Bloom and Gilman~\cite{Bloom:1970xb}, 
in which the low-energy (few GeV) cross sections averaged over
the energy intervals of the resonance structures
resemble those measured at asymptotically high energies of DIS. 
Over the past decade, duality has been verified in 
the unpolarized structure functions $F_2$ and 
$F_L$ at four-momentum-transfer-squared $Q^2$ values 
below 1~(GeV/$c$)$^2$~\cite{Niculescu:2000tk,Liang:2004tj,Psaker:2008ju,Malace:2009dg,Malace:2009kw}, 
in the proton spin asymmetry $A_1^p$ down to 
$Q^2=1.6$~(GeV/$c$)$^2$~\cite{Airapetian:2002rw}, 
in the spin structure function $g_1$ down to 
$Q^2=1.7$-$1.8$~(GeV/$c$)$^2$~\cite{Bosted:2006gp,Solvignon:2008hk},
in the helicity-dependent structure functions $H_{1/2,3/2}$~\cite{Malace:2011ad},
and for charged pion electroproduction in semi-inclusive 
scattering~\cite{Navasardyan:2006gv}.
It was speculated that duality is a universal feature of the quark-hadron
transition that should be exhibited not only in electromagnetic interactions,
but also in charged lepton scattering via the weak interactions~\cite{Carlson:1993wy}, and perhaps other processes as well.

We report here details of a PVDIS experiment that was carried out at 
the Thomas Jefferson National
Accelerator Facility (Jefferson Lab, or JLab) in 2009, JLab E08-011. During this 
experiment, PVES asymmetries on a deuterium target were measured 
at two DIS and five nucleon resonance kinematic settings.
The precision of the DIS measurement was higher than that of E122, and the kinematics
were optimized for the extraction of the $C_{2q}$ couplings. 
The DIS asymmetry and the $C_{2q}$ couplings, published 
in Ref.~\cite{Wang:2014bba}, improved over previous data by a factor of five. 
Data taken at 
resonance settings had larger uncertainties, but nevertheless provided the 
first PVES data covering the whole nucleon resonance region. The resonance
asymmetry results, published in Ref.~\cite{Wang:2013kkc}, provided 
the first
observation of quark-hadron duality on parity-violating observables.
In this archival paper we first review the formalism for PVDIS, the SLAC 
E122 experiment, 
then describe the new JLab experiment E08-011 including its apparatus, 
data analysis, and all systematic uncertainties. In addition to PVES asymmetries, 
we report asymmetry results on inclusive pion production, pair-production, 
and beam-normal asymmetries. Finally, we provide interpretations of the 
electron asymmetries in DIS and the nucleon resonance regions.

\subsection{Formalism for Parity-Violation in Electron Inelastic Scattering}\label{sec:formalism}
\begin{figure}[!htp]
 \hspace*{0.15\textwidth}
 \includegraphics[width=0.2\textwidth]{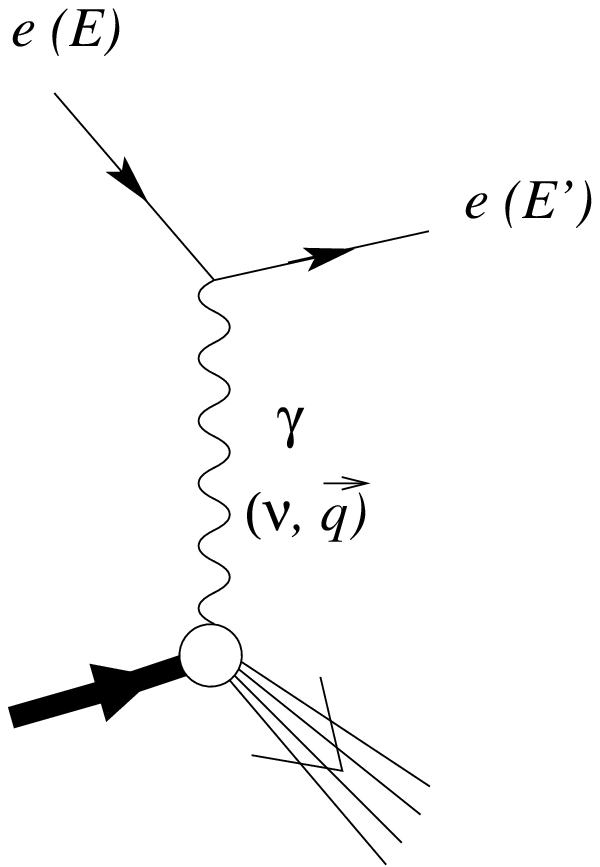}
 \hspace*{0.1\textwidth}
 \includegraphics[width=0.2\textwidth]{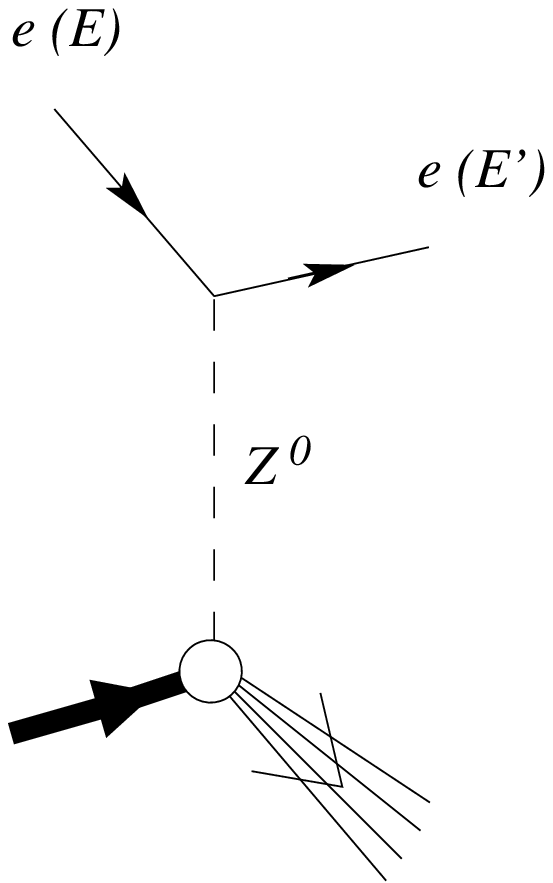}
 \caption{The electron exchanges either a virtual photon (left) or a virtual $Z^0$ (right) 
with the target. The interference
between these two processes leads to a parity-violating asymmetry between left-
and right-handed electrons.}\label{fig:escatt}
\end{figure}

For inelastic electron scattering off a nucleon or nuclear target,
the parity-violating asymmetry  
originates from the interference between photon- and $Z^0$-exchanges from 
the electron to the target (Fig.~\ref{fig:escatt}). 
This asymmetry can be written as~\cite{Cahn:1977uu}
\begin{eqnarray}
 A_{PV} &=& -{{G_FQ^2}\over{4\sqrt{2}\pi\alpha(Q^2)}}
         \left[a_1(x,Q^2)Y_1(x,y,Q^2)+a_3(x,Q^2)Y_3(x,y,Q^2)\right]~,\label{eq:Apvdis1}
\end{eqnarray}
where $G_F$ is the Fermi constant, $\alpha(Q^2)$ is the fine structure constant, 
$y=\nu/E=(E-E^\prime)/E$ is the fractional energy loss of the electron with $E$ and
$E^\prime$ the incident and the scattered electrons' energy, 
$Q^2\equiv -q^2$ is the negative of the four-momentum transferred from
the electron to the target $q$, squared:
\begin{eqnarray}
 Q^2=2EE^\prime (1-\cos\theta)~ \label{eq:qmu2}
\end{eqnarray}
with $\theta$ the electron scattering angle. 
The Bjorken scaling variable $x$ is defined as 
\begin{eqnarray}
 x&\equiv& Q^2/(2M\nu)~, \label{eq:xbj}
\end{eqnarray} 
with $M$ the proton mass. 
Another important variable is the invariant mass of the $\gamma$-nucleon 
(or $Z^0$-nucleon) system, which for a fixed nucleon target is given by 
\begin{eqnarray}
 W^2=M^2+2M\nu-Q^2~. \label{eq:wmm}
\end{eqnarray}
Typically, the region $M<W<2$~GeV is the nucleon resonance region and 
$W>2$~GeV corresponds to the DIS region.

The kinematic factors $Y_{1,3}$ are defined as 
\begin{eqnarray}
 Y_1&=&\left[\frac{1+R^{\gamma Z}}{1+R^\gamma}\right]
   \frac{1+(1-y)^2-y^2\left[1-\frac{r^2}{1+R^{\gamma Z}}\right]-xy{M\over E}}
   {1+(1-y)^2-y^2\left[1-\frac{r^2}{1+R^{\gamma}}\right]-xy{M\over E}}~\label{eq:y11}
\end{eqnarray}
and
\begin{eqnarray}
  Y_3&=&\left[\frac{r^2}{1+R^\gamma}\right]
   \frac{1-(1-y)^2}
   {1+(1-y)^2-y^2\left[1-\frac{r^2}{1+R^{\gamma}}\right]-xy{M\over E}}~,\label{eq:y13}
\end{eqnarray}
where $r^2=1+{{Q^2}\over{\nu^2}}$, 
and $R^{\gamma(\gamma Z)}(x,Q^2)$ is the ratio of 
the longitudinal to transverse virtual photon electromagnetic 
absorption cross sections ($\gamma-Z^0$ interference 
cross sections). 
With some algebra, one can express the $xyM/E$ term by $r^2$ and $y^2$ and
Eqs.(\ref{eq:y11},\ref{eq:y13}) change to (as in Ref.~\cite{Brady:2011uy}):
\begin{eqnarray}
 Y_1&=&\left[\frac{1+R^{\gamma Z}}{1+R^\gamma}\right]
   \frac{1+(1-y)^2-\frac{y^2}{2}
     \left[1+r^2-\frac{2r^2}{1+R^{\gamma Z}}\right]}
   {1+(1-y)^2-\frac{y^2}{2}\left[1+r^2-\frac{2r^2}{1+R^{\gamma}}\right]}~\label{eq:y11_prd}
\end{eqnarray}
and
\begin{eqnarray}
  Y_3&=&\left[\frac{r^2}{1+R^\gamma}\right]
   \frac{1-(1-y)^2}
   {1+(1-y)^2-\frac{y^2}{2}\left[1+r^2-\frac{2r^2}{1+R^{\gamma}}\right]}~.\label{eq:y13_prd}
\end{eqnarray}
To a good approximation $R^{\gamma Z}$ can be assumed to be equal to 
$R^{\gamma}$, resulting in $Y_1(x,y,Q^2)= 1$. 

The $a_{1,3}$ terms in Eq.~(\ref{eq:Apvdis1}) are
\begin{eqnarray}
  a_1(x)&=&2 g_A^e\frac{F_1^{\gamma Z}}{F_1^\gamma}~,\label{eq:a11}\\
  a_3(x)&=&g_V^e\frac{F_3^{\gamma Z}}{F_1^\gamma}~,\label{eq:a31}
\end{eqnarray}
where the structure functions, $F_{1,3}^{\gamma, \gamma Z}$, 
can be interpreted
in the quark-parton model (QPM) in terms of the parton 
distribution functions (PDF) $q_i(x,Q^2)$ and $\bar q_i(x,Q^2)$ of the target:
\begin{eqnarray}
  F_1^\gamma(x,Q^2)&=&{1 \over 2} \sum{Q_{q_i}^2 \left[q_i(x,Q^2)+\bar q_i(x,Q^2)\right]}~,\label{eq:F1gqpm}\\
  F_1^{\gamma Z}(x,Q^2)&=&\sum{Q_{q_i} g_V^i\left[q(x,Q^2)+\bar q_i(x,Q^2)\right]}~,\label{eq:F1gzqpm}\\
  F_3^{\gamma Z}(x,Q^2)&=&2\sum{Q_{q_i} g_A^i \left[q_i(x,Q^2)-\bar q_i(x,Q^2)\right]}~.\label{eq:F3gzqpm}
\end{eqnarray}
Here, $Q_{q_i}$ denotes the quark's electric charge and 
the summation is over the quark flavors $i=u,d,s\cdots$. 
Equations~(\ref{eq:a31},\ref{eq:F3gzqpm}) show that the $a_3(x,Q^2)$ term 
involves the chirality of the quark ($g_A^i$) and therefore is suppressed by the kinematic
factor $Y_3$ due to angular momentum conservation. 
It vanishes at the forward angle $\theta=0$ or $y=0$, and increases with $\theta$ 
or $y$ at fixed $x$. 

In most world parameterizations, it is 
common to fit the structure functions $F_2$ and $R$ simultaneously to cross-section 
data. They are related through 
\begin{eqnarray}
 F_2^{\gamma(\gamma Z)} &=& {{2xF_1^{\gamma(\gamma Z)}(1+R^{\gamma(\gamma Z)})}\over{r^2}}~,\label{eq:f2f11}
\end{eqnarray}
or equivalently:
\begin{eqnarray}
 F_1^{\gamma(\gamma Z)} &=& {{r^2F_2^{\gamma(\gamma Z)}}\over{2x(1+R^{\gamma(\gamma Z)})}}~.\label{eq:f1f21}
\end{eqnarray} 
In the QPM with the Bjorken scaling limit $Q^2\to\infty$ at fixed $x$, 
the ratios $R^{\gamma(\gamma Z)}$ are zero, and $r=1$. 
Hence one can construct the $F_{2}$ structure functions from PDFs as
\begin{eqnarray}
  F_2^\gamma(x)&=& 2x F_1^\gamma(x)=x\sum{Q_{q_i}^2 \left[q_i(x)+\bar q_i(x)\right]}~,\label{eq:f1f2gqpm}\\
  F_2^{\gamma Z}(x)&=& 2x F_1^{\gamma Z}(x)=2x\sum{Q_{q_i} g_V^i\left[q_i(x)+\bar q_i(x)\right]}~.\label{eq:f1f2gzqpm}
\end{eqnarray}
Note that the use of the approximation $F_2=2xF_1$ does not affect the $a_1$ term 
of the asymmetry, since the extra terms $r^2$ and $2x$ in the numerator $F_1^{\gamma Z}$
and the denominator $F_1^\gamma$ cancel.

For electron scattering, one defines the product of the electron and the
quark weak couplings as the effective weak coupling constants $C_{1q,2q}$. 
In leading order of one-photon and one-$Z^0$
exchanges between the electron and the target (Fig.~\ref{fig:escatt}), 
\begin{eqnarray}
 C_{1u} = 2g^e_A g^u_V~,~&& 
 C_{2u} = 2g^e_V g^u_A~,\label{eq:c1factorized}\\
 C_{1d} = 2g^e_A g^d_V,~&&
 C_{2d} = 2g^e_V g^d_A.\label{eq:c2factorized}
\end{eqnarray}
Using the appropriate electric charge and the weak isospin of quarks, 
they are related to the weak mixing angle $\theta_w$ as 
\begin{eqnarray}
 C_{1u} &=& 2g^e_A g^u_V = 2\left(-\frac{1}{2}\right)
                          \left(\frac{1}{2}-\frac{4}{3}\sin^2\theta_W\right)
          = -\frac{1}{2} + \frac{4}{3} \sin^2\theta_{W}~,\label{eq:c1ufactorized}\\
 C_{2u} &=& 2g^e_V g^u_A = 2\left(-\frac{1}{2}+2\sin^2\theta_W\right)
                          \left(\frac{1}{2}\right)
          = - \frac{1}{2} + 2 \sin^2\theta_{W}~,\label{eq:c1dfactorized}\\
 C_{1d} &=& 2g^e_A g^d_V = 2\left(-\frac{1}{2}\right)
                          \left(-\frac{1}{2}+\frac{2}{3}\sin^2\theta_W\right)
          = \frac{1}{2} - \frac{2}{3} \sin^2\theta_{W}~,\label{eq:c2ufactorized}\\
 C_{2d} &=& 2g^e_V g^d_A = 2\left(-\frac{1}{2}+2\sin^2\theta_W\right)
                          \left(-\frac{1}{2}\right)
          =  \frac{1}{2} - 2 \sin^2\theta_{W}~.\label{eq:c2dfactorized}
\end{eqnarray} 

\begin{figure}[!htp]
 \begin{center}
 \includegraphics[width=0.2\textwidth]{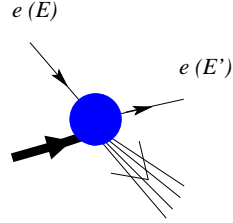}
 \end{center}
 \caption{Feynman diagram for contact interactions, used commonly to describe 
beyond-Standard-Model interactions.}\label{fig:escatt_contact}
\end{figure}

In Standard-Model-test experiments, new physics 
that can be accessed by PVES asymmetries typically cannot be described by the one-boson exchange of
Fig.~\ref{fig:escatt} and Eq.~(\ref{eq:c1factorized}-\ref{eq:c2dfactorized})
are no longer valid.  In this case, one writes~\cite{Erler:2013xha}
\begin{eqnarray}
 C_{1u} = g^{eu}_{AV}~,~&& 
 C_{2u} = g^{eu}_{VA}~,\label{eq:c1contact}\\
 C_{1d} = g^{ed}_{AV},~&&
 C_{2d} = g^{ed}_{VA},\label{eq:c2contact}
\end{eqnarray}
and the corresponding Feynman diagrams change from Fig.~{\ref{fig:escatt}} to
Fig.~\ref{fig:escatt_contact}. 
The $C_{1q},C_{2q}$ couplings therefore provide information on new contact
interactions beyond the Standard Model.
Note that even though $C_{1,2}$ cannot be factorized into an electron and 
a target vertex, their chiral property remains the same.

The formalism of inelastic PV asymmetries, Eq.~(\ref{eq:Apvdis1}), can be simplified 
as follows:
Defining $q_i^\pm(x)\equiv q_i(x)\pm\bar q_i(x)$, one has in the QPM 
\begin{eqnarray}
  a_1(x)&=&2\frac{\sum{C_{1i}Q_{q_i}q_i^+(x)}}{\sum{Q_{q_i}^2q_i^+(x)}}~,\label{eq:a1}\\
  a_3(x)&=&2\frac{\sum{C_{2i}Q_{q_i}q_i^-(x)}}{\sum{Q_{q_i}^2q_i^+(x)}}~.\label{eq:a3}
\end{eqnarray}
For an isoscalar target such as the deuteron, neglecting
effects from charm and bottom quarks, and assuming $s=\bar s$, $c=\bar c$ 
and the isospin symmetry 
that $u^p=d^n$, $d^p=u^n$ [$u,d^{p(n)}$ are the up and down quark PDF
in the proton (neutron)], the functions 
$a_{1,3}(x)$ simplify to
\begin{eqnarray}
  a_1(x)&=&\frac{6\left[2C_{1u}(1+R_C)-C_{1d}(1+R_S)\right]}{5+R_S+4R_C}~,\label{eq:a1d}\\
  a_3(x)&=&\frac{6\left(2C_{2u}-C_{2d}\right)R_V}{5+R_S+4R_C}~,\label{eq:a3d}
\end{eqnarray}
where 
\begin{eqnarray}
 R_C\equiv\frac{2(c+\bar c)}{u+\bar u+d+\bar d},~~
 R_S\equiv\frac{2(s+\bar s)}{u+\bar u+d+\bar d}, 
~~\q{and}~~
 R_V\equiv\frac{u-\bar u+d-\bar d}{u+\bar u+d+\bar d}~. \label{eq:Rpdf}
\end{eqnarray}
The asymmetry then becomes 
\begin{widetext}
\begin {eqnarray}
 A_{PV} &=&\left(\frac{3G_FQ^2}{2\sqrt{2}\pi\alpha}\right)
 \frac{2C_{1u}[1+R_C(x)]-C_{1d}[1+R_S(x)]+Y_3(2C_{2u}-C_{2d})R_V(x)}{5+R_S(x)+4R_C(x)}~.
\label{eq:Apvdis_R}
\end{eqnarray}
\end{widetext}
The factor $Y_3R_V$ is therefore crucial in accessing the $C_{2q}$.

If one neglects sea quarks completely ($R_C=R_S=0$, $R_V=1$), the deuteron becomes 
equal amount of up and down valence quarks only (the ``valence quark only'' 
picture). In this case no PDF is needed: 
\begin{eqnarray}
  a_1(x)=\frac{6}{5}\left(2C_{1u}-C_{1d}\right)~,&&
  a_3(x)=\frac{6}{5}\left(2C_{2u}-C_{2d}\right)~,\label{eq:a1a3nopdf}
\end{eqnarray}
which lead to~\cite{Beringer:2012}
\begin{eqnarray}
  A_{PV} &=&\left(\frac{3G_FQ^2}{10\sqrt{2}\pi\alpha}\right)
 \left[(2C_{1u}-C_{1d})+Y_3(2C_{2u}-C_{2d})\right]~.\label{eq:Apvdis_pdg}
\end{eqnarray}
This expression can be used to estimate how the PDFs affect 
the interpretation of the asymmetry measurement.

\subsection{Previous Data on Electron-Quark VA Coupling}

\begin{figure}
 \begin{center}
  \includegraphics[width=0.9\textwidth]{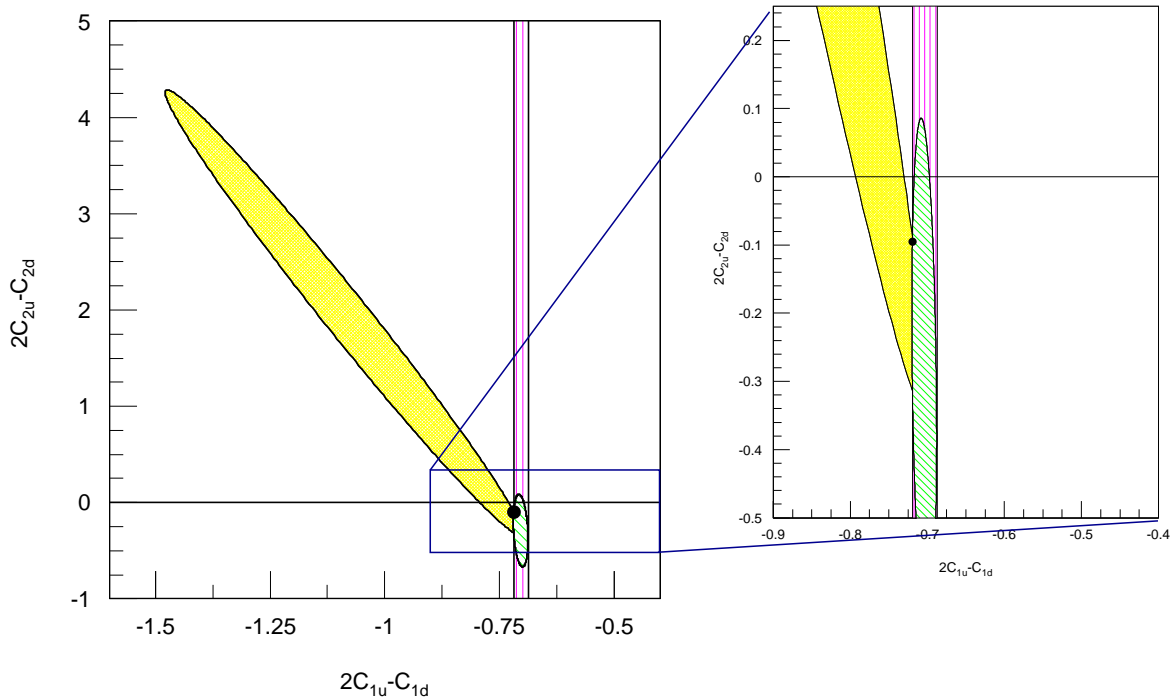}
 \end{center}
 \caption{{\it [Color online]} Previous data on $C_{2q}$. The yellow ellipse represents a 
simultaneous fit to $C_{1q}$ and $C_{2q}$ using only the 
SLAC E122 asymmetries~\cite{Prescott:1979dh} re-analyzed using the latest PDF fits 
(see Appendix~\ref{sec:app_e122}). 
The magenta vertical band represents the best $C_{1q}$ data~\cite{Androic:2013rhu}, 
and the green ellipse the combined fit of the E122 asymmetries and the 
best $C_{1q}$. The right panel shows an enlarged view with the vertical and the 
horizontal axis at the same scale. The Standard Model value is shown as the black dot, 
where the size of the dot is for visibility.}
\label{fig:c2q_before}
\end{figure}

The SLAC E122 experiment~\cite{Prescott:1978tm,Prescott:1979dh} 
was the only PVDIS measurement before the present experiment. 
During the E122 experiment, a longitudinally polarized electron beam was 
scattered from 30-cm long unpolarized proton and deuteron targets at $Q^2$
values ranging from 1.05 to 1.91 (GeV/$c$)$^2$. Four beam energies: 16.2, 17.8, 
19.4 and 22.2 GeV were used. Scattered electrons were collected in a 
magnetic spectrometer at 4$^\circ$ by integrating signals from a gas Cherenkov
detector. Data from the two highest beam energies
were published as~\cite{Prescott:1978tm}
$A_{PV}/Q^2 = (-9.5\pm 1.6)\times 10^{-5}$~(GeV/$c$)$^{-2}$.
The average $y$ value was 0.21 and the average $Q^2$ was 1.6~(GeV/$c$)$^2$. 
The value of $\sin^2\theta_W$ was extracted from the measured asymmetries. 
We re-analyzed the E122 kinematics~\cite{Prescott:1979dh}
using the latest PDF fits (see Appendix~\ref{sec:app_e122}) and extracted the 
coupling combination $2C_{2u}-C_{2d}$ and $2C_{1u}-C_{1d}$ from their asymmetry results. 
These results are shown as the yellow ellipse in Fig.~\ref{fig:c2q_before}. 
Also shown in Fig.~\ref{fig:c2q_before} is the most recent fit~\cite{Androic:2013rhu} 
to $C_{1q}$ data from all elastic PVES and Cs atomic parity 
violation experiments. 
One can see that the uncertainty on the $2C_{2u}-C_{2d}$ is nearly 
two orders of magnitude larger than on $2C_{1u}-C_{1d}$.

\newpage
\section{Apparatus}\label{sec:apparatus}

\begin{figure}[!ht]
\begin{center}
\includegraphics[width=0.7\textwidth]{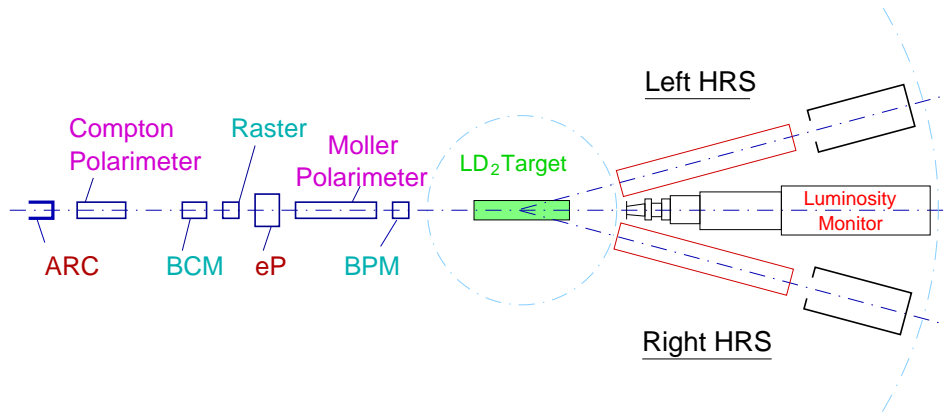}
\caption{Schematic floor plan of the 6~GeV PVDIS experiment in Hall A at JLab.  
The electron beam enters from the left, passes through a series of monitoring 
devices such as the ARC and the eP for energy measurement, Compton and M{\o}ller 
polarimeters for polarization measurement, the beam charge monitor (BCM) and
the beam position monitor (BPM), then scatters from a liquid D$_2$ target in the 
middle of the hall.  The scattered electrons were detected in the
HRS pair in inclusive mode.
}
\label{fig:floorplan}
\end{center}
\end{figure}

The experiment was performed in experimental Hall A 
at JLab.
The floor plan for Hall A is shown schematically in Fig.~\ref{fig:floorplan}.
A 105~$\mu$A longitudinally polarized electron beam was incident
on a 20-cm long liquid deuterium target, and scattered electrons were 
detected by the two
High Resolution Spectrometers (HRS)~\cite{Alcorn:2004sb} in inclusive mode.
A series of beam diagnostic devices was used to measure the beam 
energy, position, and current.
A Luminosity Monitor was located downstream from the target to monitor 
target density fluctuation and possible false asymmetries. 
For DIS measurements the beam energy used was 6 GeV, the highest 
achievable with the continuous electron beam accelerator facility (CEBAF) 
of JLab before its 12 GeV Upgrade.

The experimental techniques for measuring small asymmetries of order 
1 part per million (ppm) or less have been successfully used in  
the HAPPEx experiments~\cite{Aniol:2000at,Aniol:2004hp,Aniol:2005zf,Aniol:2005zg,Acha:2006my,Ahmed:2011vp} and the 
PREx~\cite{Abrahamyan:2012gp} experiment in JLab 
Hall A. These two experiments had maintained systematic 
uncertainties associated with beam helicity reversal at the $10^{-8}$ level.
The asymmetries sought for in this experiment were of order $10^2$ ppm with
required statistical accuracies at the $(3-4)\%$ level, which were 
two orders of magnitude 
larger than the systematic uncertainty established in the recent PVES experiments.
The main challenge of the experiment was a reliable rejection of the large pion
electro- and photo-production background (that is only present in inelastic scattering) 
while identifying electrons at high rates. 
While the standard HRS detector package and data acquisition (DAQ) system 
routinely provide high particle identification (PID) performance, 
they are based on full recording of the detector signals and are limited 
to event rates of 4 kHz. This is not sufficient for the few-hundred kHz 
rates expected for the present experiment. A new DAQ electronic system  
was built to count event rates up to 600~kHz with hardware-based 
particle identification. See Ref.~\cite{Subedi:2013jha} for a complete report 
on the DAQ design, its PID performance, deadtime effects, and the quality
of the asymmetry measurement. The standard DAQ of the HRS will be referred to 
as the HRS DAQ hereafter.

The apparatus and its effect on the measured asymmetry
are presented in this section.
The polarized electron beam will be described first 
(section~\ref{sec:app_electronbeam}), followed by
descriptions of the beam monitors (section~\ref{sec:beam_mon}),
the beam polarimetry (section~\ref{sec:polarimetry}), 
the target system (section~\ref{sec:target}),
and the spectrometers and detectors (section~\ref{sec:HRS}).

\subsection{Polarized Electron Beam}\label{sec:app_electronbeam}

The electron beam was produced from a strained superlattice GaAs/GaAsP 
photocathode illuminated by 
circularly polarized laser light~\cite{Sinclair2007}. 
The laser polarization is controlled by a Pockels cell. 
By reversing the high voltage on the Pockels cell, the sign of the laser circular 
polarization flips and the direction of the electron spin at the target is 
reversed every 33~ms~\cite{Paschke:2007zz}.
These 33-ms periods are called ``beam helicity windows'' or simply ``windows''.
Data collected in the first 0.5~ms of each window are rejected to 
allow the Pockels cell to settle. 
During this experiment, the helicity of the
electron beam was controlled by a helicity signal, and followed a 
“quartet” structure of either ``RLLR'' or ``LRRL'', with each state lasting
33~ms and the first state of each quartet selected from a 
pseudorandom sequence~\cite{Aniol:2004hp,Aniol:2005zf,Aniol:2005zg,Acha:2006my}. 
The helicity signal was sent to the data acquisition system after being 
delayed by eight helicity states (two quartets). 
This delayed helicity sequence controlled the data collection. 
The helicity signal was line-locked to the 60~Hz line,
thus ensuring a good cancellation of the power-line noise. 
 
To reduce possible systematic errors, a half-wave plate (HWP) was inserted
intermittently into the path of the polarized laser, which resulted in a 
reversal of the actual beam helicity while keeping the helicity signal sequence 
unchanged. Roughly equal statistics were 
accumulated with opposite HWP states for the measured asymmetry, 
which suppressed many systematic effects.  
The expected sign flips in the measured asymmetries between the two 
beam HWP configurations were observed.

The direction of the beam polarization could also be
controlled by a Wien filter and solenoidal lenses
near the injector~\cite{GramesWien2011}.  After accelerating, the beam was 
directed into Hall A, where its intensity, energy and trajectory on 
target were inferred from the response of several monitoring devices.
The beam monitors and the scattered electron trigger signals from the DAQ were integrated 
over the helicity window and digitized, from which raw cross-sectional asymmetries $A^\q{raw}$ 
were formed, see section~\ref{sec:Araw}. 
To keep spurious beam-induced asymmetries under control at well below 
the ppm level, careful attention was given to the design and 
configuration of the laser optics leading to the photocathode.
A specialized DAQ system (called the HAPPEx DAQ)~\cite{Aniol:2000at,Aniol:2004hp,Aniol:2005zf,Aniol:2005zg,Acha:2006my,Ahmed:2011vp} 
was used to provide feedback at the photocathode to minimize
these beam asymmetries~\cite{Paschke:2007zz}. 
Measurement of the polarization of the beam will be described
in section~\ref{sec:polarimetry} and the polarization results in section~\ref{sec:beam_pol}.

\subsection{Beam Monitoring and Rastering}\label{sec:beam_mon}

As a direct input to the asymmetry extraction, the beam intensity was measured by
two microwave cavity Beam Current Monitors (BCMs) and an Unser monitor
located 25~m upstream of the target~\cite{Alcorn:2004sb}. 
In addition, helicity correlations in the beam properties such as energy and position 
could add systematic uncertainties and widen the uncertainty of $A^\q{raw}$, 
and thus are a primary concern for parity-violation experiments.  
At JLab, the beam position is measured by ``stripline"
monitors~\cite{stripline}, each of which consists of a set of four thin wires
placed symmetrically around the beam pipe. The wires act as antennae
that provide a signal, modulated by the microwave structure of the
electron beam, that is proportional to the beam position as well as
intensity. Two such Beam Position Monitors (BPMs) are available in Hall A, located
7.524~m (BPMA) and 1.286~m (BPMB) upstream of the target center. Beam positions 
measured at BPMA and BPMB were extrapolated to provide the position and the 
incident angle at the target. An additional BPM (BPM12x) is available in the 
arc section of the beamline just before it enters the hall to monitor changes in 
the beam energy.

The electron beam at JLab has a nominal spot size of 100-200~$\mu$m 
(root-mean-square or rms value). 
To avoid over-heating the target, the beam is routinely moved at 20 kHz by a
rastering system consists of
two sets of steering magnets located 23~m upstream of the target. 
This fast rastering system can deliver beam with 
a uniform elliptical or rectangular distribution of
size between 100~$\mu$m and several mm at the target. A square distribution of approximately
$4\times 4$~mm$^2$ was used for this experiment.  
The exact correspondence between BPM signals and the actual beam position
at the target varies with beam energy and must be calibrated. 
In addition, the BPM information is not
fast enough to provide event-by-event information 
and the raster currents must be used to calculate real-time beam position 
on the target. 
Establishing the relation between BPM signals and beam positions, 
and between raster currents and the beam positions, is part of the 
BPM calibration described in section~\ref{sec:bpmcalib}.

\subsection{Beam Polarimetry}\label{sec:polarimetry}

Three beam polarimetry techniques were available for the present experiment: 
a Mott polarimeter in the injector of the linac, and a M{\o}ller
and a Compton polarimeter in Hall A. 
The Mott and the M{\o}ller measurements must be done separately 
from production data taking, while Compton measurements are non-intrusive. 
The Mott polarimeter~\cite{Price:1998xd,Price:1997qf,Price:1996up,Steigerwald-MottJLab} 
is located near the injector to the first linac where the
electrons have reached 5 MeV in energy. During the beam normal asymmetry 
$A_n$ measurement, it was used for setting up the transversely-polarized beam 
and verifying that the beam polarization was fully in the vertical direction. 
In the following we will describe the principle of only the M{\o}ller and Compton 
polarimeters.
For production runs, since the Mott polarimeter measures only the polarization 
at the injector which can differ from the beam polarization in the 
experimental hall, its results were not used directly in our analysis. 

\subsubsection{M{\o}ller Polarimeter}
\label{sec:moller_method}

A M{\o}ller polarimeter~\cite{Alcorn:2004sb} measures the beam polarization
via a measurement of the asymmetry in $\vec e-\vec e$ (M{\o}ller) scattering, 
which depends on the beam and target polarizations $P^{\rm beam}$ 
and $P^{\rm targ}_{\rm M{\o}ller}$, as well as on the M{\o}ller scattering
analyzing power $\ath_M$:
\begin{eqnarray}
\label{moller_asy}
 A_M = \sum_{i=X,Y,Z} 
  (A^\q{th}_{Mi}\cdot{}{P}^{\rm targ}_{i,\rm M{\o}ller}\cdot{}{P}^{\rm beam}_{i})~.
\end{eqnarray}
Here, $i = X,Y,Z$ defines the projections of the polarizations with 
$Z$ parallel to the beam and $OXZ$ the M{\o}ller scattering plane.
The analyzing powers $A^\q{th}_{Mi}$ depend on 
the scattering angle in the
$\vec e-\vec e$ center-of-mass (CM) frame, $\theta_{\rm CM}$, and are calculable in QED.
The longitudinal analyzing power is
\begin{eqnarray}
\label{eq:moller_apower}
\ath_{MZ} = - \frac{ \sin^2 \theta_{\rm CM} 
( 7 + \cos^2 \theta_{\rm CM}) }
{ {(3 + \cos^2 \theta_{\rm CM})}^2 }.
\end{eqnarray}
The absolute value of $\ath_{MZ}$ reaches a maximum of 7/9 
at $\theta_{\rm CM}=90^{\circ}$. 
At this angle the transverse analyzing powers are $\ath_{MX}=-\ath_{MY}=\ath_{MZ}/7$.

The M{\o}ller polarimeter target was a ferromagnetic foil
magnetized in a magnetic field of 24 mT~along its plane.
The target foil can be oriented at various angles
in the horizontal plane, providing both
longitudinal and transverse polarization
measurements.  The asymmetry was measured
at two target angles ($\pm 20^{\circ}$) 
and the average taken, which cancels contributions from 
transverse components of the beam spin and thus reduces
the uncertainties from target angle measurements.
At a given target angle, two sets of measurements
with oppositely-signed target polarizations 
were made which cancels some systematic effects
such as those from beam current asymmetries.  The M{\o}ller target
polarization was approximately 8\%.

The M{\o}ller-scattered electrons were
detected in a magnetic spectrometer 
consisting
of three quadrupoles and a dipole~\cite{Alcorn:2004sb}.
The spectrometer selects electrons in a range of
$75^{\circ} \leqslant \theta_{\rm CM} \leqslant
105^{\circ}$ and $-5^{\circ} \leqslant \phi_{\rm CM}
\leqslant 5^{\circ}$ where $\phi_{\rm CM}$ is
the azimuthal angle in the CM frame.  The detector consisted
of lead-glass calorimeter modules in two 
arms to detect the electrons in coincidence.
The M{\o}ller measurements must be performed separately from 
production runs, and each measurement takes approximately 4
hours including setting up the magnets to direct the electron
beam to the M{\o}ller target. The statistical uncertainty of the M{\o}ller
measurements is negligible compared to the approximately
2\% systematic error which is dominated by the uncertainty in
the foil polarization.

\subsubsection{Compton Polarimeter}
\label{sec:cpt_exp_meth}

The Compton polarimeter~\cite{Neyret:1999tr,Baylac:2002en,Alcorn:2004sb,Friend:2011qh}
is based on scattering of the polarized electron beam from
a polarized laser beam in a beam chicane. 
For this experiment, the beam polarization was extracted from the 
backscattered photon signals detected in a GSO (Gd$_2$SiO$_5$:Ce) 
crystal in the integrated mode~\cite{Friend:2011qh}.
Scattered electrons can be detected either in the inclusive mode or
in coincidence with the backscattered photons, but electron detection was not used
in this experiment.

The Compton asymmetry $A_C=(n^R_C-n^L_C)/(n^R_C+n^L_C)$ was measured, where
$n^R_C (n^L_C)$ refers to the scattered photon counting rate for right (left)
electron helicity normalized to the beam intensity. This asymmetry is
related to the electron beam polarization via
\begin{eqnarray}
P_e=\frac{A_C}{P_\gamma \ath_C}~,
\label{eq:a_expc}
\end{eqnarray}
where $P_\gamma$ is the photon polarization and $\ath_C$ the Compton analyzing
power.  At typical JLab energies (a few GeV), the Compton cross-section
asymmetry is only a few percent.
To compensate for the small asymmetry, a Fabry-Perot cavity~\cite{Jorda:1998gb} 
was used to amplify the photon density from a standard
low-power Nd:YaG laser ($\lambda=1064$~nm) such that high statistics can be obtained within one to a 
few hours. An average power of 1200~W
was accumulated inside the cavity with a photon beam waist of the
order of 150 $\mu$m and a photon polarization above 99\%, monitored
online at the exit of the cavity~\cite{Falletto:2000mu}. 
When extracting the beam polarization from Compton data, a GEANT4-based 
simulation~\cite{Parno:2012xa} was performed to reproduce the measured 
photon energy distribution and to extract the analyzing power. 
For the present experiment the systematic uncertainty of Compton measurement 
was approximately 1.92\% relative and was dominated by the understanding of the 
analyzing power (1.75\% relative) and the laser polarization (0.8\% relative).

\subsection{Target System}\label{sec:target}

The Hall A cryogenic target system~\cite{Alcorn:2004sb} was used for this 
experiment. We used a 20-cm long deuterium target cell 
for the main production data-taking. Solid 
targets were used for evaluating backgrounds, studying the
spectrometer optics, and checking beam centering.
The target cell and a solid target ladder 
sit in an evacuated cylindrical scattering chamber of 104~cm
diameter, centered on the pivot for the spectrometers. 
Also located inside the scattering chamber were subsystems for cooling,
temperature and pressure monitoring, target motion, gas-handling and controls.  
The scattering
chamber was maintained under a $10^{-6}$ Torr ($10^{-4}$ Pa) vacuum. The exit windows
on the scattering chamber allowed scattered particles to reach 
the spectrometers. These windows were made of 0.406-mm thick Al foil.

\begin{figure}[!htp]
 \begin{center}
 \includegraphics[width=0.4\textwidth]{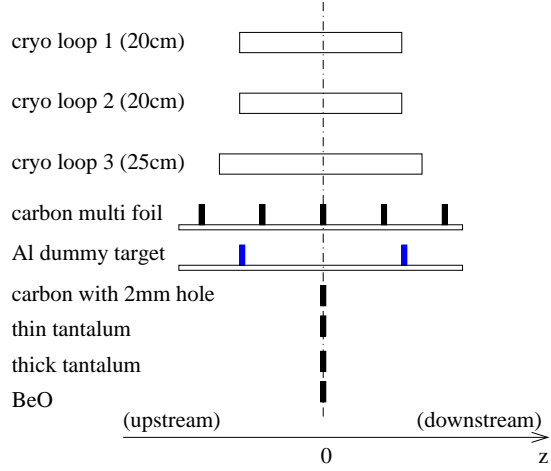}
\caption{Schematic diagram of the target ladder arrangement used during 
the experiment. The electron beam is along the horizontal direction 
(the $z$-axis) and is incident from the left on the target.
The carbon multi foils were located at $z=(-15, -7.5, 0, 7.5, 15)$~cm and 
the Al dummy foils were located at $z=(-10,10)$~cm. All other solid 
targets were located at $z=0$~cm and were about 1 inch apart in the 
vertical direction.}\label{fig:target_ladder}
 \end{center}
\end{figure}

Figure~\ref{fig:target_ladder} shows a schematic diagram of the target ladder
arrangement used during this experiment. Of the three cryogenic 
loops, only loop~1 was used for the liquid deuterium. 
It was operated at a temperature of 22~K and a
pressure of 25~psia ($1.7\times 10^{5}$ Pa), leading to a density of about 0.1676~g/cm$^3$. 
The diameter of the cell was 2.0~cm. The thicknesses of 
its walls and of the solid targets are summarized in 
Table~\ref{tab:target_data}.
\begin{widetext}
\begin{table}[!htp]
 \begin{center}
 \begin{tabular}{c|c|c|c}\hline
  Target & Position along $z$  & Purity & Thickness \\\hline
  cryo-loop 1$^a$
         & Entrance window,-10~cm &   & $0.126\pm 0.011\pm 0.003$~mm$^c$\\
         & Exit window, +10~cm   & & $0.100\pm 0.008\pm 0.003$~mm\\
     & Wall, beam left upstream  & & $0.313\pm 0.008\pm 0.003$~mm\\
     & Wall, beam left middle    & & $0.317\pm 0.002\pm 0.003$~mm\\
     & Wall, beam left downstream& & $0.323\pm 0.003\pm 0.003$~mm\\
     & Wall, beam right upstream & & $0.340\pm 0.002\pm 0.003$~mm\\
     & Wall, beam right middle   & & $0.336\pm 0.007\pm 0.003$~mm\\
     & Wall, beam right downstream&& $0.313\pm 0.008\pm 0.003$~mm\\\hline
  Carbon multi foil & (-15, -7.5, 0, 7.5, 15)~cm
                             &  99.5\% & $0.042\pm 0.001$ g/cm$^2$ (all foils)\\
  Al Dummy$^{a,b}$ & -10~cm &      & $0.359\pm 0.0003$ g/cm$^2$\\
               & +10~cm     &      & $0.367\pm 0.0003$ g/cm$^2$\\
  Carbon hole$^b$&0~cm    & 99.95\% & $0.08388\pm 0.00012$ g/cm$^2$\\
  Tantalum Thin  & 0~cm  & 99.9\%  & $0.021487\pm 0.000078$ g/cm$^2$\\
  Tantalum Thick & 0~cm  & 99.9\%  & $0.12237\pm 0.000341$ g/cm$^2$\\
  BeO            & 0~cm & 99.0\%  & $0.149\pm 0.001$ g/cm$^2$\\\hline
 \end{tabular}\\

 $^a$ All aluminum used for the cryo-target and the Al Dummy are made 
from Al 7075 T-6 plates.\\
 $^b$ Both Al Dummy and Carbon Hole targets had a 2-mm hole to calibrate
the target motion relative to the beam position. \\
 $^c$ The first error bar comes from the standard deviation of multiple 
measurements at different positions on the target, and the second error 
is from calibration of the instrument. 
\caption{Position, material, and thickness of the target system used in
this experiment. The position is defined along the
beam direction with respect to the hall center, see Fig.~\ref{fig:target_ladder}.}
\label{tab:target_data}
 \end{center}
\end{table}
\end{widetext}

When using a fluid target for electron scattering, the energy deposit of the
electron beam in the target can cause local density fluctuations. This will
add noise to the measurement that cannot be improved by increasing statistics. 
This systematic effect, often called the ``target boiling effect'' although 
it is not related to an actual phase change of the target, was measured 
at the beginning of the experiment for different beam transverse sizes and
target cooling conditions (see section~\ref{sec:boiling}). During production 
data taking, the 
transverse size of the beam was controlled such that the boiling effect
did not visibly widen the statistical uncertainty of the asymmetry measurement.

\subsection{Spectrometers, Detectors, and DAQ}\label{sec:HRS}

The Hall A high resolution spectrometers (HRS) are a pair
of identical spectrometers whose magnet system each consists of one dipole 
and three focusing quadrupoles in a $Q_1Q_2DQ_3$ sequence~\cite{Alcorn:2004sb}.
The spectrometer and their standard detector package served to select for 
and to measure the kinematics quantities $(E^\prime, \theta)$
while suppressing backgrounds originating from the target.
The spectrometers were designed to have a reasonable acceptance with excellent 
angle and momentum resolutions, high accuracy in the 
reconstructed kinematic variables of the events and 
precise normalization of the cross section.

\begin{figure}[!htp]
\begin{center}
 \includegraphics[width=0.7\textwidth]{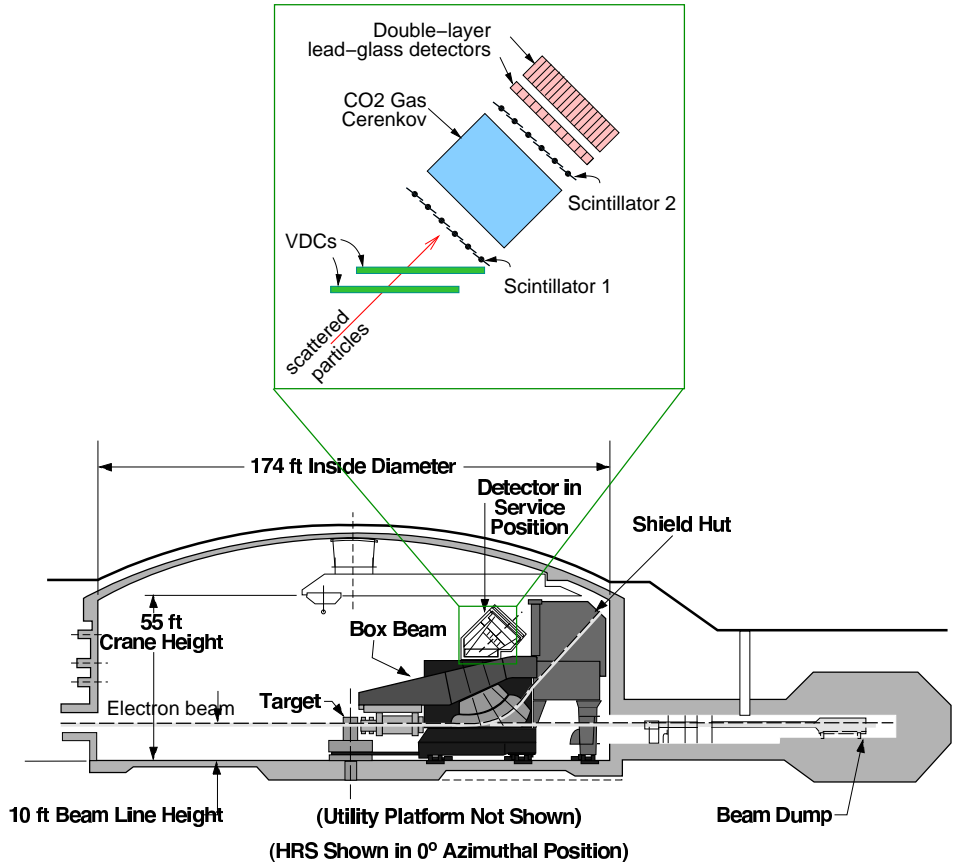}
 \caption{Bottom: Schematic diagram for the HRS in Hall A of JLab, figure taken
from Ref.~\cite{Alcorn:2004sb}. Top: Zoom-in view of the detector package in the HRS. }\label{fig:HallA}
\end{center}
\end{figure}

Figure~\ref{fig:HallA} shows a sideview of the HRS and its detector package. 
In each HRS, two layers of scintillators provide fast timing information
of the scattered particles, vertical drift chambers (VDCs) provide
tracking information, and a gas Cherenkov and a double-layered lead-glass
detector provide the particle identification (PID).

To achieve high resolution and accuracy in determining the event position, 
scattering angle and momentum, the HRS features an optics focusing system
that can be described as a simple matrix operation between the original
interaction point at the target $(x_{tg},y_{tg},\theta_{tg},\phi_{tg})$ 
(in the target coordinate system~\cite{Alcorn:2004sb,OpticsNote_Nilanga})
and the positions and angles of the particle detected at the focal plane 
$(x,\theta,y,\phi)$~\cite{Alcorn:2004sb,OpticsNote_Nilanga}, 
where the focal plane refers to the first of the four high-voltage wire planes
of the VDC. This optics matrix
varies with the beam energy and the spectrometer angle and momentum settings, 
and must be calibrated every time these conditions are changed.
The optics calibration directly affects the determination
of the $Q^2$-values of the present experiment and will be described in 
Sec.~\ref{sec:optics}.

The DAQ~\cite{Subedi:2013jha} of this experiment utilized signals from 
the two scintillator planes, 
the CO$_2$ gas Cherenkov counter and the double-layered lead glass 
detector. 
Both electron and pion triggers were formed. To better understand the counting 
deadtime of the DAQ, two sets of electronics were formed for each trigger, which
were expected to differ only in the deadtime. These two sets of triggers 
will be referred to as the ``narrow'' and the ``wide''
paths, with the narrow path exhibiting less deadtime loss.
The electron and pion triggers were sent to digital scalers 
where they were integrated over each helicity window of the electron beam.
The standard tracking detector (the VDCs) was turned off during production data 
taking because it might not endure the expected high event rates. 
During low-rate calibration runs, the VDCs
were turned on to study the efficiencies of the triggering detectors. 
Efficiencies of the electron and
pion triggers, the background contamination in each trigger, 
and the counting loss due to deadtime 
were analyzed in detail and reported in Ref.~\cite{Subedi:2013jha}.

\section{Data Analysis}
The experiment ran between October 26th and December 22nd, 2009. Data
were taken first with a 6-GeV beam at two DIS settings at $Q^2=1.085$
and $1.901$~(GeV/$c$)$^2$. These were the main
production kinematics and will be referred to as DIS\#1 and DIS\#2, 
respectively.   
Due to limitations in the spectrometer magnets, DIS\#1 was
taken only on the Left HRS, while DIS\#2 was taken on both Left and Right
HRSs.
A total of $1.02\times 10^7$ beam helicity pairs were selected to 
form the final electron sample for $Q^2=1.085$~(GeV/$c$)$^2$, and 
$2.5\times 10^7$ pairs for the $Q^2=1.901$~(GeV/$c$)$^2$ measurement.
The statistical precision achieved was 
3\% at $Q^2 = 1.1$~{(GeV/$c$)}$^2$ and 4\% at $Q^2 = 1.9$~{(GeV/$c$)}$^2$.
The systematic uncertainty achieved was smaller than $3\%$. 

Data were taken at five additional nucleon resonance settings to provide 
inputs for electromagnetic radiative corrections. Resonance setting IV was taken
with the 6 GeV beam on the left HRS, between data taking of DIS\#1 and \#2. 
Setting V was taken over a short period before IV due to difficulties in rotating
the HRS to the desired angle. It had low statistics and, with $W$ greater than 
2~GeV, was not strictly speaking in the resonance region. However we refer to it 
as setting RES V for convenience and present its result for completeness. 
Three more resonance settings (RES I, II and III) were taken with a 4.8 GeV beam 
at the end of the experiment, on either Left or Right HRS. 
For RES~I which was taken on the left HRS only, the $Q_1$ and the dipole magnets 
were set at 4.00~GeV/$c$, but its $Q_2$ and $Q_3$ were limited to 3.66~GeV/$c$ due to a power supply malfunction. 
Dedicated measurements for the beam transverse asymmetry -- also called
the normal asymmetry $A_n$ -- 
were carried out at DIS \#1 and \#2 in which the beam spin
was directed fully perpendicular to the scattering plane. 
An overview of the beam energy and spectrometer settings for each 
kinematics, the observed scattered electron rate 
and the ratio of $\pi^-/e$ rates 
is shown in Table~\ref{tab:kine_settings} 
in chronological order.

\begin{widetext}
\begin{table}[!htp]
 \begin{center}
 \begin{tabular}{c|c|c|c|c|c|c|c}\hline
  HRS &  Date      & Kine\# & $E_b$ (GeV) & $\theta_0$ & $E_0^\prime$ (GeV) 
  & $R_e$(kHz)  & $R_\pi/R_e$ \\\hline
  \multirow{7}{*}{Left} & 
    11/04-12/01/2009 & DIS\#1  & 6.0674 & $12.9^\circ$ & 3.66 & $\approx 210$&$\approx 0.5$ \\
  & 12/01-12/02/2009 & $A_n$   & 6.0674 & $12.9^\circ$  & 3.66& $\approx 210$&$\approx 0.5$ \\
  & 12/02/2009       & RES V   & 6.0674 & $14^\circ$   & 3.66 & $\approx 130$ & $<0.7$\\
  & 12/03/2009       & RES IV  & 6.0674 & $15^\circ$   & 3.66 & $\approx 80$ & $<0.6$\\
  & 12/04-12/17/2009 & DIS\#2  & 6.0674 & $20.0^\circ$ & 2.63 &$\approx 18$&$\approx 3.3$ \\
  & 12/17-12/19/2009 & RES I   & 4.8674 & $12.9^\circ$ & 4.0 & $\approx 300$ & $<0.25$ \\
  & 12/19-12/22/2009 & RES II  & 4.8674 & $12.9^\circ$ & 3.55 & $\approx 600$ & $<0.25$\\\hline

  \multirow{4}{*}{Right} & 
    11/04-12/01/2009 & DIS\#2  & 6.0674 & $20.0^\circ$ & 2.63 &$\approx 18$&$\approx 3.3$ \\
  & 12/01-12/02/2009 &  $A_n$  & 6.0674 & $20.0^\circ$ & 2.63 &$\approx 18$&$\approx 3.3$ \\
  & 12/02-12/17/2009 & DIS\#2  & 6.0674 & $20.0^\circ$ & 2.63 &$\approx 18$&$\approx 3.3$ \\
  & 12/17-12/22/2009 & RES III & 4.8674 & $12.9^\circ$ & 3.1 & $\approx 400$ & $<0.4$ \\\hline  
 \end{tabular}
 \caption{Overview of kinematics settings of this experiment and the observed
scattered electron rate $R_e$ and the charged pion to electron rate
ratio $R_\pi/R_e$. The kinematics include the beam energy $E_b$, and 
the spectrometer central angle $\theta_0$ and central momentum 
$E_0^\prime$. Measurement of the transverse asymmetry $A_n$
was performed at the production DIS settings on December 1-2. 
 For RES\#I which was taken on the left HRS only, the $Q_1$ and the dipole magnets 
were set at 4.00~GeV/$c$, but its $Q_2$ and $Q_3$ were limited to 3.66~GeV/$c$ due to a power supply malfunction. 
The electron rate $R_e$
was obtained directly from the DAQ, while the pion rate was the rate recorded by the 
DAQ corrected for trigger efficiency and background contamination.}
\label{tab:kine_settings}
 \end{center}
\end{table}
\end{widetext}

In this section the procedure for the data analysis will be described. 
The extraction of the raw asymmetries $A^\q{raw}$ from the DAQ count rates will be described first, 
followed by beam charge (intensity) normalization and its effect on the
measured asymmetry.
Then, corrections due to fluctuations in the beam position, angle 
and energy (section~\ref{sec:beamcorr}) are applied to extract the
beam-corrected raw asymmetries $A^\q{bc,raw}$. 
Results on the target boiling effect are presented next (section~\ref{sec:boiling}). 
Results on beam polarization are presented in section~\ref{sec:beam_pol}
which constitute a major normalization to the asymmetry, leading to the
preliminary physics asymmetry $A^\q{phys}_\q{prel.}$. 
Calibrations of the beam position and HRS optics are crucial 
for evaluation of the event kinematics (section~\ref{sec:optics}), 
and a full scale simulation of the HRS transport functions 
was carried out to confirm our understanding of the kinematics 
resulting from these calibrations (section~\ref{sec:ana_q2}).
Next, corrections to the preliminary physics asymmetries due to various 
backgrounds will be presented in detail (section~\ref{sec:ana_allbg}). 
Radiative corrections due to energy losses of the incident and the 
scattered electrons will be presented (section~\ref{sec:ana_radcor}), 
followed by corrections due to the higher-order $\gamma\gamma$ box
diagrams (section~\ref{sec:ana_box}). After all corrections are applied, the
preliminary physics asymmetries become the final physics asymmetry results
presented in section~\ref{sec:results_asym}.

\subsection{Forming Raw Asymmetries}\label{sec:Araw}
The scattered electrons and pions were counted by the DAQ for
each 33~ms helicity window. The response of each beam monitor, 
including the BCM and all BPMs, was digitized and integrated over the
same helicity windows and recorded. 
For each window pair $i$, the pair-wise raw electron 
cross-section asymmetry $A^\q{raw}_i$
in each HRS was computed from the the DAQ counts $c^{+(-)}_i$
normalized to the integrated beam intensity $I^{+(-)}_i$ 
in the positive (negative) helicity window:
\begin{eqnarray}
  A_i^\q{raw} &=& \left({\frac{{{c^+_i}\over{I^+_i}}-{{c^-_i}\over{I^-_i}}}
   {{{c^+_i}\over{I^+_i}}+{{c^-_i}\over{I^-_i}}}}\right)~.\label{eq:Araw_pair}
\end{eqnarray}
If the noise from beam fluctuations and the target boiling effect is negligible, the
uncertainty is given by the purely statistical value:
\begin{eqnarray}
  \delta A_{i,\q{stat}}^\q{raw} 
   &=& \sqrt{1\over{c^+_i+c^-_i}}~.\label{eq:dAraw_pair}
\end{eqnarray}
If a total of $n$ window pairs have been collected, the average raw 
asymmetry $A^\q{raw}$ was formed by
\begin{eqnarray}
 A^\q{raw}=\langle A_{i}^\q{raw}\rangle \equiv
  \frac{\sum_{i=1}^n {A_{i}^\q{raw}/(\delta A_{i,\q{stat}}^\q{raw})^2}}
  {\sum_{i=1}^n 1/(\delta A_{i,\q{stat}}^\q{raw})^2}~,
\end{eqnarray}
and its statistical uncertainty is 
\begin{eqnarray}
  \delta A_\q{stat}^{\q{raw}} &=& \sqrt{1\over{N^++N^-}}
 \approx \frac{\delta A_{i,\q{stat}}^\q{raw}}{\sqrt{n}}~,\label{eq:dAraw_Npairs}
\end{eqnarray}
where $N^{\pm}=\sum_{i=1}^n c_i^\pm$ refer to the total electron counts from 
the $n$ window pairs and the approximation is valid if the beam current
remains constant during the data taking.

When forming raw asymmetries, loose requirements were imposed on the
beam quality: 
periods with low beam current or with the energy measured in BPM12x 
differing by more than $10\sigma$ from its nominal value were rejected, 
removing about 10\%\ of the total data sample. 
No beam-helicity-dependent cuts were applied.
The uncertainty in $A^{\q{raw}}$ could be enlarged 
by helicity-dependent fluctuations in the beam intensity, position, angle, 
energy, and target boiling, causing a non-statistical contribution to the measurement. 
Therefore, 
an important criterion for a successful asymmetry measurement is 
to control non-statistical noise to a negligible level, which 
ensures that the main source of the uncertainty is 
the well-understood statistical fluctuation, and minimizes the run time.

\subsection{Beam Intensity Normalization, Beam Corrections, and Their Systematic Fluctuations}
\label{sec:beamcorr}
For all PVES experiments at JLab, the polarized beam and the target were 
designed such that the fluctuations in the helicity difference in 
the signal between a pair of successive windows were
dominated by scattered electron counting 
statistics. 
An example of possible non-statistical contributions is a window-to-window 
relative beam intensity asymmetry $A_I\equiv(I^+-I^-)/(I^++I^-)$ 
with an uncertainty $\delta(A_{I})$. 
During the PVDIS experiment, $A_I$ for a 30~ms beam window at a 
100~$\mu$A current was measured to be below $4\times 10^{-5}$, with $\delta A_I$ 
between $2\times 10^{-4}$ and $2\times 10^{-3}$ depending on the 
quality of the laser and the beam tune. At a 1-MHz 
counting rate the counting statistics for each 66-ms beam helicity pair 
is $\delta (A^\q{raw}_i)=0.00387$ [Eq.~(\ref{eq:dAraw_pair})]. The 
actual value was larger because the rate was lower than 1~MHz (Table~\ref{tab:kine_settings}). 
Therefore, the small 
$\delta(\asy_I)$ of the polarized beam at JLab guaranteed 
$\delta(\asy_I)\ll \delta A_{i,\q{stat}}^\q{raw}$ for this experiment.
Thanks to the feedback control to the laser at the polarized source, 
the cumulative average for $A_I$ throughout the experiment 
was below 0.1 ppm. 

Beam properties other than the intensity do not enter the direct asymmetry evaluation, 
but they might affect the asymmetry measurement. To study how such beam properties
affect the measured asymmetry, we first write Eq.~(\ref{eq:Araw_pair}) as
\begin{eqnarray}
  A_{i}^\q{raw} &\approx& \left(\frac{c^+-c^-}{c^++c^-}\right)_i 
  - \left(\frac{I^+-I^-}{I^++I^-}\right)_i
 = A_{i,c}^\q{raw} - \left(\frac{1}{I^++I^-}\right)_i \Delta I_i~,
\end{eqnarray}
where $A_{i,c}^\q{raw}$ is the raw count asymmetry and $\Delta I_i\equiv (I^+-I^-)_i$. 
This approximation is valid for $A_I\ll 1$ which was true as stated in the previous paragraph. 
Similarly, the raw asymmetry might be affected by 
fluctuations in beam energy, position and angle. These beam-related corrections (bc) can be
parametrized as:
\begin{eqnarray}
(A^\q{bc,raw})_i = A_i^\q{raw}
-\sum_j{\left[ {\alpha_j(\Delta X_j)_i}\right]}.
\end{eqnarray}
Here, $X_j$ denote beam parameters such as energy, position and
angle, $\Delta X_j\equiv X_j^+-X_j^-$ their corresponding helicity
fluctuation, and $\alpha_j$ their 
coefficients that depend on the kinematics of the specific
reaction being studied as well as the detailed spectrometer and
detector geometry of the experiment.

The five BPMs equipped during
this experiment: BPMA-X (horizontal), BPMA-Y (vertical), BPMB-X,Y,
and BPM12x allowed measurements of the relative change in the beam 
energy, position and angle within one helicity window pair. 
One can then write
\begin{eqnarray} 
(A^\q{bc,raw})_i = A_i^\q{raw}
-\sum_j{\left[ {c_j(\Delta x_j)_i}\right]}~,~\label{eq:Acorr}
\end{eqnarray}
where $x_j$ is the beam position measured by the five BPMs 
(BPMAX,Y, BPMBX,Y, BPM12x) and $c_j \equiv \alpha_j\partial X_j/\partial x_j$.
It is worth noting that this approach of making
corrections window by window automatically accounts for occasional
random instabilities in the accelerator.%

If one corrects the pair-wise asymmetry for the beam fluctuations based 
on Eq.~(\ref{eq:Acorr}), the resulting asymmetry averaged over a certain 
number of helicity pairs can be written as
\begin{eqnarray}
 A^\q{bc,raw} &\equiv& \langle A_i^\q{bc,raw}\rangle
 = \langle A_i^\q{raw}\rangle -\sum_j{c_j\langle \Delta x_j\rangle}
 = A^\q{raw} - \sum_j \Delta\asy_{x_j} 
\label{eq:Acorr_ditreg}
\end{eqnarray}
where 
$\Delta\asy_{x_j}\equiv c_j\langle (x_j^+-x_j^-)_i\rangle$ 
represents the correction needs to be applied
to the raw asymmetry due to helicity-dependent fluctuation in $x_j$.

For this experiment, the values of $c_j$ were obtained using two methods: The first one is
called the ``dithering'' method~\cite{Aniol:2004hp}, in which the beam 
position, angle, 
and energy were modulated periodically during data taking. The values of $c_j$ 
were then calculated from the 
resulting variation in the measured asymmetry recorded for each 
of the five BPM variables.
The energy of the beam was varied by applying a control voltage to a 
vernier input on a cavity in the accelerator's South Linac. The beam 
positions and angles were modulated using seven air-core corrector coils 
in the Hall A beamline upstream of the dispersive arc~\cite{Aniol:2004hp}. 
Because these modulation periods represent quality data, they were included
in the production data sample with the appropriate corrections made. 
In the second method the values of $c_j$ were evaluated utilizing only natural fluctuations
of the beam position, angle, and energy. This is called the ``regression'' method.  The
difference in the corrected asymmetry between the dithering and the 
regression method was used as the uncertainty in the beam-corrected raw asymmetries 
$A^\mathrm{bc,raw}$.

To control the beam position differences at BPMA and BPMB, 
the feedback system controlled
by the HAPPEx DAQ made adjustments of the circular 
polarization of the laser beam. The resulting beam position differences were
in the range $(0.01-0.1) \mu$m at the target for the majority of the data taking period. Based on the measured $c_j$
values this resulted in $\Delta A_{x_j}$ in the range $(0.1-1)$~ppm. The cumulative
averages for $\Delta A_{x_j}$ were found to be below 0.1~ppm integrated over the whole experiment.
The measured asymmetry was found to be much less sensitive to beam energy fluctuations
than to those of the beam position. 
Table~\ref{tab:beamcorrerr} shows the corrections due to fluctuations in 
the five measured beam positions using the dithering method. 
The beam-corrected asymmetries based on both the dithering and regression 
methods, $A^\q{bc,raw}_\q{dit}$ and $A^\q{bc,raw}_\q{reg}$, are shown in Table~\ref{tab:beamcorr}.
The narrow and the wide paths of the DAQ produced very similar results, with 
slight differences in their event collection due to DAQ deadtime and
different timing alignment between electronic modules, resulting in a slightly 
better PID performance of the wide-paths~\cite{Subedi:2013jha}. In addition, 
dithering and regression methods are in principle equivalent. Still, 
the narrow-path asymmetry results with the beam corrections applied using the
dithering method were used to produce the physics results of the present
experiment because of the smaller deadtime. 

\begin{widetext}
\begin{table}[!ht]
  \begin{center}
    \begin{tabular}{|c|c|c|c|c|c|c|}
      \hline\hline
      Monitor      & \multicolumn{2}{c|}{Left DIS\#1}        &   \multicolumn{2}{c|}{Left DIS\#2}        &    \multicolumn{2}{c|}{Right DIS\# 2} \\ 
                   & \multicolumn{2}{c|}{$\Delta A_\q{dit}$ (ppm)}   &   \multicolumn{2}{c|}{$\Delta A_\q{dit}$ (ppm)}   &   \multicolumn{2}{c|}{$\Delta A_\q{dit}$ (ppm)} \\ \hline
   DAQ path& narrow & wide   & narrow & wide  & narrow & wide\\ \hline
   BPM4AX  &  0.173 &  0.179 &  0.513 & 0.569 & -0.172 & -0.182\\
   BPM4AY  &  0.001 & -0.010 &  0.286 & 0.262 & -0.021 & -0.027\\
   BPM4BX  & -0.152 & -0.159 & -0.368 &-0.430 &  0.226 &  0.237 \\
   BPM4BY  & -0.028 & -0.020 & -0.262 &-0.243 & -0.008 & -0.003 \\
   BPM12x  &  0.000 &  0.000 &  0.024 & 0.022 & -0.003 & -0.003  \\ \hline
   Total   & -0.006 & -0.010 &  0.193 & 0.180 &  0.022 &  0.022 \\ \hline
     \hline
    \end{tabular}
    \begin{tabular}{|c|c|c|c|c|c|c|c|c|c|c|}
      \hline\hline
      Monitor      & \multicolumn{2}{c|}{RES I}        &   \multicolumn{2}{c|}{RES II}        &    \multicolumn{2}{c|}{RES III}        &    \multicolumn{2}{c|}{RES IV}        &    \multicolumn{2}{c|}{RES V} \\ 
                   & \multicolumn{2}{c|}{$\Delta A_\q{dit}$ (ppm)}   &   \multicolumn{2}{c|}{$\Delta A_\q{dit}$ (ppm)}   & \multicolumn{2}{c|}{$\Delta A_\q{dit}$ (ppm)}   &   \multicolumn{2}{c|}{$\Delta A_\q{dit}$ (ppm)}   &   \multicolumn{2}{c|}{$\Delta A_\q{dit}$ (ppm)} \\ \hline
   DAQ path& narrow & wide   & narrow & wide  & narrow & wide
           & narrow & wide   & narrow & wide  \\ \hline
   BPM4AX  & -0.175 & -0.178 &  0.313 & 0.320 & -0.013 &  0.000 &-1.004 &-1.192 &-3.708 &-3.631\\
   BPM4AY  &  0.230 &  0.224 &  0.096 & 0.107 &  0.047 &  0.046 & 0.328 & 0.328 & 0.400 & 0.317\\
   BPM4BX  &  0.369 &  0.375 & -0.568 &-0.582 &  0.020 & -0.005 & 1.398 & 1.596 & 4.754 & 4.603\\
   BPM4BY  & -0.139 & -0.133 & -0.132 &-0.143 & -0.038 & -0.037 &-0.235 &-0.250 &-0.265 &-0.183\\
   BPM12x  & -0.010 & -0.011 &  0.045 & 0.045 & -0.005 & -0.005 & 0.002 & 0.003 &-0.035 &-0.036\\ \hline
   Total   &  0.275 &  0.277 & -0.246 &-0.253 &  0.011 & -0.001 & 0.489 & 0.485 & 1.146 & 1.070\\ \hline
     \hline
    \end{tabular}
 \end{center}
\caption{Corrections to DIS (top) and resonance (bottom) 
asymmetries evaluated using the dithering method, 
$\Delta A_\q{dit}$. The ``narrow'' and ``wide'' refer to the DAQ trigger 
type~\cite{Subedi:2013jha}. 
The corrections were applied as 
$A^\q{bc,raw}_\q{dit}=A^\q{raw}-\Delta A_\q{dit}$ [Eq.~(\ref{eq:Acorr_ditreg})]. 
}
\label{tab:beamcorrerr}
\end{table}
\end{widetext}
\begin{widetext}
\begin{table}[!ht]
\begin{center}
\begin{tabular}{|c|c|c|c|}
\hline\hline
          & Left DIS\#1       & Left DIS\#2        & Right DIS\#2     \\ \hline
$A^\q{raw}$, narrow (ppm) 
          & $-78.4\pm 2.7$  & $-140.5\pm 10.4$ & $-139.9\pm 6.6$\\ 
$A^\q{bc,raw}_\q{dit}$, narrow (ppm)
          & $-78.5\pm 2.7$  & $-140.3\pm 10.4$ & $-139.8\pm 6.6$\\ 
$A^\q{bc,raw}_\q{reg}$, narrow (ppm)
          & $-78.5\pm 2.7$  & $-140.5\pm 10.4$ & $-140.3\pm 6.6$\\ 
$\vert A^\q{bc,raw}_\q{dit}-A^\q{bc,raw}_\q{reg}\vert$, narrow (ppm)
          & 0.1 & 0.2 & 0.5 \\\hline
$A^\q{raw}$, wide (ppm) 
          & $-78.2\pm 2.7$  & $-140.3\pm 10.4$ & $-140.9\pm 6.6$\\ 
$A^\q{bc,raw}_\q{dit}$, wide (ppm)
          & $-78.3\pm 2.7$  & $-140.1\pm 10.4$ & $-140.9\pm 6.6$\\ 
$A^\q{bc,raw}_\q{reg}$, wide (ppm)
          & $-78.3\pm 2.7$  & $-140.3\pm 10.4$ & $-141.4\pm 6.6$\\  
$\vert A^\q{bc,raw}_\q{dit}-A^\q{bc,raw}_\q{reg}\vert$, wide (ppm)
          & 0.1 & 0.1 & 0.5 \\\hline\hline
\hline
\end{tabular}
\begin{tabular}{|c|c|c|c|c|c|}
\hline\hline
          & Left RES I       & Left RES II        & Right RES III     & Left RES IV & Left RES V\\ \hline
$A^\q{raw}$, narrow (ppm) 
          & $-55.4\pm 6.8$  & $-63.5\pm 5.9$ & $-54.4\pm 4.5$ & $-104.5\pm 15.3$ & $-69.0\pm 21.3$\\ 
$A^\q{bc,raw}_\q{dit}$, narrow (ppm)
          & $-55.1\pm 6.8$  & $-63.8\pm 5.9$ & $-54.4\pm 4.5$ & $-104.0\pm 15.3$ & $-67.9\pm 21.3$\\ 
$A^\q{bc,raw}_\q{reg}$, narrow (ppm)
          & $-55.2\pm 6.8$  & $-63.6\pm 5.9$ & $-54.6\pm 4.5$ & $-104.3\pm 15.3$ & $-68.6\pm 21.2$\\
$\vert A^\q{bc,raw}_\q{dit}-A^\q{bc,raw}_\q{reg}\vert$, narrow (ppm)
          & $0.1$  & $0.2$ & $0.2$ & $0.3$ & $0.7$\\ 
 \hline
$A^\q{raw}$, wide (ppm) 
          & $-54.9\pm 6.8$  & $-63.6\pm 5.9$ & $-54.0\pm 4.5$ & $-105.0\pm 15.3$ & $-69.0\pm 21.5$\\ 
$A^\q{bc,raw}_\q{dit}$, wide (ppm)
          & $-54.6\pm 6.8$ & $-63.9\pm 5.9$ & $-54.0\pm 4.5$ & $-104.6\pm 15.3$ & $ -67.9\pm 21.5$ \\
$A^\q{bc,raw}_\q{reg}$, wide (ppm)
          & $-54.6\pm 6.8$ & $-63.7\pm 5.9$ & $-54.2\pm 4.5$ & $-104.9\pm 15.2$ & $ -68.7\pm 21.4$ \\
$\vert A^\q{bc,raw}_\q{dit}-A^\q{bc,raw}_\q{reg}\vert$, wide (ppm)
          & $0.1$ & $0.2$ & $0.2$ & $0.3$ & $0.8$\\ 
\hline
\hline
\end{tabular}
\end{center}
\caption{Measured raw asymmetries from the narrow and the wide 
triggers after applying corrections from 
beam energy and position changes using
the dithering and the regression methods. The asymmetry errors shown are 
statistical only. 
The differences between the two corrected asymmetries, 
$\vert A^\q{bc,raw}_\q{dit}-A^\q{bc,raw}_\q{reg}\vert$,
were used as the uncertainty from beam corrections. 
The dithering-corrected asymmetries were used in further analysis, 
although dithering and regression methods are in principle equivalent. 
The narrow and the wide paths of the DAQ produced very similar results, with 
slight differences in their event collection due to DAQ deadtime and
different timing alignment between electronic modules. The narrow-path
asymmetry results ($A^\q{bc,raw}_\q{dit}$, narrow) were used in 
further analysis to produce the physics results 
because of their smaller deadtime~\cite{Subedi:2013jha}.
%
}\label{tab:beamcorr}
\end{table}
\end{widetext}

Compared to the uncertainties from counting statistics, one can see that 
overall the corrections due to beam fluctuation 
were quite small, and their uncertainties are negligible.
The asymmetry measurement was completely dominated by the counting statistics 
of the scattered electrons~\cite{Subedi:2013jha}.

\subsection{Target boiling effect on the measured asymmetry}\label{sec:boiling}

As described in section~\ref{sec:target}, the electron beam deposited energy
in the liquid deuterium target and caused additional noise to 
the measurement. 
This target boiling effect would manifest 
itself as an increase in the 
standard deviation of the measured pair-wise asymmetry $A^\q{raw}$ above
that expected from the counting statistics of 
Eq.~(\ref{eq:dAraw_pair},\ref{eq:dAraw_Npairs}).
Rastering the beam to larger transverse sizes reduces the beam heating and thus
the boiling effect. 

\begin{figure}[!htp]
 \begin{center}
 \includegraphics[width=0.45\textwidth]{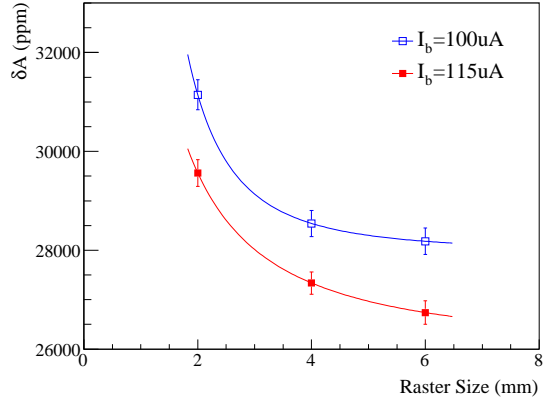}
 \caption{Measured standard deviation of the pair-wise asymmetries at 
kinematics DIS\# 2, for various raster sizes and two beam currents $100$ 
and $115$~$\mu$A. 
The curves show the results of the fit $\delta A = p_0 x^{p_1}+p_2$ where $x$ is the 
raster size in mm. The parameter $p_2$ represents the purely statistical 
fluctuation that depends only on the beam current and not the raster size, 
while the term $p_0 x^{p_1}$ is an empirical term that describes the size 
of target boiling. 
The fit results for 100~$\mu$A are
 $p_0=(1.77\pm 1.94)\times 10^4$, $p_1=-2.48\pm 1.85$, $p_2=27973.0 \pm 681.7$;
and for 115~$\mu$A are 
 $p_0=(9.40\pm 3.78)\times 10^3$, $p_1=-1.37\pm 1.09$, $p_2=25941.0\pm 1433.4$.
At a raster size of $4\times 4$~mm$^2$ ($x=4$), the boiling noise is 
at the level of $569$~ppm for 100~$\mu$A and $1407$~ppm for 115~$\mu$A, 
and is negligible compared to the value from purely statistical fluctuations. 
}\label{fig:targ_boiling_raster}
 \end{center}
\end{figure}

Studies of the target 
boiling effect was performed. 
For each measurement a Gaussian was fitted to the distribution of the pair-wise asymmetries 
with $\delta A$ given by the fitted width. 
Figure~\ref{fig:targ_boiling_raster} shows the measured $\delta A$, 
taken at kinematics DIS \#2 for various raster sizes at two beam currents 
100 and 115~$\mu$A.

Results of $\delta A$ in Fig.~\ref{fig:targ_boiling_raster} were 
fitted with the functional form 
$p_0 x^{p_1}+p_2$ where $x$ is the raster size in mm. 
The parameter $p_2$ represents the purely statistical 
fluctuation that depends only on the beam current, 
while the term $p_0 x^{p_1}$ is an empirical term that describes the size 
of target boiling. 
Using the approximate electron rate (Table~\ref{tab:kine_settings}), the purely statistical 
uncertainty for 66-ms wide beam helicity pairs is 0.029 at 100~$\mu$A and 0.027 at 115~$\mu$A. 
The fit results for $p_2$ agree with the expectation very well. 
The fit results for $p_0$ and $p_1$ show that the broadening due to boiling at a $4\times 4$~mm$^2$ 
raster size, $p_0 x^{p_1}$ with $x=4$, is at the level of $569$~ppm for 100~$\mu$A
and $1407$~ppm for 115~$\mu$A.
This is quite small compared to the value from purely statistical fluctuations ($p_0\sim 10^4$~ppm), 
and thus the boiling effect did not contribute significantly to the uncertainty of 
the asymmetry measurement.

\begin{figure}[!htp]
 \includegraphics[width=0.45\textwidth]{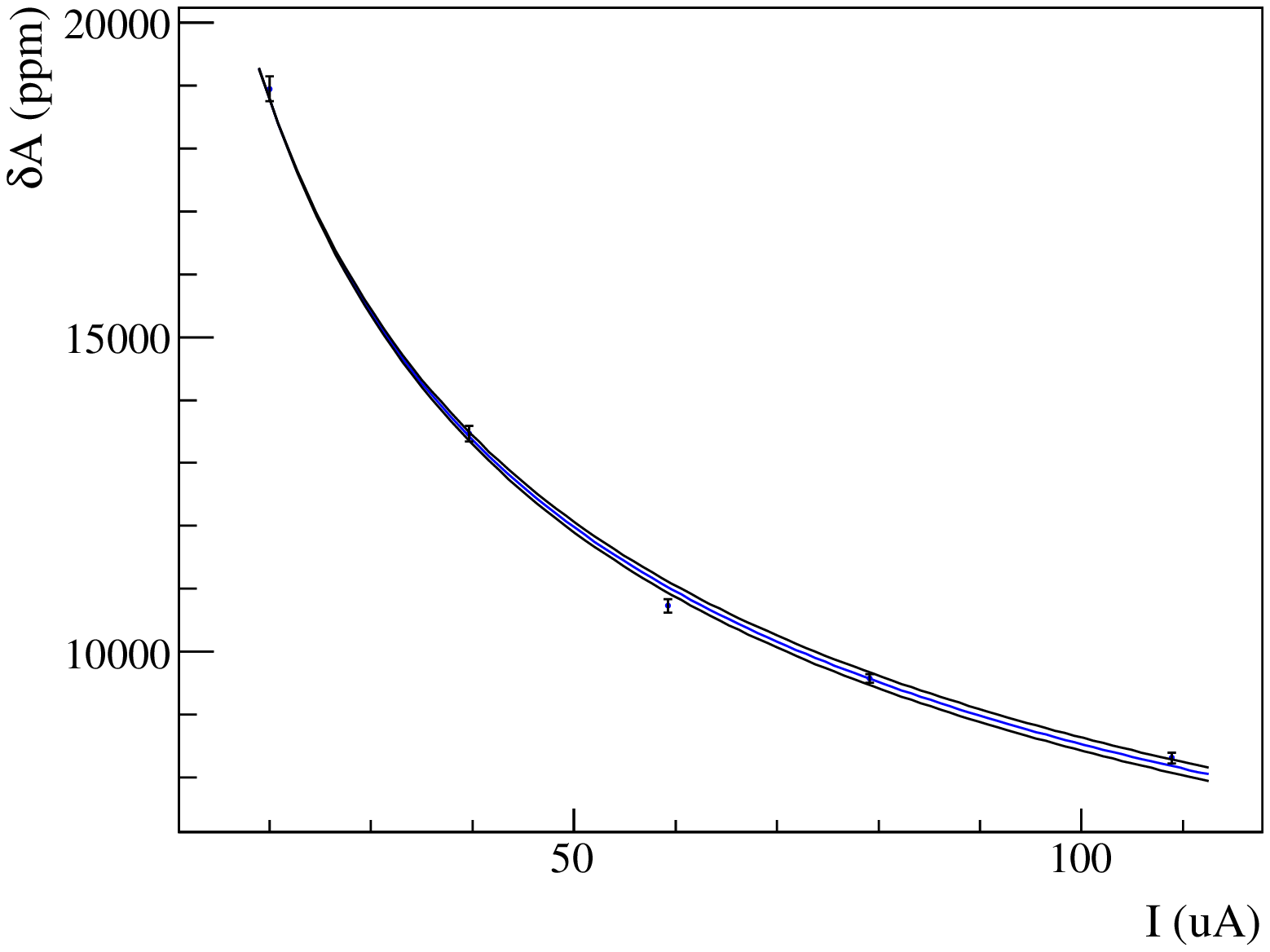}
 \includegraphics[width=0.45\textwidth]{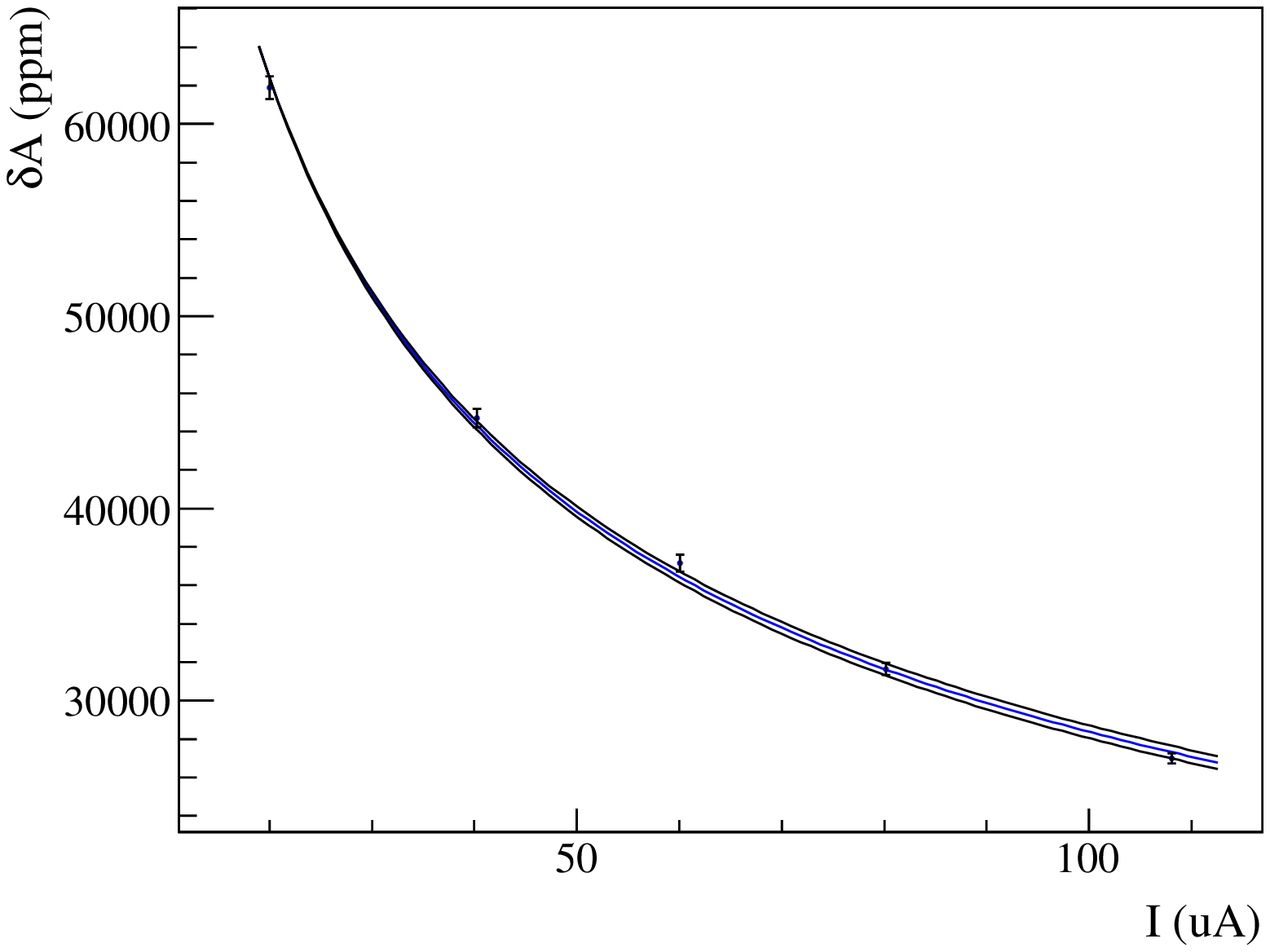}
 \caption{{\it [Color online]} Measured standard deviation of the pair-wise asymmetries at
various beam currents for DIS\# 1 (left) and \# 2 (right), 
with a $4\times 4$~mm$^2$ square raster. The curves 
show the results of the fit $\delta A\propto I^{p_3}$ and its error band. The fit
results are $p_3=0.4900\pm 0.0076$ and $p_3=0.4897\pm 0.0072$ 
for DIS\# 1 and \# 2 respectively. These results are in good agreement
with pure counting statistics ($\delta A\propto\sqrt{I}$). 
}\label{fig:targ_boiling_I}
\end{figure}

Figure~\ref{fig:targ_boiling_I} shows the measured $\delta A$ for
various beam currents $I$ performed with a $4\times 4$~mm$^2$ square raster. 
If the measurement is dominated by statistical uncertainty, one expects
$\delta A\propto\sqrt{I}$. Fit results of the measured $\delta A$ 
indeed agree very well with this expectation, indicating 
that boiling effects at the running condition of this experiment was negligible.

\subsection{Beam Polarization}\label{sec:beam_pol}

As described in the previous section, the electron raw asymmetry was first
corrected for the beam intensity and other beam-related properties such 
as position, angle and energy. The resulting asymmetry $A^\q{bc,raw}$ is
then referred to as the measured asymmetry, $A^\q{meas}$, and must be 
corrected for the beam polarization $P_e$:
\begin{eqnarray}
A^\q{phys}_\q{prel.}=A^\q{meas}/P_e~, \label{eq:a_exp}
\end{eqnarray}
to obtain the preliminary physics asymmetry $A^\q{phys}_\q{prel.}$. Both 
Compton and M{\o}ller polarimeters described in section~\ref{sec:polarimetry} 
were used.
 
\begin{figure}[!ht]
\centering
\includegraphics[width=0.5\textwidth]{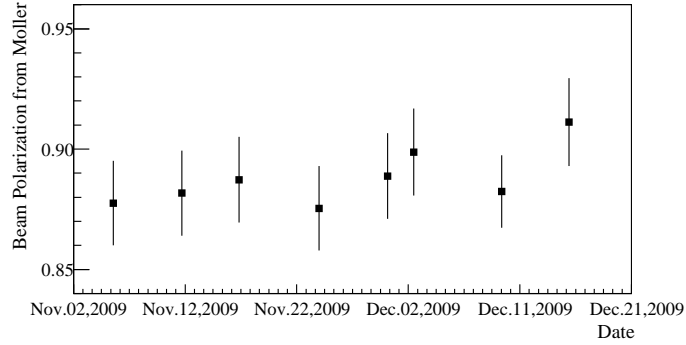}
\caption{Polarization results from the M{\o}ller polarimeter measurements
taken with a beam energy of 6.067~GeV. 
The error bars represent the quadratic sum of the statistical and systematic errors. 
However, for each measurement the statistical uncertainty was in the order of 0.1\%, much smaller than the systematic error. 
An additional measurement was
done with a beam energy of 4.867~GeV at the end of the run period, which 
gave a similar polarization.}
\label{fig:moller}
\end{figure}

\begin{figure}[!ht]
\centering
\includegraphics[width=0.5\textwidth]{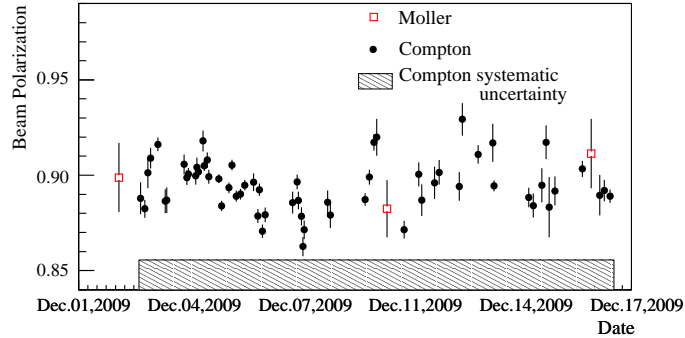}
\caption{Comparison between Compton (black solid circles) 
and M{\o}ller (red open squares) measurements taken during the time
period when both polarimeters were available. The beam energy was 6.067~GeV. 
The error bars for M{\o}ller represent the quadratic sum of the statistical and systematic errors, 
with the statistical error is smaller than the systematic by one order of magnitude. 
For Compton measurement, the statistical error are plotted with the data 
points and the systematic error ($1.92\%$ relative) are plotted along the horizontal axis. 
A constant fit to Compton measurements
gave an average of 89.45\% while the average of M{\o}ller results was $88.74\%$.}
\label{fig:compton}
\end{figure}

During our experiment, the M{\o}ller polarimeter was available 
the entire time, while the Compton polarimeter initially suffered from 
a high background and only produced results in the last three weeks of 
the 2-month 6-GeV run period.  The Compton polarimeter was also not 
available during the 4.8-GeV run period.  Figure~\ref{fig:moller} shows 
the M{\o}ller polarimetry measurements taken with the 6~GeV beam. 
During the three weeks when both polarimeters were functioning, 
the average beam polarization from constant fits is $88.74\%$ 
for M{\o}ller and $89.45\%$ for Compton. The results from the two
polarimeters are compared in Fig.~\ref{fig:compton}.  Note that the
beam polarization can fluctuate over time due to motion of the laser
position on the photocathode and photocathode aging.

The experimental asymmetries were corrected for the beam polarization as 
follows:
\begin{enumerate}
\item{When the Compton polarimeter was not available 
(before Dec. 2nd and after Dec. 17th, 2009), 
only M{\o}ller results were used. Each M{\o}ller result was used 
until the next measurement was available.}

\item{ When there were both Compton and M{\o}ller measurements (from 
Dec. 2nd to Dec. 17th, 2009), the Compton data were averaged first for the 
time interval between two M{\o}ller measurements, then was averaged with 
the corresponding M{\o}ller measurement from the beginning of the interval. 
The averages were weighted by the statistical error. 
The systematic uncertainty of the combined polarization was obtained from 
that of each method as 
\begin{eqnarray}
\left(\Delta P_b/P_b\right)_{syst, combined}
 =1/\sqrt{\left(\Delta P_b/P_b\right)_{syst, compton}^{-2}+\left(\Delta P_b/P_b\right)_{syst, moller}^{-2}}, 
\label{eq:Compton_Moller_syst_combined}\end{eqnarray}
thus was smaller than the systematic uncertainty of either polarimetry.
Each combined result was used until a next M{\o}ller measurement was available.}

\item {The beam polarization was corrected run by run for DIS\#1 and \#2. 
For resonance kinematics, the run period was short and a single correction 
was used for each kinematics.}
\end{enumerate}
The average beam polarization corrections are shown in 
Table~\ref{tab:polresults} for all kinematics.
\begin{widetext}
\begin{table}[!htp]
  \begin{center}
    \begin{tabular}{c|c|c|c}
      \hline
           & Left DIS\#2 & Right DIS\#2 & RES IV and V\\ \hline
     Combined $P_e$ (syst.) & $(89.29\pm 1.19)\%$ & $(88.73\pm 1.50)\%$ &$(89.65\pm 1.24)\%$ \\ 
      \hline
    \end{tabular}\\
  \medskip
    \begin{tabular}{c|c|c}
      \hline
           & Left DIS\#1  & RES I, II and III \\ \hline
      M{\o}ller $P_e$ (syst.) & $(88.18\pm 1.76)\%$  & $(90.40\pm 1.54)\%$\\ 
      \hline
    \end{tabular}
  \end{center}
\caption{Average beam polarization $P_e$ for each kinematics. 
These are either the combined
results of Compton and M{\o}ller measurements (top), or results from M{\o}ller alone (bottom),
depending on which polarimeter was available during the corresponding 
run period.  For DIS\#1 and \#2 the corrections were applied run-by-run 
and the statistically-averaged value of $P_e$ is shown. The uncertainties shown here 
are dominated by the systematic uncertainty, which for the combined results were
obtained using Eq.~(\ref{eq:Compton_Moller_syst_combined}). For all resonance 
kinematics which had short running period, a single value was used for
each setting.}
\label{tab:polresults}
\end{table}
\end{widetext}

\subsection{Calibration of the HRS Optics}\label{sec:optics}

To accurately determine the kinematics $(Q^2, x, W)$ of each event, one must
determine the absolute beam position on the target, and 
reconstruct the vertex position, the scattering 
angle and the scattered electron's momentum. 
These are provided by beam position calibration and the HRS
optics calibration, as described below. 

\subsubsection{Beam Position Calibration}\label{sec:bpmcalib}

\begin{figure}[!ht]
 \begin{center}
  \includegraphics[width=0.4\textwidth]{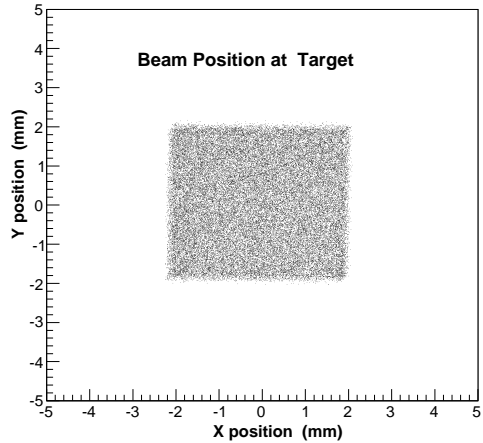}
 \end{center}
\caption{Calibrated beam spot distribution at the target.
}\label{fig:raster}
\end{figure}

As described in Sec.~\ref{sec:beam_mon}, the beam position information for each
event was obtained from the raster current rather than from the delayed
BPM information.
Calibrations between the raster current and the beam position thus became 
necessary. The BPM calibration can be described as:
\begin{eqnarray}
 \q{bpm~}~x &=&\langle\q{bpm~offset~}x\rangle+\langle\q{raster~current~}x\rangle\times\frac{\sigma_{\q{bpm},x}}{\sigma_\q{raster~current}}~,\\
 \q{bpm~}~y &=&\langle\q{bpm~offset~}y\rangle+\langle\q{raster~current~}y\rangle\times\frac{\sigma_{\q{bpm},y}}{\sigma_\q{raster~current}}~.
\end{eqnarray}
Figure~\ref{fig:raster} shows the 
beam spot distributions projected to the target using the calibrated 
BPMA and BPMB information. 

\subsubsection{Optics Calibration Procedure and the Resulting 
Uncertainties in $Q^2$ Determination}
The trajectory and momentum of each electron detected was determined
by calibration of the transport functions (optics) of each HRS. 
During optics calibration runs, the VDCs were turned on to provide precise 
information on the particle trajectory, from which the hit position 
and angles at the focal plane 
$(x,\theta,y,\phi)$ can be determined~\cite{Alcorn:2004sb,OpticsNote_Nilanga}.
The next step is to reconstruct the interaction position, angle, and 
momentum at the target from these focal plane variables, i.e., to 
determine the inverse of the HRS optical transport matrix. In practice, 
instead of a matrix operation, a set of tensors up to the 5th order
were used to calculate the target variables from the focal plane values. 

\begin{figure}[!htp]
 \begin{center}
  \includegraphics[width=0.45\textwidth]{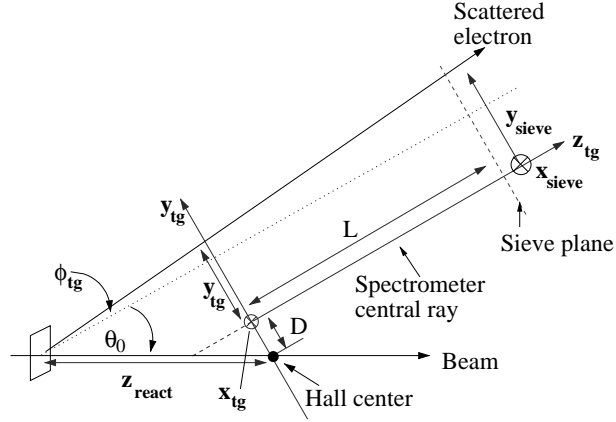}
 \end{center}
 \caption{Topview of the target coordinate system (TCS) $(x_{tg},y_{tg},z_{tg})$ and 
the sieve plane coordinate system $(x_\q{sieve},y_\q{sieve})$. 
The $z_{tg}$ axis is along the HRS central ray, the $y_{tg}$ axis is pointing to the left, the $x_{tg}$ axis is pointing vertically down, 
and the origin of the TCS is the point along the HRS central ray that is the closest to 
the Hall center.
The $\phi_{tg}$ is the tangent of the in-plane angle and $\theta_{tg}$ is the 
tangent of the out-of-plane angle (not shown) w.r.t. the spectrometer 
central ray.  
The sieve plane is located at a drift distance $L=1.12$~m from the TCS origin, 
with the $y_\q{sieve}$ axis pointing to the left of the spectrometer entrance and
the $x_\q{sieve}$ axis pointing vertically down. The pointing of the HRS, $D$,
describes how much the HRS central ray misses the Hall center, 
and $\theta_0$ is the angle of the HRS central ray w.r.t. the beamline. 
Figure reproduced from Refs.~\cite{Alcorn:2004sb,OpticsNote_Nilanga}}
\label{fig:tg_coord}
\end{figure}

The target coordinates of the scattering event, $(x_{tg},y_{tg},\theta_{tg},\phi_{tg})$,
are defined in the target coordinate system (TCS)~\cite{OpticsNote_Nilanga} 
with respect to the spectrometer central ray direction, see 
Fig.~\ref{fig:tg_coord}. Here the angles $\theta_{tg}$ and $\phi_{tg}$ 
refer to the tangent of the vertical and horizontal angles relative to 
the HRS central ray.
The spectrometer pointing $D$ is the distance at which the spectrometer 
misses the Hall center in the direction perpendicular to the spectrometer 
central ray. The sieve plane corresponds to the entrance of the spectrometer 
which is located at $L=1.12$~m from the TCS origin. The particle hit 
position and the angles at the sieve plane can be directly calculated 
from the focal plane variables. 

The calibration procedure involves three separate steps: 
\begin{enumerate}
\item The vertex position along the beam, 
$z_\q{react}$, 
is related to $y_{tg}, \phi_{tg}$ in the TCS as well as the pointing $D$ of the 
spectrometer. The vertex calibration was done by taking data
on the multi-foil carbon target with known foil positions. The foil positions were
determined from data using the HRS optics matrix, the focal plane variables, and 
$D$. 
The precision on $z_\q{react}$ in the direction perpendicular to the 
spectrometer central ray is given by
\begin{eqnarray}
 &&\Delta (z_\q{react}\sin\theta_0) = 
 \sqrt{(\Delta z_\q{foil}\sin\theta_0)^2 + (\Delta z_\q{foil~data}\sin\theta_0)^2 + (\Delta D)^2}~.\label{eq:optics_zreact}
\end{eqnarray}
Here $\Delta z_\q{foil}=\pm 2.5$~mm is the uncertainty of the actual 
foil position caused by possible shifts of the target ladder during the 
target cool-down. 
The quantity $\Delta z_\q{foil~data}$ is 
the discrepancy in oil positions obtained from calibration data and the expected values. 
If the discrepancy is found to be consistent with zero, the value 
$\pm 0.1$~mm is used.
The uncertainty $\Delta D$ can be obtained from a spectrometer 
pointing survey with a typical precision of $\pm 0.5$~mm. If a survey 
was not available, the value of $D$ can be derived from surveys 
performed at a previous spectrometer angle setting. In this case, one compares
the multi-carbon-foil data before and after the spectrometer rotation: if 
the observed shifts in $z$ in all foil positions can be explained consistently 
by a global change in $D$, then the shift is added to the value of $D$ from the
previous survey and the uncertainty of $D$ is taken as $\pm 0.5$~mm. 
If neither carbon foil data nor a survey was available, $\Delta D$
is taken to be $\pm 5$~mm which is the limit of how much the spectrometer 
can physically miss the Hall center. 
At last, the uncertainty in the scattering angle due to the vertex calibration is
\begin{eqnarray}
 \Delta\phi_{tg}  =\Delta (z_\q{react}\sin\theta_0)/L~.~\label{eq:optics_D}
\end{eqnarray}

\item The scattering angles, $\theta_{tg},\phi_{tg}$, were calibrated by  
inserting a so-called ``sieve slit'' plate -- a 0.5-mm thick tungsten 
plate with an array of pinholes -- at the entrance of the spectrometer.  
Reconstruction of hole positions depends on the angle 
elements of the optical matrix. The angle uncertainties from sieve slit
calibrations are:
\begin{eqnarray}
 \Delta\theta_{tg}=\sqrt{(\Delta x_\q{hole})^2+(\Delta x_\q{hole~data})^2}/L~,\label{eq:optics_sieve_x}\\
 \Delta\phi_{tg}  =\sqrt{(\Delta y_\q{hole})^2+(\Delta y_\q{hole~data})^2}/L~,\label{eq:optics_sieve_y}
\end{eqnarray}
where the in-plane angle $\phi_{tg}$ affects the scattering angle $\theta$ directly,
while the out-of-plane angle $\theta_{tg}$ affects $\theta$ only in the second order
and the effect is small.
The quantities $\Delta x_\q{hole}$, $\Delta y_\q{hole}$ are uncertainties in the actual hole
position in the sieve plane. The most straightforward way to determine $x_\q{hole},y_\q{hole}$ is 
by a survey of the sieve slit plate. The survey uncertainty is 
$\pm 0.5$~mm for both directions. However survey was not always done for each kinematic
setting. Past experience has shown 
that the horizontal position $y_\q{hole}$ is highly reproducible, 
to $\pm 0.1$~mm, and the vertical position $x_\q{hole}$ is 
reproducible to $\pm 0.5$~mm due to the fact that this is the direction 
in which the sieve plate is moved into or out of the HRS entrance. 
Thus if no survey was available, results from earlier surveys were used with 
these additional uncertainties added. 
The quantities $\Delta x_\q{hole~data}$, $\Delta y_\q{hole~data}$ are the discrepancy
between the hole position obtained from calibration data and the expected values. 
In the case where no sieve slit calibration data was taken, the angle calibration 
of a preceeding experiment can be used based on the high reliability of the HRS. 
In this case, an additional $\pm 0.5$~mrad of uncertainty should be added to both 
$\Delta\theta_{tg}$, $\Delta\phi_{tg}$ to account for possible changes in the optics.

\item The most precise way to calibrate the momentum is to
use elastic scattering from a carbon target or the proton inside a water target.
With a water target, the relative momentum $\delta\equiv dp/p$ with $p$
the HRS central momentum setting can be determined to $\pm 1\times 10^{-4}$.
Due to the high beam energy used, elastic measurement was not possible for 
the present experiment. However, water target calibration was performed during the 
preceding experiment (HAPPEx-III)~\cite{Ahmed:2011vp}. The HAPPEx-III water 
calibration results were used for the present experiment with an uncertainty 
$\delta=\pm 5\times 10^{-4}$ thanks to the established high stability of the HRS
magnets and transport system.
\end{enumerate}

The three calibration steps described above are assumed to be 
independent from each other, i.e., matrix elements related to position 
reconstruction have little dependence on those related to angle reconstruction,
etc. For all calibrations, the optics tensor coefficients
were determined from a $\chi^2$ minimization procedure in which the events were
reconstructed as close as possible to the known position of the corresponding
foil target or the sieve-slit hole. 

\subsubsection{Optics Calibration Results}

\begin{figure}[!htp]
  \includegraphics[width=0.5\textwidth]{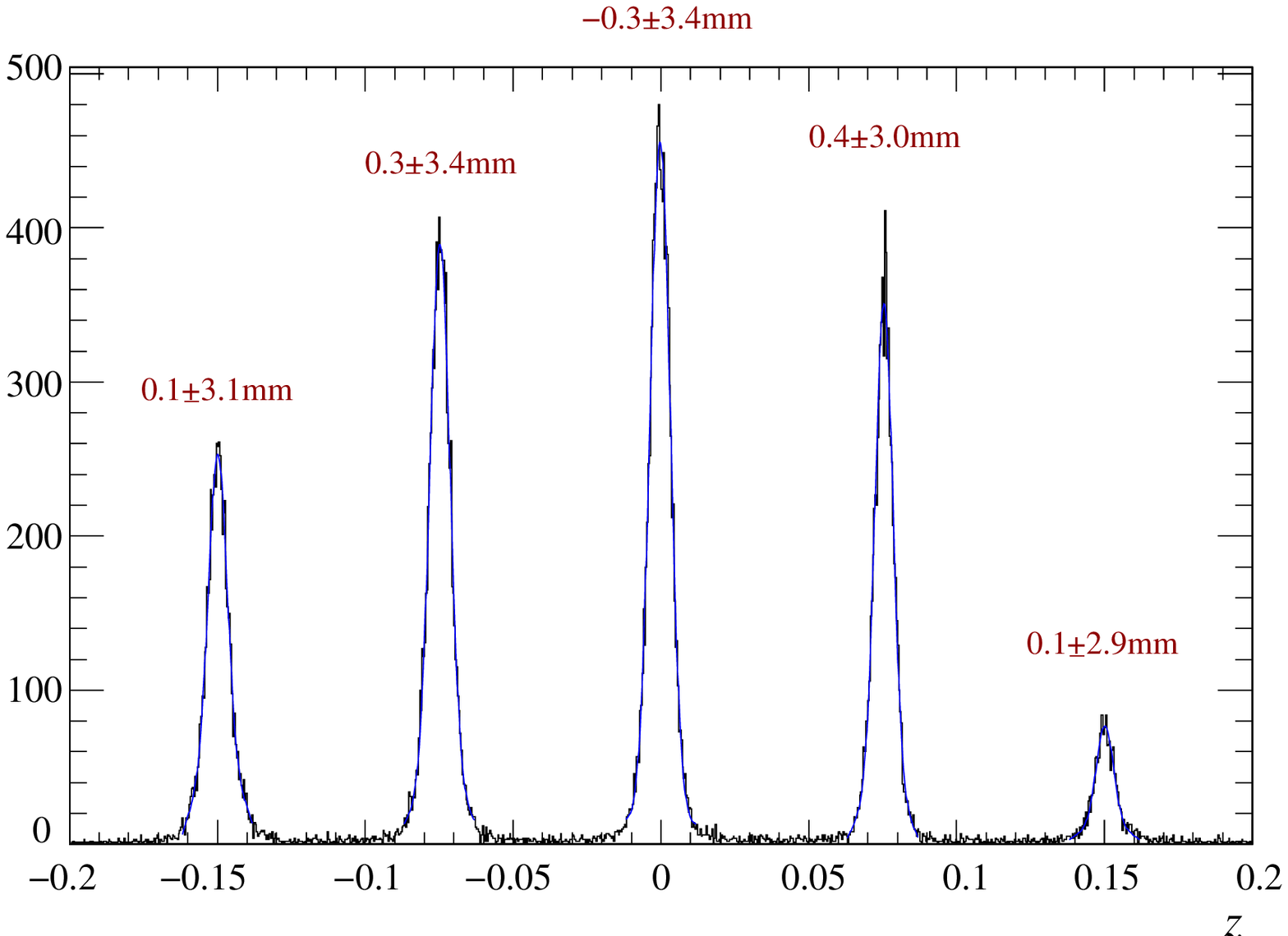}
  \includegraphics[width=0.4\textwidth]{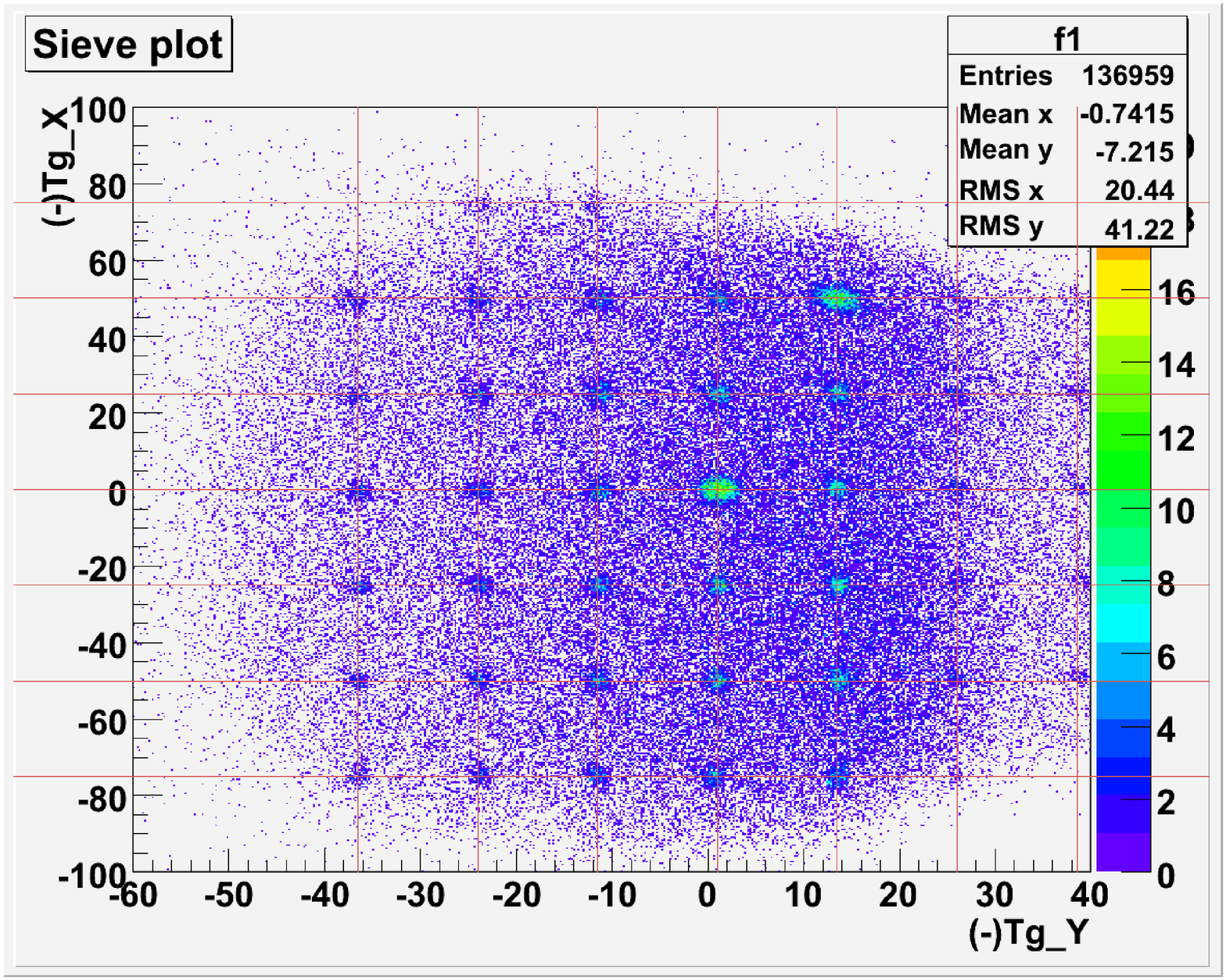} 
  \caption{Left: vertex reconstruction for Left DIS\#1. The number
above each foil is $\Delta z_\q{foil~data}$, defined as how much the observed 
foil position misses the expected value. 
For all foils we have $\Delta z_\q{foil~data}\leqslant 0.4$~mm. 
Right: reconstruction of the sieve hole positions for Left RES~II. 
The data are shown as scattered points and are compared to the expected positions (grids),. 
No obvious discrepancy is seen. 
The axes are oriented such that the sieve hole pattern is as if viewed 
when facing the spectrometer entrance.
Two of the sieve holes are larger than others to allow identifying
the center and the orientation of the sieve plate. 
}\label{fig:Loptics}
\end{figure}

During the PVDIS experiment, there were seven kinematics settings in total 
with one of them carried out on both Left and Right HRS, thus there were a total
of eight HRS+kinematics combinations: 
Left HRS DIS~\#1, Left and Right HRS DIS~\#2, Left HRS Resonance (RES)~I, 
Left HRS RES II, Right HRS RES III, Left HRS RES IV, and Left HRS RES V.
Either vertex or angle calibrations, or both, were carried out for all eight
settings except RES IV and V. 
The vertex calibration for Left DIS\#1 and the angle calibration
results for Left RES II are shown in Fig.~\ref{fig:Loptics}.

\subsubsection{$Q^{2}$ Uncertainties}

The $Q^2$ of each event was calculated using Eq.~(\ref{eq:qmu2}). The uncertainty in $Q^2$ is 
determined by the uncertainties in $\theta$, $E$ and $E^\prime$, but is dominated by the scattering angle uncertainty. 
The scattering angle is calculated as~\cite{Alcorn:2004sb}:
\begin{eqnarray} 
 \theta = \cos^{-1}\left(\frac{\cos\theta_0-\phi_{tg}\sin\theta_0}
              {\sqrt{1+\theta_{tg}^2+\phi_{tg}^2}}\right)~,
\end{eqnarray}
thus calibration of the horizontal angle $\phi_{tg}$ dominates the angle uncertainty. 
The total uncertainty on the scattering angle is the combination of 
the vertex calibration Eqs.~(\ref{eq:optics_zreact}-\ref{eq:optics_D}) 
and $\Delta \phi_{tg}$ from the angle calibration:
\begin{eqnarray}
 \Delta\theta\approx \sqrt{(\Delta D/L)^2+(\Delta z_\q{foil}\sin\theta_0/L)^2+
  (\Delta z_\q{foil~data}\sin\theta_0/L)^2+(\Delta \phi_{tg})^2}~.
\end{eqnarray}
where $\Delta \phi_{tg}$ is either from Eq.~(\ref{eq:optics_sieve_y}) if
a sieve slit calibration was available, or from previous calibrations with
a 0.5~mrad additional uncertainty added. Here  
the drift distance is $L=1.12$~m as shown in Fig.~\ref{fig:tg_coord}.

For some settings during PVDIS, there were both angle and vertex calibrations (Left RES I and II), 
or only the vertex but not the angle calibration (Left DIS\#1, Left DIS\#2, Right DIS\#2,
Right RES III), or neither (Left RES IV and V). 
For both vertex and angle calibrations, the 
optics database and some survey results from the HAPPEx-III experiment that 
ran immediately before this experiment were used. 
 For RES\#I which was taken on the left HRS only, the $Q_1$ and the dipole magnets 
were set at 4.00~GeV/$c$, but its $Q_2$ and $Q_3$ were limited to 3.66~GeV/$c$ due to a power supply malfunction. 
This added complexity to the optical calibration for RES\#I but did not affect the 
HRS acceptance and the quality of the optical calibration results. 
Taking all uncertainties into account, the uncertainty in $Q^2$ due to HRS optics calibration
is summarized in Table~\ref{tab:optics_Q2}.

\begin{widetext}
\begin{table}[!htp]
\begin{tabular}{|p{5.6cm}|c|c|c|c|c|c|c|c|}
\hline 
HRS             & \multicolumn{6}{c|} {Left HRS}                & \multicolumn{2}{c|}{Right HRS}\\\hline
Kinematics      & DIS\#1 &RES V &RES IV  & DIS\#2 & Res I & Res II & DIS\#2 & Res III\\
\hline 
$\theta_0(^\circ)$ & 12.9   & 14.0   & 15.0   & 20     & 12.9   & 12.9   & 20     & 12.9  \\
$E_b$ (GeV)     & 6.067 & 6.067 & 6.067 & 6.067 & 4.867 & 4.867 & 4.867 & 4.867 \\
$E_0'$ (GeV)      & 3.66   & 3.66   & 3.66   & 2.63 & 4.0$^a$  & 3.66   & 2.63   & 3.1   \\\hline 
HRS pointing
     survey?    & Y      & N      & N      & Y      & N      & N      & Y      & N  \\
$\delta D$ 
 (survey)(mm)   & 0.5    &        &        & 0.5    &        &        & 0.5    &     \\
Carbon multi foil data
  available?    & Y      & N      & N      & Y      & Y      & Y      & Y      & Y   \\
$\delta D$ 
 (from data, no survey) (mm) &       &        &        &        & 0.5    & 0.5    & 0.5    & 0.5 \\
$\delta D$ (no survey,
 no data)(mm)    &       & 5.0    & 5.0    &        &        &        &        &     \\
$\delta z_\q{foil~data}$
  (mm)  & 0.4    & N/A   & N/A    & 0.4    & 2.0    & 0.3    & 0.7    & 1.1 \\
$\delta z_\q{foil}$
        & 2.5    & N/A    & N/A    & 2.5    & 2.5    & 2.5    & 2.5    & 2.5   \\
$\Delta\theta$ from vertex calibration (mrad), Eq.~(\ref{eq:optics_D})
                & 0.676  & 4.464  & 4.464  & 0.893  & 0.779  & 0.672  & 0.901  & 0.704 \\\hline 

sieve survey    & N      & N      & N      & N      & N      & N      & N      & N  \\
sieve data      & N      & N      & N      & N      & Y      & Y      & N      & N  \\
$\Delta x_\q{hole}$, 
 from
prior survey (mm)& 0.51  & 0.51   & 0.51   & 0.51   & 0.51   & 0.51   & 0.51   & 0.51\\
$\Delta x_\q{hole~data}$ 
 (mm)    & 0.1   & N/A    & N/A    & 0.1    & 0.1    & 0.1    & 0.1    & 0.1\\
additional $\Delta\phi_{tg}$ (mrad) 
                 &0.5$^b$&0.5$^b$ &0.5$^b$  &0.5$^b$ & none    & none   & 0.5$^c$& 0.5$^c$ \\  
$\Delta\theta$ from angle calibration (mrad), Eq.~(\ref{eq:optics_sieve_y})
                & 0.682  & 0.676  & 0.676  & 0.682  & 0.464  & 0.464  & 0.676 & 0.676   \\\hline 
Total $\Delta\theta$ (mrad)
                & 0.960   & 4.515 & 4.515  & 1.124  & 0.907  & 0.816  & 1.134 & 0.976 \\
\hline 
Total $\Delta\theta/\theta$ (\%)
                & 0.426   & 1.848 & 1.725  & 0.322  & 0.403  & 0.363  & 0.325 & 0.434 \\
\hline 
$\Delta E_0^\prime/E_0^\prime$
                & \multicolumn{8}{c|}{$5\times 10^{-4}$} \\\hline
Total $\Delta Q^2/Q^2$ (\%)$^d$
                & 0.853   & 3.696 & 3.449  & 0.644  & 0.805  & 0.725  & 0.650 & 0.867\\
\hline
\end{tabular}\\

$^a$ For RES\#I which was taken on the left HRS only, the $Q_1$ and the dipole magnets 
were set at 4.00~GeV/$c$, but its $Q_2$ and $Q_3$ were limited to 3.66~GeV/$c$ due to a power supply malfunction; \\
$^b$ Due to using sieve calibration taken at Left RES\#I;\\
$^c$ Due to using optics database from HAPPEx-III;\\
$^d$ Including uncertainties due to both scattering angle $\Delta\theta$ and momentum $\Delta E'$, but is dominated by the former.
\caption{Uncertainty in $Q^2$ determination derived from optics calibration. For each HRS, the kinematics
are shown from left to right in the chronological order.}\label{tab:optics_Q2}
\end{table}
\end{widetext}

\subsection{HRS Simulations}\label{sec:ana_q2}

For the present experiment, a simulation package called ``HAMC'' (Hall A Monte Carlo) was used to 
simulate the transport function and the acceptance of HRS. The simulation was then used 
to calculate the effect of electromagnetic radiative corrections and particle identification
efficiency. 
To ensure that HAMC works correctly, we simulated the kinematics $(Q^2,W,x)$ of the 
scattering, and it is expected that the 
simulated values should agree with the measured ones within the uncertainty of the 
optics calibration, Table~\ref{tab:optics_Q2}.  

In HAMC, events were
generated with a uniform distribution along the beam direction and within a given
raster size and the solid angle 
$d\Omega = \sin(\theta) \hskip 0.02in d\theta \hskip 0.02in d\phi$,
then transported through the HRS magnets using a set of polynomials that model 
the electrons' trajectories through the magnetic fields. 
For RES \#I, a separate set of polynomials were developed for the mismatching fields
of $Q_2$ and $Q_3$.
Events that passed all magnet entrance
and exit apertures fall within the HRS acceptance and are recorded. 
An average energy loss of of 3~MeV was used for the incident electron beam 
to account for the effect of traversing all material along the
beamline to the target center. 
Multiple scattering in the target material, energy loss due to external and internal Brehmstrahlung 
and ionization loss, and the $200~\mu$m resolution of the VDC wires were also taken into account in HAMC. 
The physical differential cross section 
${d^2\sigma}/({dE^\prime d\Omega})$ and the parity-violating asymmetry were calculated 
using the MSTW PDF parametrization for each simulated event.  

Because the DAQ used in the present experiment relied on hardware-based PID, PID calibration
runs were carried out daily to monitor the detector and the DAQ performance. It was found that 
the electron efficiency varied with the particle's hit position in the vertical (dispersive)
direction on the lead-glass detector. This variation could cause a shift in the $Q^2$ value
of the measurement and must be incorporated into HAMC. In HAMC, the hit position 
on the lead-glass detector was calculated from the focal plane coordinates, such that the 
PID efficiency measured from data can be applied to each simulated event. 
The efficiency could drift due to electronic module malfunction 
and drifts in the discriminator thresholds. For most of kinematics, such a drift was
gradual and daily calibrations were sufficient to correct for its effect.

\begin{figure}[!htp]
\begin{center}
\includegraphics[width=0.5\textwidth]{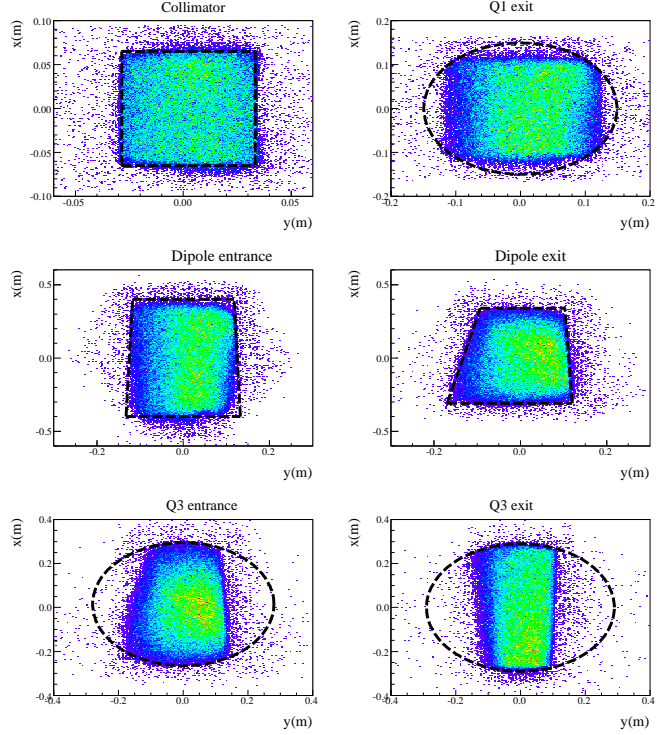}
\caption{Fine-tuning of the HRS acceptance in HAMC. Event distributions from data
are plotted at the collimator (entrance of the HRS $Q_1$), 
$Q_1$ exit, entrances and exits of the dipole and $Q_3$. 
From these distribution, the best estimate of the position and the size of the 
apertures were determined (black dashed lines and curves). These were then used 
as aperture or acceptance cuts in HAMC. The axes are oriented such that the distributions 
are as viewed along the particle trajectory, with $y$ the horizontal and $x$ the vertical
(dispersive) directions, respectively.}\label{fig:hamc_acc}
\end{center}
\end{figure}

In general, the acceptance of the HRS is defined by combining the opening 
geometry of the intermediate apertures, whose nominal settings were documented in Ref.~\cite{Alcorn:2004sb}. 
The real acceptance however can be different from the nominal settings. The HRS acceptance of the
simulation was fine-tuned by matching these apertures to the cross-section-weighted event distributions
obtained from data. This process is illustrated in Fig.~\ref{fig:hamc_acc}.

Once all magnet apertures were optimized, the kinematics $(Q^2,x)$ were 
calculated from HAMC using Eqs.~(\ref{eq:qmu2},\ref{eq:xbj}), 
the beam energy $E$ (minus 3~MeV as mentioned earlier), and the $E^\prime$ and
the scattering angles of the simulated events. 
Similarly, we calculated the $(Q^2,x)$ values from data using the vertex
coordinates $(x_{tg},y_{tg},\theta_{tg},\phi_{tg})$ reconstructed from the
detected focal plane variables, based on HRS transport functions. 
The agreement between the HAMC $(Q^2,x)$ and those reconstructed from the data
thus provides a measure of how well the simulation works.

\begin{figure}[!htp]
\begin{center}
\includegraphics[width=0.65\textwidth]{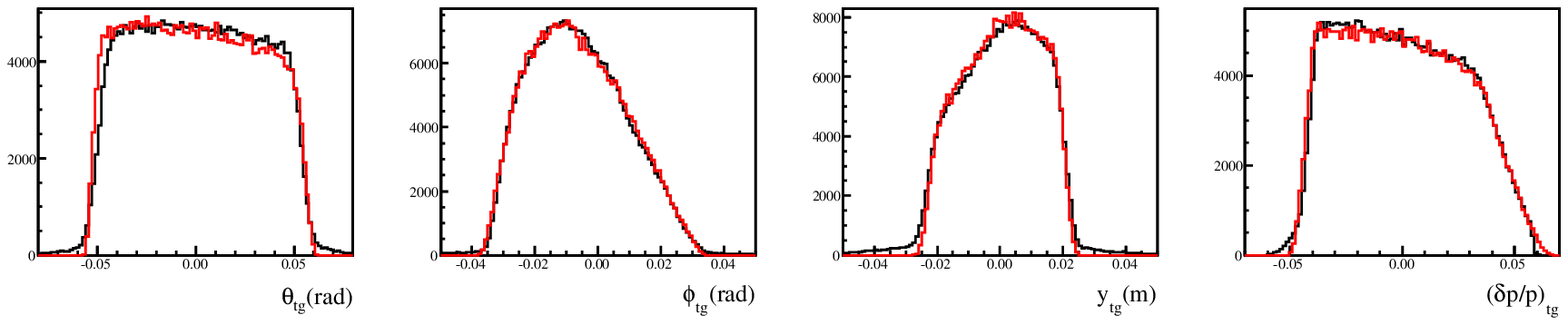}\\
\includegraphics[width=0.65\textwidth]{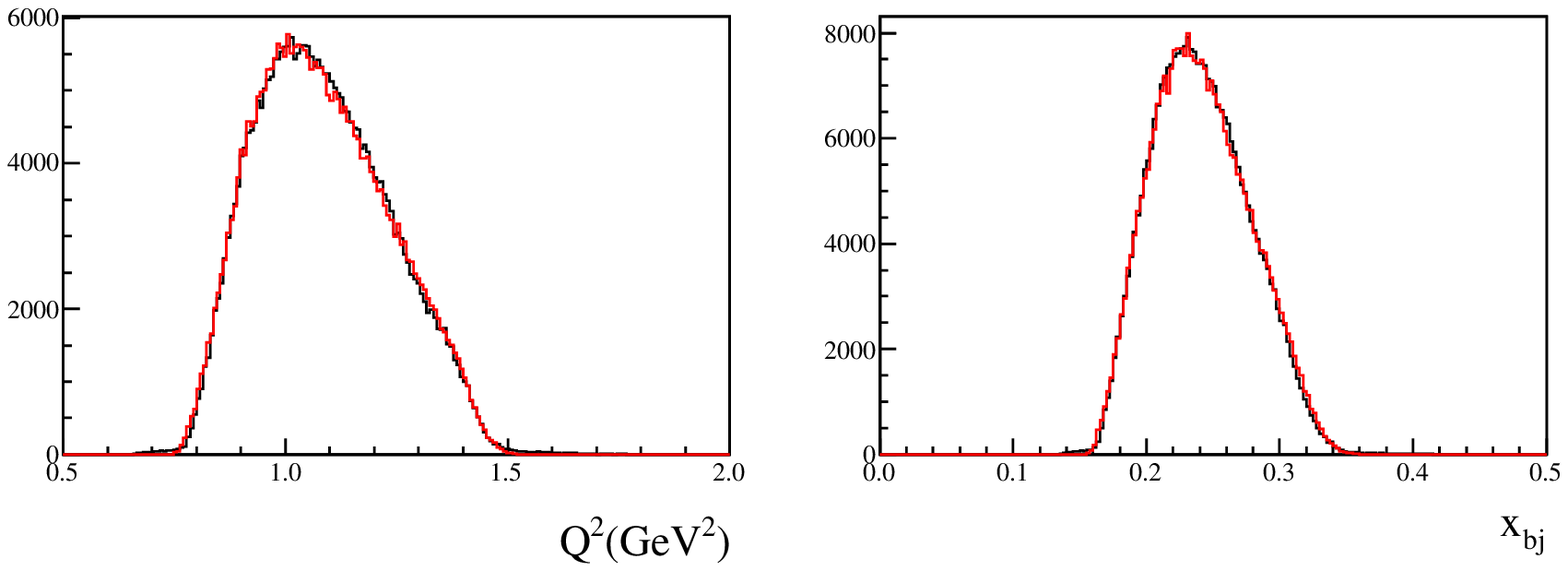}\\
\includegraphics[width=0.65\textwidth]{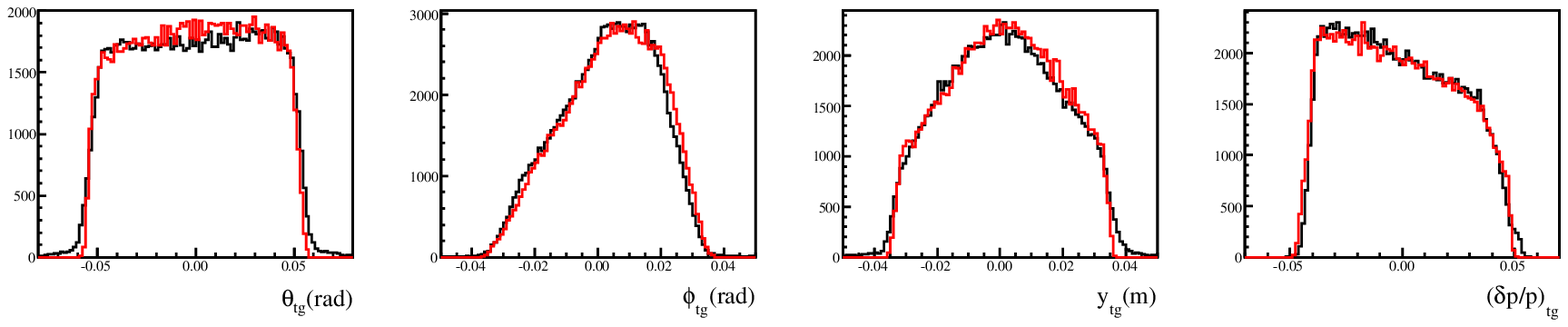}\\
\includegraphics[width=0.65\textwidth]{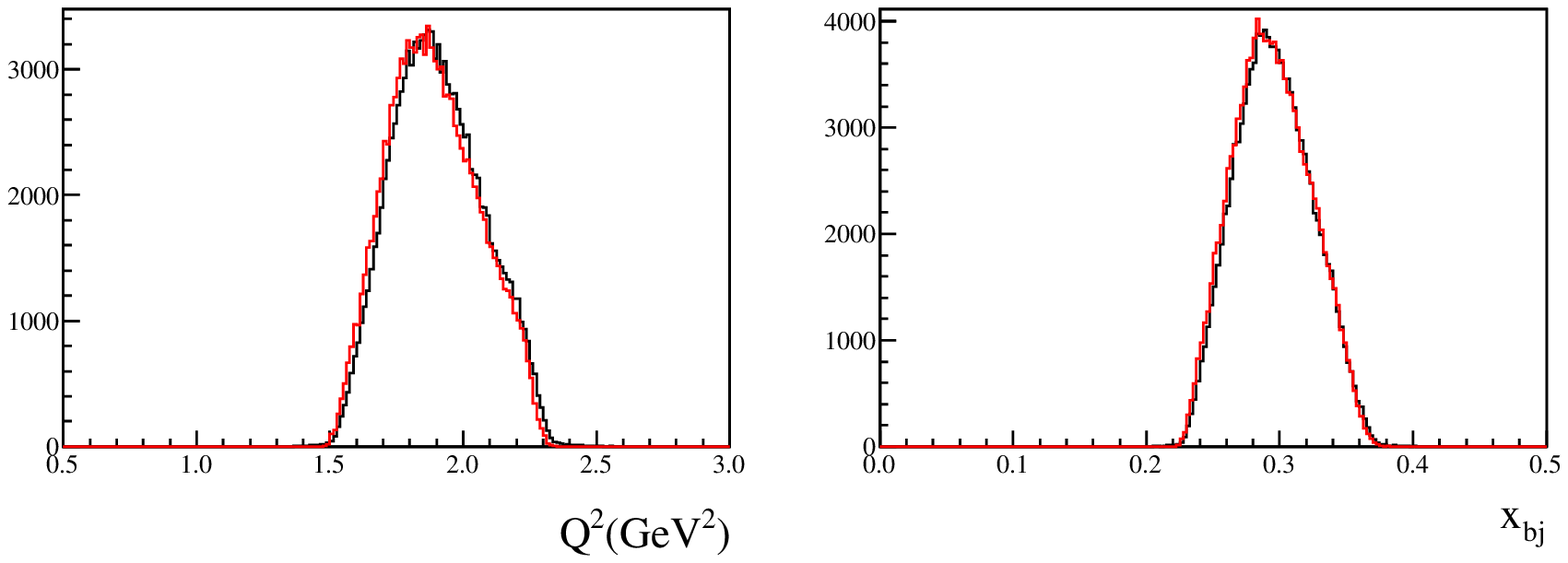}\\
\caption{Comparison between HAMC (red) and data (black). From top to bottom: 
target variables -- $\theta_\q{tg}$, $\phi_\q{tg}$, $y_\q{tg}$ and 
$(\delta p/p)_\q{tg}$ -- for Left HRS DIS\#1; $Q^2$ and $x$ for Left HRS 
DIS\#1; target variables for Right HRS DIS\#2; $Q^2$ and $x$ for Right HRS 
DIS\#2. 
}\label{fig:hamc_comp}
\end{center}
\end{figure}

Figure~\ref{fig:hamc_comp} shows comparisons between data and simulation 
for all four target variables, $Q^2$ and $x$, for Left HRS DIS \#1 and 
Right HRS DIS \#2. A summary of the comparison for all kinematics is
given in Table~\ref{tab:hamc_q2_allkine}.
The observed differences in $Q^2$ are consistent with the uncertainties shown
in Table~\ref{tab:optics_Q2} for most of the kinematics. For RES III, there is 
a two-standard-deviation disagreement in $Q^2$, but is still negligible compared
to the statistical uncertainty at this kinematics. In addition, since we interpret the
asymmetry results at the measured $Q^2$, not the simulated value, this disagreement
does not affect the final result or its uncertainty evaluation and interpretation.

\begin{table}[!htp]
 \begin{center}
 \begin{tabular}{|c|c|c|c|c|c|c|c|} \hline
   \multirow{2}{*}{Kinematics} & \multicolumn{3}{c|}{HAMC} & \multicolumn{3}{c|}{data} & relative \\\cline{2-7}
              & $\langle Q^2\rangle$ & $\langle x\rangle$ & $\langle W^2\rangle$ & $\langle Q^2\rangle$ & $\langle x\rangle$ & $\langle W^2\rangle$ & difference \\                         & (GeV/$c$)$^2$ &  & GeV$^2$ &  (GeV/$c$)$^2$ &  & GeV$^2$ & in $Q^2$\\\hline
   Left HRS DIS\#1       &$1.084$&$0.241$&$4.294$&$1.085$&$0.241$&$4.297$&$<0.1\%$ \\ 
   Left+Right HRS DIS\#2 &$1.892$&$0.294$&$5.424$&$1.901$&$0.295$&$5.430$&$0.5\%$ \\  
   Left HRS RES~I        &$0.956$&$0.571$&$1.600$&$0.950$&$0.571$&$1.595$&$0.6\%$ \\  
   Left HRS RES~II       &$0.832$&$0.336$&$2.528$&$0.831$&$0.335$&$2.530$&$0.1\%$\\   
   Right HRS RES~III     &$0.745$&$0.225$&$3.443$&$0.757$&$0.228$&$3.450$&$1.6\%$\\   
   Left HRS RES~IV       &$1.456$&$0.324$&$3.925$&$1.472$&$0.326$&$3.923$&$1.1\%$\\   
   Left HRS RES~V        &$1.268$&$0.282$&$4.109$&$1.278$&$0.283$&$4.122$&$0.8\%$\\   
 \hline
 \end{tabular}
 \end{center}
 \caption{Comparison of $Q^2$, $x$, and $W^2$ between HAMC and data for all kinematics. 
The Left and the Right DIS\#2 have been combined. The difference in $Q^2$ between 
HAMC and data is smaller than Table~\ref{tab:optics_Q2} for most of the kinematic 
settings. }\label{tab:hamc_q2_allkine}
\end{table}

\subsection{Background Analysis}\label{sec:ana_allbg}

In this section we analyze all backgrounds that affect the extracted 
PV electron asymmetry. 
Assuming each background has an asymmetry $A_i$ and affects the electron sample
with a fraction $f_i$, the correction can be applied as
\begin{eqnarray}
A^\q{phys} &=& {{\left(\frac{A^\q{bc,raw}}{P_b}-
\sum_{i} A_{i}f_{i}\right)}\over{1-\sum_{i} f_{i}}}~,\label{eq:Aphys_cor1}
\end{eqnarray}
where $A^\q{bc,raw}$ is the measured asymmetry with helicity-dependent
beam corrections applied, and $P_b$ is the beam longitudinal polarization
presented in section~\ref{sec:beam_pol}.
When all $f_i$ are small with $A_i$ comparable to or no larger than 
$A^\q{bc,raw}$, one can define 
\begin{eqnarray}
 \bar f_i &=& f_i(1-\frac{A_i}{A^\q{bc,raw}}P_b)\label{eq:fbar}
\end{eqnarray}
and approximate 
\begin{eqnarray}
 A^\q{phys} &\approx& \frac{A^\q{bc,raw}}{P_b}\Pi_i\left(1+\bar f_i\right)~,
 \label{eq:Aphys_cor}
\end{eqnarray}
i.e., all background corrections can be treated as multiplicative. 
As can be seen from Eq.~(\ref{eq:Aphys_cor}), the order of the corrections 
is flexible and the corrections can be applied to the measured asymmetry 
${A^\q{bc,raw}}$ before normalizing to the beam polarization. 
The uncertainty of the correction $\bar f_i$ causes directly 
a relative uncertainty on the electron asymmetry
\begin{eqnarray}
  \frac{\Delta A_e}{A_e} &=& \Delta \bar f_i.
\end{eqnarray}
Some effects, such as charged pion and pair-production background, 
are very small such that corrections [Eq.~(\ref{eq:Aphys_cor})] are not necessary. 
For those cases only the uncertainty $\Delta\bar f_i$ or $\Delta A_e/A_e$ is presented.
The prescription of Eq.~(\ref{eq:Aphys_cor}) was also used for the treatment of 
the $Q^2$-uncertainty and radiative corrections (sections~\ref{sec:ana_q2},
\ref{sec:ana_radcor} and \ref{sec:ana_box}).

\subsubsection{Charged Pion Background}

Charged pions are produced in decays of nucleon resonances created by 
electron scattering off nucleon or nuclear targets. Simulations have shown that 
for the pions to have the same momentum as DIS electrons, the 
parent nucleon resonance must have been produced at a lower $Q^2$ than DIS events, 
thus typically cause a smaller parity-violating asymmetry than DIS electrons. 
This has been confirmed by the asymmetry of the pion triggers measured during the
experiment. The charged pion background thus reduces the magnitude of the 
measured asymmetry, and the effect is the largest if the charged pions did not carry
asymmetry at all.
Furthermore, the high particle identification performance of the
DAQ limited the pion contamination in the electron trigger to the level of
$f_{\pi/e}<2\times 10^{-4}$ and $<4\times 10^{-4}$ for the three DIS kinematics
and the five resonance kinematics, respectively~\cite{Subedi:2013jha}. 
Due to the small contamination, no correction to the measured
electron asymmetries was made. The total systematic uncertainty on the measured electron 
asymmetry due to pion contamination and pion asymmetry is:
\begin{eqnarray}
 \left({{\Delta A_e}\over{A_e}}\right)_{\pi^-}
   = \sqrt{\left(\Delta f_{\pi/e}\right)^2 
    + \left(f_{\pi/e}\frac{\vert{A_\pi}\vert+\Delta A_\pi}{A_e}\right)^2}~,
    ~\label{eq:pionbg}
\end{eqnarray}
where $f_{\pi/e}$ and $\Delta f_{\pi/e}$ are the event fraction of the electron 
trigger that is from actual pions and its uncertainty, $A_\pi$ is the measured 
pion asymmetry with $\Delta A_\pi$ its uncertainty, and $A_e$ is the measured 
electron asymmetry. The term $\vert A_\pi\vert+\Delta A_\pi$ corresponds to 
how much the pion asymmetry could differ from zero at the 68.3\% confidence 
level. As inputs to the background correction, the extraction of pion 
asymmetries is described below. 

\bigskip
\noindent
{\bf pion asymmetry measurement}

The PID performance of both electron and pion triggers of the DAQ was reported 
in Ref.~\cite{Subedi:2013jha}. To properly extract pion asymmetries from the 
trigger, one must account for the effect of electron contamination in the pion 
triggers, $f_{e/\pi}$. Because $f_{e/\pi}$ was relatively high and the electron 
asymmetries are larger than those of pions, corrections were applied to the 
asymmetries extracted from the pion triggers using
\begin{eqnarray}
  A_\pi^\q{meas} = \frac{A_{\pi,\q{dit}}^\q{bc,raw}
              -f_{e/\pi}A_{e,\q{dit}}^\q{bc,raw}}{1-f_{e/\pi}}~,
\end{eqnarray}
where $A_{\pi,\q{dit}}^\q{bc,raw}$ and $A_{e,\q{dit}}^\q{bc,raw}$ are 
asymmetries extracted from pion and electron triggers, respectively, with 
beam corrections applied using the dithering method. 
Then the measured pion asymmetries were normalized with the beam polarization, giving 
physics asymmetry results for pion inclusive production:
\begin{eqnarray}
  A_\pi^\q{phys} = \frac{A_\pi^\q{meas}}{P_b}~.
\end{eqnarray}
Results for pion asymmetries in the DIS and resonance kinematics are given 
in Tables~\ref{tab:results_Api_dis} and \ref{tab:results_Api_res}.
As described in Ref.~\cite{Subedi:2013jha}, the narrow-path triggers of the DAQ had
smaller counting deadtime than the wide-path triggers, but slightly lower PID
performance. As a result the narrow pion triggers had more electron contamination
than the wide triggers and requires a larger correction, which causes a larger 
uncertainty in the extracted pion asymmetry. 
\begin{widetext}
\begin{table}[!htp]
\begin{center}
 \begin{tabular}{c|c|c|c}\hline
  HRS, Kinematics          & Left DIS\#1 & Left DIS\#2 & Right DIS\#2\\\hline
  \multicolumn{4}{c}{narrow path}\\\hline
  $A_{\pi,\q{dit}}^\q{bc,raw}\pm \Delta A_{\pi,\q{dit}}^\q{bc,raw}$(stat.) (ppm)
  & $-57.3\pm 8.0$    & $-26.0\pm 14.9$  & $-21.5\pm 4.2$ \\
  $f_{e/\pi}\pm \Delta f_{e/\pi}$(total) 
  & $0.2653\pm 0.0603$  & $0.0331\pm 0.0034$ & $0.0103\pm 0.0013$ \\
  $A_\pi^\q{meas}\pm \Delta A_\pi^\q{meas}$ (total) (ppm)    
  & $-48.8\pm 14.0$   & $-22.0\pm 21.4$  & $-20.3\pm 6.0$ \\
  $A_\pi^\q{phys}\pm \Delta A_\pi^\q{phys}$ (total) (ppm)    
  & $-55.3\pm 15.9$   & $-24.6\pm 24.0$  & $-22.9\pm 6.8$ \\\hline
  \multicolumn{4}{c}{wide path}\\\hline
  $A_{\pi,\q{dit}}^\q{bc,raw}\pm \Delta A_{\pi,\q{dit}}^\q{bc,raw}$(stat.) (ppm)
  & $-49.6\pm 7.7$    & $-27.0\pm 14.9$  & $-21.4\pm 4.2$ \\
  $f_{e/\pi}\pm \Delta f_{e/\pi}$(total)             
  & $0.2176\pm 0.0573$  & $0.0281\pm 0.0037$ & $0.0091\pm 0.0013$ \\
  $A_\pi^\q{meas}\pm \Delta A_\pi^\q{meas}$ (total) (ppm)    
  & $-41.3\pm 12.8$   & $-23.7\pm 21.4$  & $-20.3\pm 6.0$\\
  $A_\pi^\q{phys}\pm \Delta A_\pi^\q{phys}$ (total) (ppm)    
  & $-46.8\pm 14.6$   & $-26.5\pm 24.0$  & $-22.9\pm 6.8$\\\hline
 \end{tabular}
 \caption{For DIS kinematics: 
beam-corrected pion asymmetries $A_{\pi,\q{dit}}^\q{bc,raw}$ 
with their statistical uncertainties,
electron contamination in the pion triggers $f_{e/\pi}$, pion 
asymmetry results after being corrected for electron contamination 
$A_\pi^\q{meas}$, and physics asymmetry results for pion inclusive
production $A_\pi^\q{phys}$.  As described in Ref.~\cite{Subedi:2013jha}, the
narrow-path triggers had higher electron contamination, thus required a larger
correction and had a larger uncertainty in the extracted pion asymmetry.}\label{tab:results_Api_dis}
\end{center}
\end{table}

\begin{table}[!htp]
 \begin{center}
 \begin{tabular}{c|c|c|c|c|c}\hline
  HRS                     & Left RES I & Left RES II & Right RES III & Left RES IV & Left RES V  \\\hline
  \multicolumn{6}{c}{narrow path}\\\hline
  $A_{\pi,\q{dit}}^\q{bc,raw}\pm \Delta A_{\pi,\q{dit}}^\q{bc,raw}$(stat.) (ppm)
  & $-44.2\pm 40.1$& $-69.8\pm 26.5$ & $-17.1\pm 8.5$  & $21.8\pm 47.7$  & $-46.7\pm 64.0$\\
  $f_{e/\pi}\pm \Delta f_{e/\pi}$(total)              
  &{\small $0.4114\pm 0.0201$}& $0.3155\pm 0.0163$& $0.0849\pm 0.0030$& $0.1852\pm 0.0073$& $0.1871\pm 0.0077$ \\
  $A_\pi^\q{meas}\pm \Delta A_\pi^\q{meas}$ (total) (ppm)  
  & $-33.7\pm 88.6$& $-73.2\pm 48.8$ & $-13.5\pm 12.7$ & $52.2\pm 76.2$  & $-41.5\pm 102.4$ \\
   $A_\pi^\q{phys}\pm \Delta A_\pi^\q{phys}$ (total) (ppm)  
  & $-37.3\pm 98.0$& $-81.0\pm 54.0$ & $-14.9\pm 14.0$ & $58.2\pm 85.0$  & $-46.3\pm 114.2$ \\\hline
  \multicolumn{6}{c}{wide path}\\\hline
  $A_{\pi,\q{dit}}^\q{bc,raw}\pm \Delta A_{\pi,\q{dit}}^\q{bc,raw}$(stat.) (ppm)
  & $-45.4\pm 39.4$& $-69.2\pm 26.1$ & $-18.3\pm 8.5$   &$30.9\pm 47.6$   & $-51.0\pm 64.9$ \\
  $f_{e/\pi}\pm \Delta f_{e/\pi}$(total)                  
  & $0.3423\pm 0.0231$& $0.2409\pm 0.0200$& $0.0633\pm 0.0060$& $0.1661\pm 0.0080$& $0.1598\pm 0.0086$  \\
  $A_\pi^\q{meas}\pm \Delta A_\pi^\q{meas}$ (total) (ppm)   
  & $-39.8\pm 74.9$ & $-71.0\pm 43.7$& $-15.8\pm 12.4$  & $58.8\pm 74.7$  & $-47.7\pm 101.4$  \\
  $A_\pi^\q{phys}\pm \Delta A_\pi^\q{phys}$ (total) (ppm)   
  & $-44.0\pm 82.9$ & $-78.5\pm 48.4$& $-17.5\pm 13.7$  & $65.6\pm 83.3$  & $-53.2\pm 113.1$  \\\hline\hline
 \end{tabular}
 \caption{For resonance kinematics: 
beam-corrected pion asymmetries $A_{\pi,\q{dit}}^\q{bc,raw}$ 
with their statistical uncertainty,
electron contamination in the pion triggers $f_{e/\pi}$, pion 
asymmetry results after being corrected for electron contamination 
$A_\pi^\q{meas}$, and physics asymmetry results for pion inclusive
production $A_\pi^\q{phys}$.  As described in Ref.~\cite{Subedi:2013jha}, the
narrow-path triggers had higher electron contamination, thus required a larger
correction and had a larger uncertainty in the extracted pion asymmetry.}\label{tab:results_Api_res}
 \end{center}
\end{table}

\end{widetext}

\bigskip
\noindent
{\bf electron asymmetry uncertainty due to pion contamination}
The measured pion and electron asymmetries are listed in 
Tables~\ref{tab:pionbg_dis} and~\ref{tab:pionbg_res} for the two DIS
and the five resonance kinematics, respectively, together with the 
total uncertainty due to pion contamination in the electron asymmetry 
as calculated with Eq.~(\ref{eq:pionbg}). 
The values listed for the pion contamination in the electron triggers $f_{\pi/e}$ 
and the electron contamination
in pion triggers $f_{e/\pi}$ and their total uncertainties are from Ref.~\cite{Subedi:2013jha}.
The narrow-path triggers have larger uncertainty due to charged pion background because
of the slightly lower pion rejection performance. Overall, the uncertainty 
due to charged pion background is very low, at the $10^{-4}$ level for all kinematics.

\begin{widetext}
\begin{table}[!htp]
\begin{center}
 \begin{tabular}{c|c|c|c}\hline
  HRS, Kinematics          & Left DIS\#1 & Left DIS\#2 & Right DIS\#2\\\hline
  \multicolumn{4}{c}{narrow path}\\\hline
  $A_\pi^\q{meas}\pm \Delta A_\pi^\q{meas}$ (total) (ppm)    
  & $-48.8\pm 14.0$   & $-22.0\pm 21.4$  & $-20.3\pm 6.0$ \\
  $A_{e,\q{dit}}^\q{bc,raw}\pm A_{e,\q{dit}}^\q{bc,raw}$ (stat.) (ppm) 
  & $-78.5\pm 2.7$    & $-140.3\pm 10.4$ & $-139.8\pm 6.6$ \\
  $f_{\pi/e}\pm\Delta f_{\pi/e}$ (total) ($\times 10^{-4}$) 
  & $(1.07\pm 0.24)$    & $(1.97\pm 0.18)$   & $(1.30\pm 0.10)$ \\
  $\left({\Delta A_e}\over{A_e}\right)_{\pi^-,n}$ 
  & $0.89\times 10^{-4}$ & $0.63\times 10^{-4}$& $0.27\times 10^{-4}$ \\\hline
  \multicolumn{4}{c}{wide path}\\\hline
  $A_\pi^\q{meas}\pm \Delta A_\pi^\q{meas}$ (total) (ppm)    
  & $-41.3\pm 12.8$   & $-23.7\pm 21.4$  & $-20.3\pm 6.0$\\
  $A_{e,\q{dit}}^\q{bc,raw}\pm \Delta A_{e,\q{dit}}^\q{bc,raw}$ (stat.) (ppm) 
  & $-78.3\pm 2.7$    & $-140.2\pm 10.4$ & $-140.9\pm 6.6$ \\
  $f_{\pi/e}\pm\Delta f_{\pi/e}$ (total) ($\times 10^{-4}$)
  & $(0.72\pm 0.22)$    & $(1.64\pm 0.17)$   & $(0.92\pm 0.13)$ \\
  $\left({\Delta A_e}\over{A_e}\right)_{\pi^-,w}$
  & $0.54\times 10^{-4}$ &$0.55\times 10^{-4}$ & $0.21\times 10^{-4}$ \\\hline
 \end{tabular}
 \caption{For DIS kinematics: pion 
asymmetry results $A_\pi^\q{meas}$, beam-corrected electron raw asymmetry $A_e^\q{bc,raw}$, 
pion contamination in electron triggers $f_{\pi/e}$, 
and total uncertainties on the electron asymmetry results  
due to pion background $\left(\Delta A_e/A_e\right)_{\pi^-,n}$ 
and $\left(\Delta A_e/A_e\right)_{\pi^-,w}$, all at the $10^{-4}$ level.}\label{tab:pionbg_dis}
\end{center}
\end{table}

\begin{table}[!htp]
 \begin{center}
 \begin{tabular}{c|c|c|c|c|c}\hline
  HRS                     & Left RES I & Left RES II & Right RES III & Left RES IV & Left RES V  \\\hline
  \multicolumn{6}{c}{narrow path}\\\hline
  $A_\pi^\q{meas}\pm \Delta A_\pi^\q{meas}$ (total) (ppm)  
  & $-33.7\pm 88.6$& $-73.2\pm 48.8$ & $-13.5\pm 12.7$ & $52.2\pm 76.2$  & $-41.5\pm 102.4$ \\  $A_{e,\q{dit}}^\q{bc,raw}\pm \Delta A_{e,\q{dit}}^\q{bc,raw}$ (stat.)  
  & $-55.1\pm 6.8$ & $-63.8\pm 5.9$  & $-54.4\pm 4.5$  & $-104.0\pm 15.3$& $-67.9\pm 21.3$ \\
  $f_{\pi/e}\pm\Delta f_{\pi/e}$ (total) ($\times 10^{-4}$)
  & $(0.79\pm 0.11)$ & $(2.40\pm 0.20)$   & $(3.82\pm 0.23)$   & $(0.26\pm 0.03)$  & $(0.45\pm 0.03)$ \\
  $\left({\Delta A_e}\over{A_e}\right)_{\pi^-,n}$ 
  &$1.75\times 10^{-4}$&$4.60\times 10^{-4}$&$1.85\times 10^{-4}$& $0.32\times 10^{-4}$& $0.96\times 10^{-4}$   \\\hline
  \multicolumn{6}{c}{wide path}\\\hline
  $A_\pi^\q{meas}\pm \Delta A_\pi^\q{meas}$ (total) (ppm)   
  & $-39.8\pm 74.9$ & $-71.0\pm 43.7$& $-15.8\pm 12.4$  & $58.8\pm 74.7$  & $-47.7\pm 101.4$  \\
  $A_{e,\q{dit}}^\q{bc,raw}\pm \Delta A_{e,\q{dit}}^\q{bc,raw}$ (stat.) (ppm)  
  & $-54.6\pm 6.8$  & $-63.9\pm 5.9$ & $-54.0\pm 4.5$ & $-104.6\pm 15.3$  & $-67.9\pm 21.5$ \\
  $f_{\pi/e}\pm\Delta f_{\pi/e}$ (total) ($\times 10^{-4}$)
  & $(0.54\pm 0.15)$  & $(1.50\pm 0.25)$  & $(2.14\pm 0.48)$  & $(0.22\pm 0.03)$    & $(0.32\pm 0.04)$ \\
  $\left({\Delta A_e}\over{A_e}\right)_{\pi^-,w}$ 
  & $1.13\times 10^{-4}$& $2.71\times 10^{-4}$&$1.22\times 10^{-4}$&$0.28\times 10^{-4}$&$0.71\times 10^{-4}$\\\hline\hline
 \end{tabular}
 \caption{For resonance kinematics: pion 
asymmetry results $A_\pi^\q{meas}$, beam-corrected electron raw asymmetry $A_e^\q{bc,raw}$, 
pion contamination in electron triggers $f_{\pi/e}$, 
and total uncertainties on the electron asymmetry results  
due to pion background $\left(\Delta A_e/A_e\right)_{\pi^-,n}$ and $\left(\Delta A_e/A_e\right)_{\pi^-,w}$, all at the $10^{-4}$ level.}\label{tab:pionbg_res}
 \end{center}
\end{table}

\end{widetext}

\subsubsection{Pair Production Background}

The pair production background results from nucleon resonance production when the 
resonance decays into neutral pions ($\pi^0$) that then decay into $e^+e^-$ pairs. 
Pair production from bremsstrahlung photons is not significant 
in the kinematics of this experiment because pair production is highly forward-peaked. 
Therefore, one expect that the effect from pair-production background to have a similar 
as that from charged pions and the prescription of Eq.~(\ref{eq:pionbg}) can be used by 
replacing $A_\pi$ with $A_{e^+}$ and $f_{\pi/e}$ with the fractional contribution of pair 
production to the main electron trigger $f_{e^+/e^-}$.
For the pair-production asymmetry, we expect it to be determined by the $\pi^0$ 
photo- and electroproduction and thus comparable to that of the charged pion asymmetry. 
The contamination factor $f_{e^+/e^-}$
was determined for the two DIS kinematics by reversing the HRS polarity and measure the rate of 
positrons from the $\pi^0$ decay. 
Due to the low rate of positron events the HRS DAQ could be used for these studies with
the VDC and a well-understood PID. However, the statistical uncertainties in the positron asymmetry
were quite large due to the very low positron rate. 
Moreover, the $\pi^+$ contamination in the positron trigger was quite high, estimated to be 
11\% and 20\% for the Left DIS\#1 and Right DIS\#2, respectively, assuming the 
PID performance of the detector does not depend on the sign of the particles' charge. 
The measured asymmetry of the pair-production background could not be corrected for
the $\pi^+$ contamination due to the lack of knowledge on the $\pi^+$ asymmetry.

Asymmetries extracted from positive polarity runs are shown in 
Table~\ref{tab:Apositron} without corrections for the $\pi^+$ background 
or beam polarization. 
\begin{table}[!htp]
 \begin{center}
\begin{tabular}{c|c|c}\hline
  HRS                 &{Left DIS\#1} & {Right DIS\#2}\\\hline
  $A_{e^+}^\q{raw}$ (ppm), narrow & $723.2\pm 1154.7$(stat.) & $1216.0\pm 1304.5$(stat.) \\
  $A_{e^+}^\q{raw}$ (ppm), wide   & $742.4\pm 1151.5$(stat.) & $1199.0\pm 1304.5$(stat.)\\\hline 
\end{tabular}
 \end{center}
 \caption{Raw positron asymmetry results. No correction for the beam
position, energy, and polarization, or the 
$\pi^+$ background was made.}\label{tab:Apositron}
\end{table}

Because the statistical uncertainties in the positron asymmetry are so large, 
we relied on the fact that $\pi^0$ must have similar asymmetries as 
$\pi^-$. We assume the $\pi^0$ asymmetry to be no larger than twice that of the $\pi^-$ 
asymmetry and estimated the uncertainty in the electron asymmetry due to pair production to be:
\begin{eqnarray}
 \left({{\Delta A_e}\over{A_e}}\right)_\q{pair} = \sqrt{\left(\Delta f_{e^+/e^-}\right)^2+\left(f_{e^+/e^-}\frac{\Delta{A_{e^+}}}{A_e}\right)^2}~,~\label{eq:posbg}
\end{eqnarray}
where $\Delta A_{e^+}$ describes how much $A_{e^+}$ differs from zero and the 
value $2(\vert A_{\pi^-}\vert +\Delta A_{\pi^-})$ was used. Results for $f_{e^+/e^-}$ 
and their statistical uncertainties are shown in Table~\ref{tab:positronbg}, and a 
$30\%$ uncertainty was used for $\Delta f_{e^+/e^-}$ to account for 
possible systematic effects in positron identification due to the high 
$\pi^+$ background in the rate evaluation.
Results for the electron asymmetry uncertainty due to pair production background 
are also shown in Table~\ref{tab:positronbg}.
\begin{widetext}
\begin{table}[!htp]
\begin{tabular}{c|c|c|c}\hline
  HRS & Left DIS\#1 & Left DIS\#2  & Right DIS\#2\\\hline
  $f_{e^+/e^-}\pm\Delta f_{e^+/e^-}$ (stat.)
      & $(2.504\pm 0.007)\times 10^{-4}$ 
      & $(5.154\pm 0.001)\times 10^{-3}$ 
      & $(4.804\pm 0.001)\times 10^{-3}$\\\hline
  $\left({\Delta A_e}\over{A_e}\right)_\q{pair, narrow}$ 
      & $4.1\times 10^{-4}$ & $3.5\times 10^{-3}$ & $2.3\times 10^{-3}$ \\\hline
  $\left({\Delta A_e}\over{A_e}\right)_\q{pair, wide}$ 
      & $3.5\times 10^{-4}$ & $3.7\times 10^{-3}$ & $2.3\times 10^{-3}$ \\\hline
\end{tabular}
 \caption{Results for pair production (positron) contamination in the electron trigger
$f_{e^+/e^-}$ and its statistical 
uncertainty, and the total uncertainty 
on electron asymmetry due to pair production background, 
$\left({\Delta A_e}\over{A_e}\right)_\q{pair}$. Only DIS kinematics 
are shown. The errors shown for $f_{e^+/e^-}$ are statistical only, and a 
30\% systematic uncertainty on $f_{e^+/e^-}$ was used in the evaluation of 
$\Delta A_e\over A_e$.}\label{tab:positronbg}
\end{table}
\end{widetext}

There was no measurement for the pair production rate for any resonance kinematics. 
The value $3\times 10^{-3}$ (the average of the uncertainty at DIS\#2) 
was used as the relative uncertainty 
due to pair production for all resonance asymmetry results.
This is a conservative estimate because the $\pi^-/e$ rate ratios for resonance 
settings were similar to DIS \#1 and are about one order of magnitude smaller
than that of DIS\#2 (see Table~\ref{tab:kine_settings}),

\subsubsection{Target EndCap Corrections}

Electrons scattered off the target aluminum endcaps (Al 7075) cannot be separated from 
those scattered off the liquid deuterium. 
The parity-violating asymmetries from aluminum and the alloying elements differ slightly 
from that of deuterium and a correction must be made. Because the Al 7075 alloy is
made of $\approx 90\%$ aluminum, we calculate the effect from the aluminum
asymmetry below, and the effect from other non-isoscalar elements ($\approx 6\%$ Zn
and $\approx 1.4\%$ Cu) was estimated to be $<8\%$ of that of Al.
Based on 
Eqs.~(\ref{eq:Apvdis1}-\ref{eq:F3gzqpm}), the value of parity-violating (PV) 
asymmetry from $e-$Al scattering was calculated as
\begin{eqnarray}
  A_\q{Al} &=& \frac{13A_p\sigma_p+14A_n\sigma_n}{13\sigma_p+14\sigma_n}~,
 \label{eq:Aal}
\end{eqnarray}
where $\sigma_{p(n)}$ is the cross section and $A_{p(n)}$ is the PV asymmetry 
for scattering off the proton (neutron). The cross sections $\sigma_{p(n)}$ were
calculated using a fit to world resonance and DIS data~\cite{Bosted:2007xd}.
The asymmetries  $A_{p(n)}$ were calculated using Eq.~(\ref{eq:Apvdis_pdg}):
\begin{widetext} 
\begin{eqnarray}
  A_p &=& \left(-\frac{3 G_FQ^2}{2\sqrt{2}\pi \alpha}\right)
    \frac{Y_1\left[2C_{1u}(u^++c^+)-C_{1d}(d^+ +s^+)\right]
         +Y_3\left[2C_{2u}(u^-)-C_{2d}(d^-)\right]}
         {4(u^++c^+)+(d^++s^+)}~,\label{eq:Apv_proton}\\
  A_n &=& \left(-\frac{3 G_FQ^2}{2\sqrt{2}\pi \alpha}\right)
    \frac{Y_1\left[2C_{1u}(d^++c^+)-C_{1d}(u^++s^+)\right]
         +Y_3\left[2C_{2u}(u^-)-C_{2d}(d^-)\right]}
      {4(d^++c^+)+(u^++s^+)}~,\label{eq:Apv_neutron}
\end{eqnarray}
\end{widetext}
with $u^\pm\equiv u\pm \bar u$, $d^\pm\equiv d\pm\bar d$, $s^+\equiv s+\bar s$ 
and $c^+\equiv c+\bar c$.

The actual aluminum asymmetries $A_\q{Al}$ may differ from the values calculated using 
Eq.~(\ref{eq:Aal}) due to effects such as 
resonance structure (for resonance kinematics), 
and nuclear effects similar to the EMC effect
of the unpolarized, parity-conserving structure functions $F_{1,2}$~\cite{Aubert:1983xm}. 

The EMC effect on aluminum was studied by several experiments~\cite{Gomez:1993ri,Stein:1975yy,Rock:2001ps}, 
and data on various nuclei were extrapolated to infinite nuclear matter~\cite{Sick:1992pw}. 
(For a recent review of EMC effects see Ref.~\cite{Malace:2014uea}.) 
For the two DIS kinematics ($x=0.2-0.3$) the EMC effect for Al is approximately 3\%.  
A conservative relative uncertainty of $10\%$ was used for $A_\q{Al}$ 
in the DIS kinematics. 
For resonance kinematics, the EMC effect for Al is in the range $(3-14)\%$, 
and even larger for higher $x$ values. 
On the other hand, the measured electron asymmetry
at all five resonance kinematics were found to be in good agreement
(at the 10-15\% level) with the values calculated using PDFs ~\cite{Wang:2013kkc}, and 
we expect that the uncertainty in $A_\q{Al}$ due to resonance 
structure cannot exceed this level.
Adding the nuclear and the resonance effects in quadrature, 
a $20\%$ relative uncertainty was used for $A_\q{Al}$ in the resonance kinematics.

The fractional event rate from the aluminum endcaps, $\alpha_\q{Al/D}$, was 
calculated as
\begin{eqnarray}
 \alpha_\q{Al/D} &=& \eta_\q{Al/D} \frac{\sigma_\q{Al}}{\sigma_\mathrm{D}}\label{eq:fendcap}
\end{eqnarray}
where $\eta_\q{Al/D}$ is the ratio of the endcap to liquid deuterium 
thicknesses, and ${\sigma_\q{Al}}/{\sigma_\mathrm{D}}$ is the Al to deuterium per-nucleon cross-sectional ratio
from previous measurements~\cite{Gomez:1993ri,Stein:1975yy,Rock:2001ps} without the isoscalar correction. 
The target used for this experiment had entrance and exit endcaps measured 
to be $0.126\pm 0.011\pm 0.003$~mm and $0.100\pm 0.008\pm 0.003$~mm thick, respectively 
(see Table~\ref{tab:target_data}), with the first error bar from 
the standard deviation of multiple measurements at different positions on the 
endcap, and the second error from calibration of
the instrument. The ratio $\eta_\q{Al/D}$ is 
$\eta_\q{Al/D}=(0.126+0.100)$~mm$\times (2.7$~g/cm$^3)/(20$~cm$\times 0.167$~g/cm$^3)=1.827\%$ with an uncertainty
of $\Delta\eta_\q{Al/D}=0.115\%$. 

The correction to the electron PVDIS asymmetry was applied as
\begin{eqnarray}
  A^\q{Al-corrected}_e &=& A_e(1+\bar f_\q{Al}),\\
 \q{with}~
  \bar f_\q{Al} &=& -(\alpha_\q{Al/D})\frac{A_\q{Al}-A_{D}}{A_D}. 
 \label{eq:Aendcapcorr}
\end{eqnarray}
The total uncertainty due to target endcaps is 
\begin{eqnarray}
  \left(\frac{\Delta A_e}{A_e}\right)_\q{Al} 
 &=&\sqrt{\left(\Delta\alpha_\q{Al/D}\frac{A_{Al}-A_{D}}{A_D}\right)^2
 + \left[(\delta_{A_\q{Al}})\alpha_\q{Al/D}\right]^2} \label{eq:Aendcaperr}
\end{eqnarray}
where $\alpha_\q{Al/D}$ is from Eq.~(\ref{eq:fendcap}), 
$\Delta\alpha_\q{Al/D}
=(\Delta\eta_\q{Al/D}/\eta_\q{Al/D})\alpha_\q{Al/D}
=0.063\alpha_\q{Al/D}$, $A_\q{Al}$ from Eqs.(\ref{eq:Aal}-\ref{eq:Apv_neutron}), 
$A_D$ from Eq.~(\ref{eq:Apvdis_R}), 
and $\delta_{A_\q{Al}}$ is the maximal relative 
difference in the Al vs. D$_2$ PV asymmetries caused by
an EMC-like medium modification effect and resonance structures. 
As stated above, the values $\delta_{A_\q{Al}}= 10\%$ for DIS
and $= 20\%$ for resonance kinematics were used. 
Results for the endcap correction $\bar f_\q{Al}$ and 
the uncertainty on the corrected electron asymmetry are
listed in Table~\ref{tab:albg}. As one can see, the correction due to aluminum is
at the $10^{-4}$ level. The effect from other non-isoscalar alloying elements in 
Al 7075 was estimated to be at the $10^{-5}$ level and was neglected in the analysis. 
\begin{widetext}
\begin{table}[!htp]
 \begin{center}
\begin{tabular}{c|c|c|c|c|c|c|c}\hline
  Kinematics           & DIS\#1    & DIS\#2    & RES~I    & RES~II    & RES~III    & RES~IV & RES~V\\\hline
  $(A_\q{Al}-A_D)/A_D$   & $0.567\%$ & $0.727\%$ & $1.335\%$ & $0.800$   & $0.510$   & $0.799$&  $0.691$ \\
  $\alpha_\q{Al/D}$      & $2.0\%$  & $2.0\%$   & $2.0\%$  & $2.0\%$  & $2.0\%$  & $2.0\%$ & $2.0\%$ \\
  $\bar f_\q{Al}$
    ($\times 10^{-4}$)   & $-1.2$    & $-1.5$   & $-2.7$  & $-1.6$  & $-1.0$  & $-1.6$ & $-1.4$\\
  $\left({\Delta A_e}/{A_e}\right)_\q{Al}$ 
                        & $0.24\%$  & $0.24\%$ & $0.43\%$ & $0.43\%$ & $0.43\%$ & $0.43\%$ & $0.43\%$\\\hline
\end{tabular}
 \end{center}
 \caption{Target endcap correction for all kinematics. 
Shown here are the relative differences between calculated Al and D$_2$ asymmetries, 
$(A_\q{Al}-A_D)/A_D$, the fractional event rate from Al endcaps 
$\alpha_\q{Al/D}$, corrections applied to measured electron 
asymmetries $\bar f_\q{Al}$ using Eq.~(\ref{eq:Aendcapcorr}), 
and the relative uncertainty in the corrected electron asymmetry 
due to endcap corrections $(\Delta A_e/A_e)_\q{Al}$ using Eq.~(\ref{eq:Aendcaperr}). 
Here, the Al and D$_2$ asymmetries were calculated using Eqs.~(\ref{eq:Aal},\ref{eq:Apv_proton},\ref{eq:Apv_neutron})
and the MSTW2008 NLO PDF~\cite{Martin:2009iq}. 
Corrections from other non-isoscalar alloying elements in Al 7075 was estimated to be
at the $10^{-5}$ level or smaller, and thus were neglected in the analysis.}\label{tab:albg}
\end{table}
\end{widetext}

Events were also taken on a thick, ``dummy'' target consisting of two  
aluminum foils with their thickness approximately 10 times that of the liquid 
deuterium cell. The thickness was chosen such that the total radiation
length of the dummy target matches that of the liquid D$_2$ target. However, 
due to limited beam time, the asymmetry uncertainty collected from the 
aluminum dummy target was not precise enough to reduce the systematic 
uncertainty due to target endcaps.

\subsubsection{Beam Transverse Asymmetry Correction}

\begin{figure}[!htp]
 \begin{center} 
  \includegraphics[scale=1.0]{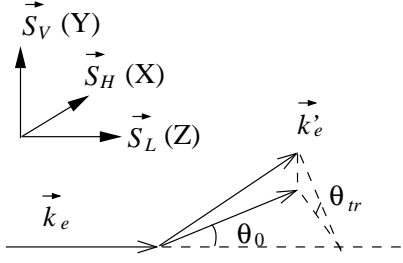}
 \end{center}
 \caption{Kinematics of the beam normal asymmetry background. The incident 
and the scattered electrons' momenta are $\vec k_e$ and $\vec k_e^\prime$, 
and $\vec S_{V,H,L}$ denote respectively the incident electron's spin polarization
components in the vertical, horizontal, and longitudinal directions. 
The central scattering angle setting of the spectrometer is $\theta_0$ and the 
scattered electron's momentum
has an out-of-plane angle denoted by $\theta_{tr}$.}\label{fig:ATkine}
\end{figure}

Transverse asymmetry background, also called the beam normal asymmetry background,
describes the effect of the electron beam spin polarized in the direction
normal to the scattering plane defined by the momentum vectors of the incident and the scattered 
electrons $\vec k_e$ and $\vec k_e^\prime$~\cite{Abrahamyan:2012cg}. 
This beam normal asymmetry is parity-conserving and must be 
treated as a background of the measurement. Calculations at the pure partonic 
level show that this asymmetry is between 0.1-0.2~ppm at the kinematics
of this experiment, but mechanisms beyond the parton level can enhance the 
asymmetry by 1-2 orders of magnitude~\cite{AApv}. 
The contribution from the beam normal asymmetry $A_n$ to the measured asymmetry 
can be expressed as
\begin{eqnarray}
  \delta A = (A_n) \vec S\cdot \hat k_n~&\q{with}&\vec k_n\equiv \hat k_e\times \hat k_e^\prime~~\q{and}~
\hat k_n=\vec k_n/\vert\vec k_n\vert~,
\end{eqnarray}
where $A_n$ is the beam-normal asymmetry and $\vec S$ is the beam polarization 
vector. Denoting $\theta_0$ the central scattering angle of the spectrometer 
and $\theta_{tr}$ the vertical angle of the scattered electron w.r.t. the nominal
setting of the spectrometer (see Fig.~\ref{fig:ATkine}), 
one has $\hat k_e=(0,0,1)$ and 
$\hat k_e^\prime = (\sin\theta_0\cos\theta_{tr}, \sin\theta_0\sin\theta_{tr},\cos\theta_0)$, 
giving $\vec k_n=(-\sin\theta_0\sin\theta_{tr},\sin\theta_0\cos\theta_{tr},0)$
and $\hat k_n = (-\sin\theta_{tr},\cos\theta_{tr},0)$, thus
\begin{eqnarray}
 \delta A &=& {A_n}\left[-S_H\sin\theta_{tr}+S_V\cos\theta_{tr}\right]~,\label{eq:AT1}
\end{eqnarray}
where $S_{V,H,L}$ are respectively the electron polarization components 
in the vertical (perpendicular to the nominal scattering plane defined 
by the electron beam and the central ray of the spectrometer), horizontal 
(within the nominal plane but transverse to the beam), and 
longitudinal directions. The value of $S_L$ is thus the beam longitudinal polarization $P_b$.
During the experiment the beam spin components were controlled 
to $\vert S_H/S_L\vert\leqslant 27.4\%$ 
and $\vert S_V/S_L\vert\leqslant 2.5\%$ 
and the average value of $\theta_{tr}$ was 
found from data to be less than 0.01~rad. Therefore the beam
vertical spin dominates this background: 
\begin{eqnarray}
 \left(\Delta A_e\right)_{A_n}\approx A_n S_V\cos\theta_{tr}\approx A_n S_V\leqslant(2.5\%)P_b A_n~.
\end{eqnarray}
 
During the experiment, the size of the beam normal asymmetry $A_n$ was 
measured for DIS kinematics during dedicated ``transverse runs'' where the 
beam was fully polarized in the vertical direction, $S_H^T=S_L^T\approx 0$ and 
$S_V=P_{b0}^T$ where the superscript $T$ stands for transverse 
asymmetry measurement and $P_{b0}^T$ is the maximum beam polarization during such measurement. 
Asymmetries measured during these runs are thus
$A_n^\q{meas}=A_n P_{b0}^{T}$. 
Because the maximum beam polarization 
is the same for production and transverse asymmetry running, one has 
$P_{b0}^T=S_0\equiv\sqrt{S_L^2+S_V^2+S_H^2}
=\sqrt{1+(0.274)^2+(0.025)^2}S_L=1.037 S_L$
and the total uncertainty in the electron asymmetry can be calculated as
\begin{eqnarray}
 \left(\frac{\Delta A_e}{A_e}\right)_{A_n}
  &=& \frac{A_nS_V}{A_e^\q{meas}} = \frac{({A_n^\q{meas}}/{S_0})S_V}{A_e^\q{meas}}
=  \frac{A_n^\q{meas}}{A_e^\q{meas}}\frac{S_V}{S_0}\leqslant 2.4\% \frac{A_n^\q{meas}}{A_e^\q{meas}}~.
\end{eqnarray}
For DIS kinematics, we denote 
$\delta A_n^\q{meas}$ as how much $A_n$ could differ 
from zero to account for the uncertainty of the $A_n$ measurement, and write
\begin{eqnarray}
 \left(\frac{\Delta A_e}{A_e}\right)_{A_n, \q{DIS}} &\leqslant& 
  2.4\% \frac{\delta A_n^\q{meas}}{A_e^\q{meas}}~.
\end{eqnarray}
If the measured $A_n$ is consistent with zero, 
the statistical uncertainty of the measurement $\Delta A_n^\q{meas}$(stat.)  
is taken as $\delta A_n^\q{meas}$, otherwise the value of
$(\vert A_n^\q{meas}\vert +\Delta A_n^\q{meas})$ is used as
$\delta A_n^\q{meas}$. 

Results for the beam transverse asymmetry measurements are shown in 
Table~\ref{tab:ATbg} for the two DIS kinematics along with the 
resulting uncertainty on the electron PVDIS asymmetry due to beam 
transverse polarizations.
\begin{table}[!htp]
 \begin{center}
\begin{tabular}{c|c|c}\hline
  Kinematics                  & Left DIS\#1& Right DIS\#2 \\
  $Q^2$ (GeV/$c$)$^2$  & $1.085$ & $1.907$ \\\hline
  $A_n^\q{meas}\pm\Delta A_n^\q{meas}$ (stat.) (ppm, narrow)
                       & $-24.15\pm 15.05$ & $23.49\pm 44.91$ \\
  $A_e^\q{meas}$ (ppm, narrow)       & $78.45$ & $-139.97$ \\
  $\left({\Delta A_e}\over{A_e}\right)_{A_n,~\q{narrow}}$ & $1.18\%$ & $0.76\%$\\\hline
  $A_n^\q{meas}\pm\Delta A_n^\q{meas}$ (stat.) (ppm, wide)
                       & $-24.66\pm 15.01$    & $24.60\pm 44.90$\\
  $A_e^\q{meas}$ (ppm, wide)       & $78.27$ & $-140.67$ \\
  $\left({\Delta A_e}\over{A_e}\right)_{A_n,~\q{wide}}$ & $1.20\%$ & $0.76\%$\\\hline
\end{tabular}
 \end{center}
 \caption{The measured beam transverse asymmetry together with
the resulting uncertainty on the electron asymmetry. 
The dithering-corrected values were used for both $A_e^\q{meas}$ 
and $A_n^\q{meas}$.  For DIS\#2, the electron asymmetry is the 
combined value from the Left and the Right HRS.
}\label{tab:ATbg}
\end{table}

Beam transverse asymmetry measurements were not performed for the resonance kinematics. 
However, $A_n$ measured in the DIS region has a similar $Q^2$ dependence 
and magnitude as that measured in previous elastic electron scattering from the proton 
and heavier nuclei~\cite{Abrahamyan:2012cg}. 
This indicates the size of $A_n$ to be determined predominantly by $Q^2$, and 
that the response of the target (elastic vs. DIS) only affects $A_n$ at higher orders. 
Based on this observation, we used Ref.~\cite{Abrahamyan:2012cg} to calculate $A_n$ 
for all resonance kinematics. We found $A_n$ to be between 
$-38$ and $-80$~ppm depending on the value of $Q^2$, and are always smaller than that of 
the electron asymmetry. Therefore the uncertainty due
to $A_n$ was estimated for resonance kinematics as
\begin{eqnarray}
 \left(\frac{\Delta A_e}{A_e}\right)_{A_n, \q{RES}} 
  &\approx& \left\vert{\frac{A_nS_V}{A_e^\q{meas}}}\right\vert 
   =\left\vert\frac{S_VA_n}{P_b A_e^\q{phys}}\right\vert
  \leqslant \left\vert{S_V/P_b}\right\vert= \left\vert{S_V/S_L}\right\vert = 2.5\%~.
\end{eqnarray}

\subsubsection{Target Purity, Density Fluctuation and Other False Asymmetries}
\label{sec:syst_purity}

The liquid deuterium used contained~\cite{target:purity} $1889$~ppm HD (hydrogen deuteride), 
$<100$~ppm H$_2$, 
$4.4$~ppm N$_2$, $0.7$~ppm O$_2$, $1.5$~ppm CO, $<1$~ppm methane and $0.9$~ppm CO$_2$.  
The only non-negligible effect 
on the measured asymmetry comes from the proton in HD. 
Since the proton asymmetry as given by Eq.~(\ref{eq:Apv_proton})
differs from the asymmetry of the deuteron by no more than $\pm (15-30)\%$, the proton 
in HD contributes an uncertainty of 
$\left(\Delta{A_e}/A_e\right)_\q{HD}<0.06\%$ to the measured electron 
asymmetry.

\subsubsection{Rescattering and Poletip Scattering Background}\label{sec:syst_background}

In this section, two kinds of backgrounds from rescattering inside
the HRS spectrometers are considered. The first is due to electrons 
from outside the HRS momentum acceptance which rescatter into
the detector.
The second effect is called ``poletip scattering'', which refers to 
electrons which scattered from polarized electrons (M{\o}ller scattering) 
in the magnetized iron in the HRS dipoles. 
These backgrounds are suppressed by a factor of 10
compared to the estimates given in 
Ref.~\cite{Aniol:2004hp} because of our
trigger threshold for the lead-glass detector.

Using Eq.~(\ref{eq:fbar}), the correction to our asymmetry for both 
cases can be written as  
\begin{eqnarray}
\bar f_\mathrm{rs}&=& - \frac{f_\mathrm{rs}\Delta A}{A^\mathrm{meas}}~,
\end{eqnarray}
where $f_\mathrm{rs}$ is the fraction of the rescattering 
background and $\Delta A=A^\mathrm{bgr}-A^\mathrm{meas}$ 
is the difference between the background's asymmetry 
and the measured asymmetry.
The correction can be evaluated by integrating over the energy that contribute 
to this background: 
\begin{eqnarray}
f_\mathrm{rs} \Delta A = \frac{1}{\Delta E_\mathrm{{HRS}}} \int_\mathrm{{outside}}
dE \frac{\ P_\mathrm{{rs}}(E)P_\mathrm{thr}{\left(\frac{d\sigma}{d\Omega dE}\right)_\mathrm{{outside}}}
   (A^\mathrm{bgr}-A^\mathrm{meas})}
   {\left(\frac{d\sigma}{d\Omega dE}\right)_\mathrm{{inside}}}~, \label{eq:scattcorr1}
\end{eqnarray} 
where $\Delta E_{\mathrm{HRS}}$ is the HRS energy acceptance, 
$P_\mathrm{{rs}}$ is the rescattering probability that describes the relative
contribution of rescattered events among all events that reach the 
detectors, $P_\mathrm{thr}$ is the probability for rescattered events that reach
the detectors to pass the trigger threshold and cause an electron trigger,
and ${\left(\frac{d\sigma}{d\Omega dE}\right)_\mathrm{{{inside}}{{{(outside)}}}}}$ is the 
scattering cross section inside (outside) the HRS acceptance. 
The integration is done from just outside the spectrometer acceptance 
(beyond $\pm 4\%$) to up to $\pm 20\%$ of the nominal 
setting $E_0^\prime$. The upper limit 
of $20\%$ is used because the function $P_\mathrm{{rs}}(E)$ 
becomes negligible beyond this range.

\begin{figure}[!htp]
 \begin{center} 
  \includegraphics[width=0.5\textwidth]{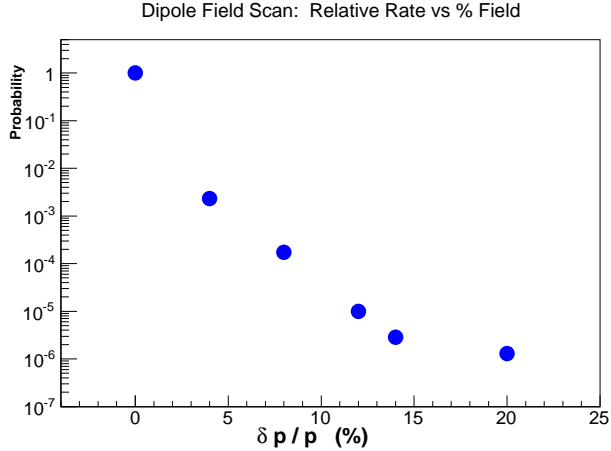}
 \end{center}
 \caption{The function $P_\mathrm{{rs}}(E)$ determined from 
HAPPEx data.}
\label{fig:rescatt_pvdis}
\end{figure}

The rescattering probability $P_\mathrm{{rs}}(E)$ was measured by the 
HAPPEx experiment~\cite{Aniol:2004hp}, and the results are shown in Fig.~\ref{fig:rescatt_pvdis}.
The probability drops to below $10^{-3}$ just outside the HRS acceptance ($4\%$)
and quickly to $10^{-6}$ at $20\%$. 
Although only the positive detune ($\delta p/p>0$) was measured, we 
assumed the distribution is symmetric around the nominal momentum of the
spectrometer. 
The trigger threshold factor $P_\mathrm{thr} \approx 0.1$ is 
estimated from the location of the trigger threshold for
our lead glass detector. The parity-violating asymmetry
scales with $Q^2$ and we found that $\bar f_\mathrm{rs} \ll 2\times 10^{-5}$.

In Ref.~\cite{Aniol:2004hp} an upper bound for the poletip
scattering effect was found.
Using that analysis, and without accounting for the further
suppression by our trigger thresholds, we estimate that
\begin{eqnarray}
\bar f_\mathrm{pole-tip}  
 < \frac{0.3~{\mathrm{ppm}}}{A^\mathrm{meas}}~.
\end{eqnarray}

Because the effects from rescattering and pole-tip scattering are both small,
no correction to the asymmetry was made and these two effects were 
counted as additional systematic uncertainties.

\subsection{Electromagnetic Radiative Corrections}\label{sec:ana_radcor}

Electrons undergo radiative energy losses due to interactions such as 
internal and external
bremsstrahlung and ionization loss, both before and after the
scattering. This causes two effects on the measurement: 1) There 
is a small beam depolarization effect associated with the energy loss
of incident electrons; 2) the energy loss of both incident
and scattered electrons would cause 
a difference between the kinematics reconstructed from the detected
signals and what really happened at the interaction 
point. We discuss these two effects separately.

\subsubsection{Beam Depolarization Effect in Bremsstrahlung}

The depolarization of electron from bremsstrahlung radiation was calculated based 
on Ref.~\cite{Olsen:1959zz} and the formalism is provided in Appendix~\ref{sec:app_depol}. 
We define a depolarization correction 
\begin{eqnarray}
  f_\q{depol} = \frac{\langle A_e D\rangle}{\langle A_e\rangle}
\end{eqnarray}
where $D$ is the beam depolarization factor (with zero depolarization corresponding to $D=100\%$)
and the average of a quantity $\langle a\rangle$ ($a=A_e$ or $A_eD$) is taken over the 
spectrometer acceptance and the cross section $\sigma$:
\begin{eqnarray}
  \langle a\rangle \equiv \frac{\int_\mathrm{HRS} a\cdot\sigma\cdot\q{(acceptance)}}
     {\int_\mathrm{HRS} \sigma\cdot\q{(acceptance)}}~.\label{eq:hamc_average}
\end{eqnarray}
The measured asymmetry should be corrected as
\begin{eqnarray}
 A^\q{depol-corrected} = A^\q{meas}_e(1+\bar f_\q{depol})~,
\end{eqnarray}
where $\bar f_\q{depol}\equiv(1/f_\q{depol})-1
\approx\langle A_e\rangle/\langle A_eD\rangle-1$.
An HAMC simulation was done to determine the value of $\bar f_\q{depol}$ and the results are
shown in Table~\ref{tab:beamdpol}.

\begin{table}[!htp]
 \begin{center}
 \begin{tabular}{c|c|c|c|c|c|c|c} \hline\hline
 Kinematics & DIS\#1 & DIS\#2 & RES~I & RES~II & RES~III & RES~IV & RES~V \\\hline
 $\bar f_\q{depol}$ & $0.096\%$ & $0.209\%$ & $0.005\%$ & $0.028\%$ & $0.093\%$ & $0.061\%$ & $0.081\%$ \\\hline\hline
 \end{tabular}
 \end{center}
 \caption{Beam depolarization correction $\bar f_\q{depol}$ for all kinematics.}\label{tab:beamdpol}
\end{table}

\subsubsection{Corrections for Vertex versus Detected Kinematics}

Due to energy losses of the electrons, the kinematics at the interaction 
vertex is not the same as those calculated from the initial beam energy 
and the electron's momentum detected by the spectrometer. This effect is 
illustrated in Fig.~\ref{fig:HAMCkin}:
\begin{figure}[!htp]
 \begin{center}
  \includegraphics[width=0.35\textwidth]{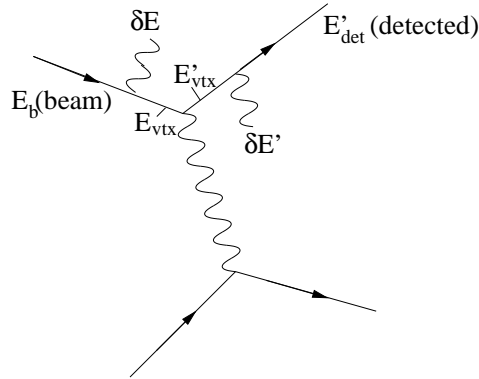}
  \caption{Kinematics used in HAMC to correct energy losses $\delta E$ 
and $\delta E^\prime$ 
for the incoming and outgoing electrons respectively.
The kinematics reconstructed from the data corresponds to $E_\q{beam}$
and $E_\q{det}^\prime$, while the vertex kinematics corresponds to $E_\q{vtx}$ and $E_\q{vtx}^\prime$.}\label{fig:HAMCkin}
 \end{center}
\end{figure}
since the shift between detected and vertex kinematics relies heavily on the experimental setup,
it is desired to correct the measured asymmetry for this effect such that the corrected 
values can be compared to theoretical expectations in an unambiguous way.
This correction factor is defined as:
\begin{eqnarray}
1+\bar f_\q{rc} &=& 
  \frac{A(\langle Q_\q{det}^2\rangle, \langle x_\q{det}\rangle)}
       {\langle A(Q_\q{vtx}^2, x_\q{vtx})\rangle}~,\label{eq:frc}
\end{eqnarray}
and is applied to the measured asymmetry as:
\begin{eqnarray}
A_e^\q{rad-corrected}= A_e^\q{meas} (1+\bar f_{rc})~.
\end{eqnarray}
Here $A(\langle Q_\q{det}^2\rangle, \langle x_\q{det}\rangle)$ 
is the asymmetry calculated at the cross-section- and acceptance-weighted 
values [see Eq.~(\ref{eq:hamc_average})] of $Q^2_\q{det}$ and 
$x_\q{det}$, evaluated from the initial beam energy and the detected 
electrons momentum, and $\langle A(Q_\q{vtx}^2, x_\q{vtx})\rangle$ 
is the asymmetry still averaged over all detected electrons following 
Eq.~(\ref{eq:hamc_average}), but now calculated using the vertex kinematics 
$Q_\q{vtx}^2$ and $x_\q{vtx}$ of each event. 
Since the value $\langle A(Q_\q{vtx}^2, x_\q{vtx}^2)\rangle$ is the 
expected value of what was actually measured in the experiment 
($A_e^\q{meas}$), the result $A_e^\q{rad-corrected}$ can be 
treated as the value corresponding to 
$\langle Q^2_\q{det}\rangle$ and $\langle x_\q{det}\rangle$. 
The value of $A_e^\q{rad-corrected}$ can thus be compared with
theoretical calculations evaluated at $\langle Q^2_\q{det}\rangle$ 
and $\langle x_\q{det}\rangle$ to extract physics results.

The radiative correction was evaluated using HAMC which calculates both 
the numerator and the denominator of Eq.~(\ref{eq:frc}). 
Therefore, we expect that any small imperfection in the understanding of the 
HRS acceptance or cross-section calculation, such as that indicated by 
the 2 standard-deviation disagreement in $Q^2$ between HAMC 
and data for RES III, would cancel out to the first 
order, and does not lead to a larger uncertainty in the radiative correction
for this kinematics. 
The treatment of radiative effects was based on the prescription of 
Mo \& Tsai~\cite{Mo:1968cg}. The detailed procedure is described below.

For each simulated event, the scattering angle $\theta$ and the momentum of 
the scattered electron $E^\prime_\q{vtx}$ at the vertex were generated randomly.
The energy loss of incoming and outgoing electrons $\delta E$ and $\delta E^\prime$ were
then calculated using the formula given on page 5-7 of Ref.~\cite{ref:hamc}, which 
includes external bremsstrahlung, internal bremsstrahlung using the effective
radiator formula, and ionization loss. Next, the incoming electron's energy at the 
vertex is calculated as $E_\q{vtx}=E_b-\delta E$ where $E_b$ is the (fixed) initial beam energy
and the detected momentum of the scattered electron calculated as
$E^\prime_\q{det}=E^\prime_\q{vtx}-\delta E^\prime$. 
If $\theta$ and $E^\prime_\q{det}$ fell within the spectrometer acceptance,
the cross section and the PV asymmetry were calculated using both the detected 
$(E_b, E_{det}, \theta)$ and the vertex kinematics $(E_\q{vtx},E'_\q{vtx},\theta)$ 
and were stored.

\begin{figure}[!htp]
 \begin{center}
  \includegraphics[width=0.5\textwidth]{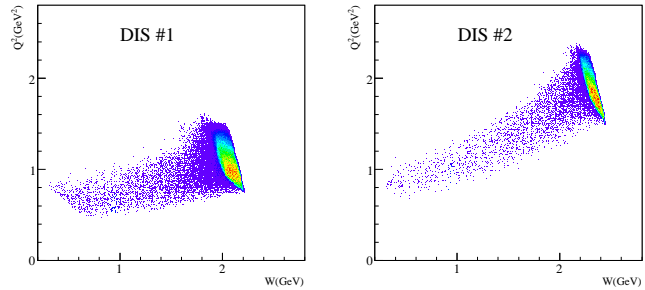}
  \caption{Simulated vertex kinematics of the two DIS kinematics \#1
(left) and \#2 (right).}\label{fig:diskine}
 \end{center}
\end{figure}

The vertex kinematics $(Q^2_\q{vtx},W_\q{vtx})$ calculated using 
$(E_\q{vtx},\theta,E_\q{vtx}^\prime)$ is shown in Fig.~\ref{fig:diskine} 
for the two DIS kinematics. One can see that the vertex
kinematics of an event could fall into one of the following categories: 
$e$-$^2$H elastic ($W<M$ with $M$ the proton mass, quasi-elastic 
($W\approx M$), nucleon resonances ($M\lesssim W<2$~GeV), 
and DIS ($W>2$~GeV).
To evaluate the PV asymmetries for different vertex kinematics, the 
following prescription was used:
\begin{enumerate}
\item For $e-^2$H elastic scattering, the method from the SAMPLE 
experiment~\cite{Beise:2004py} was used, 
where the cross section was based on Ref.~\cite{Abbott:2000ak} 
and the PV asymmetry was based on a simple model that compares well to 
the calculation of Ref.~\cite{Pollock:1990uv}. 
The strange magnetic form factor $G_M^s$ in this method was taken to be zero.

\item For quasi-elastic scattering, the cross section and the asymmetry 
were calculated using the elastic scattering formula and elastic form factors 
for the neutron and the proton [see Section VII of Ref.~\cite{Aniol:2004hp}], 
then smeared for their Fermi motion following the algorithm of 
Ref.~\cite{Bosted:2007xd}. 
The quasi-elastic (qe) PV asymmetry was then calculated as
$A_d^\q{qe} = {(A_p^\mathrm{el}\sigma_p^\mathrm{el} + A_n^\mathrm{el}\sigma_n^\mathrm{el})}
/{(\sigma_p^\mathrm{el} + \sigma_n^\mathrm{el})}$ 
where $A_{p(n)}^\mathrm{el}$ and $\sigma_{p(n)}^\mathrm{el}$ are the elastic
asymmetry and cross section for the proton (neutron), respectively.

\item For the nucleon resonance region ($1\lesssim W<2$~GeV), the cross 
section was based on Ref.~\cite{Bosted:2007xd}, and the asymmetries were
 calculated from three models: one theoretical model for the 
$\Delta(1232)$~\cite{Matsui:2005ns}, a second theoretical model that covers 
the whole resonance region~\cite{Gorchtein:2011mz}, and one 
``cross-section-scaling model'' 
where $A_\q{res}={{\sigma_\q{res}}\over{\sigma_\q{dis}}} A_\q{dis}$ 
was used. Here $A_\q{dis}$ was calculated from Eqs.~(\ref{eq:Apvdis1},
\ref{eq:y11},\ref{eq:y13},\ref{eq:a11},\ref{eq:a31},\ref{eq:F1gqpm},
\ref{eq:F1gzqpm},\ref{eq:F3gzqpm})
with MSTW2008 
PDFs~\cite{Martin:2009iq}, $\sigma_\q{dis}$ 
was calculated using the NMC fit of $F_2$~\cite{Arneodo:1995cq} 
structure functions and $R$ from Ref.~\cite{Bosted:2007xd},
and $\sigma_\q{res}$ was from Ref.~\cite{Bosted:2007xd} which exhibits
distinct resonance structures;
The cross-section-scaling model was used only when the theoretical models do not cover the 
kinematics of a particular event.
\item For DIS ($W>2$~GeV), the cross section was calculated using Bosted's 
fits~\cite{Bosted:2007xd} and the PV asymmetry was calculated using 
Eqs.~(\ref{eq:Apvdis1},
\ref{eq:y11},\ref{eq:y13},\ref{eq:a11},\ref{eq:a31},\ref{eq:F1gqpm},
\ref{eq:F1gzqpm},\ref{eq:F3gzqpm})
with MSTW2008 
PDFs~\cite{Martin:2009iq}. 
For $R$ in Eq.~(\ref{eq:y13}) again Ref.~\cite{Bosted:2007xd} was used. 
\end{enumerate}

\begin{figure}[!htp]
 \begin{center}
  \includegraphics[width=0.35\textwidth]{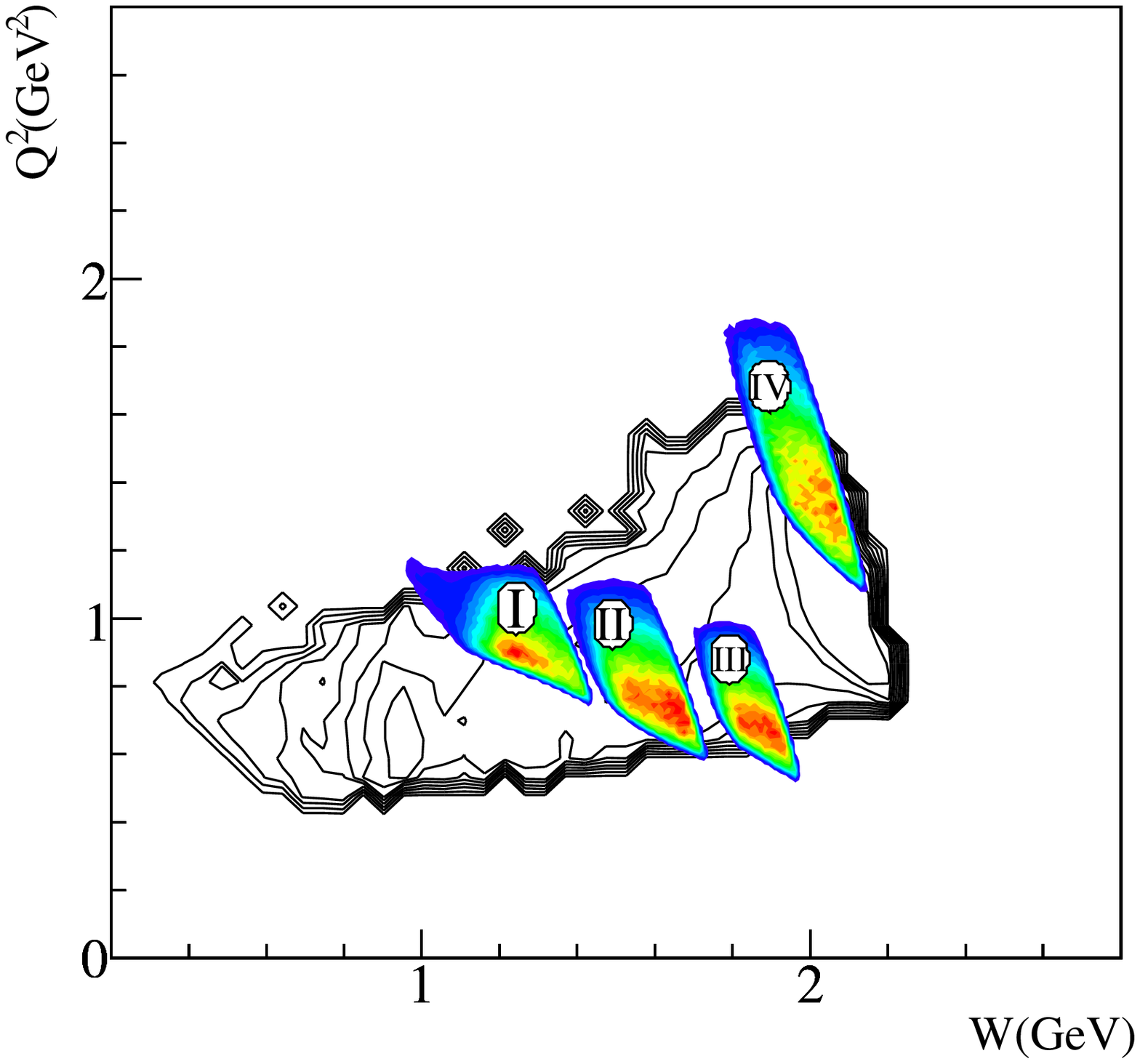}
  \includegraphics[width=0.35\textwidth]{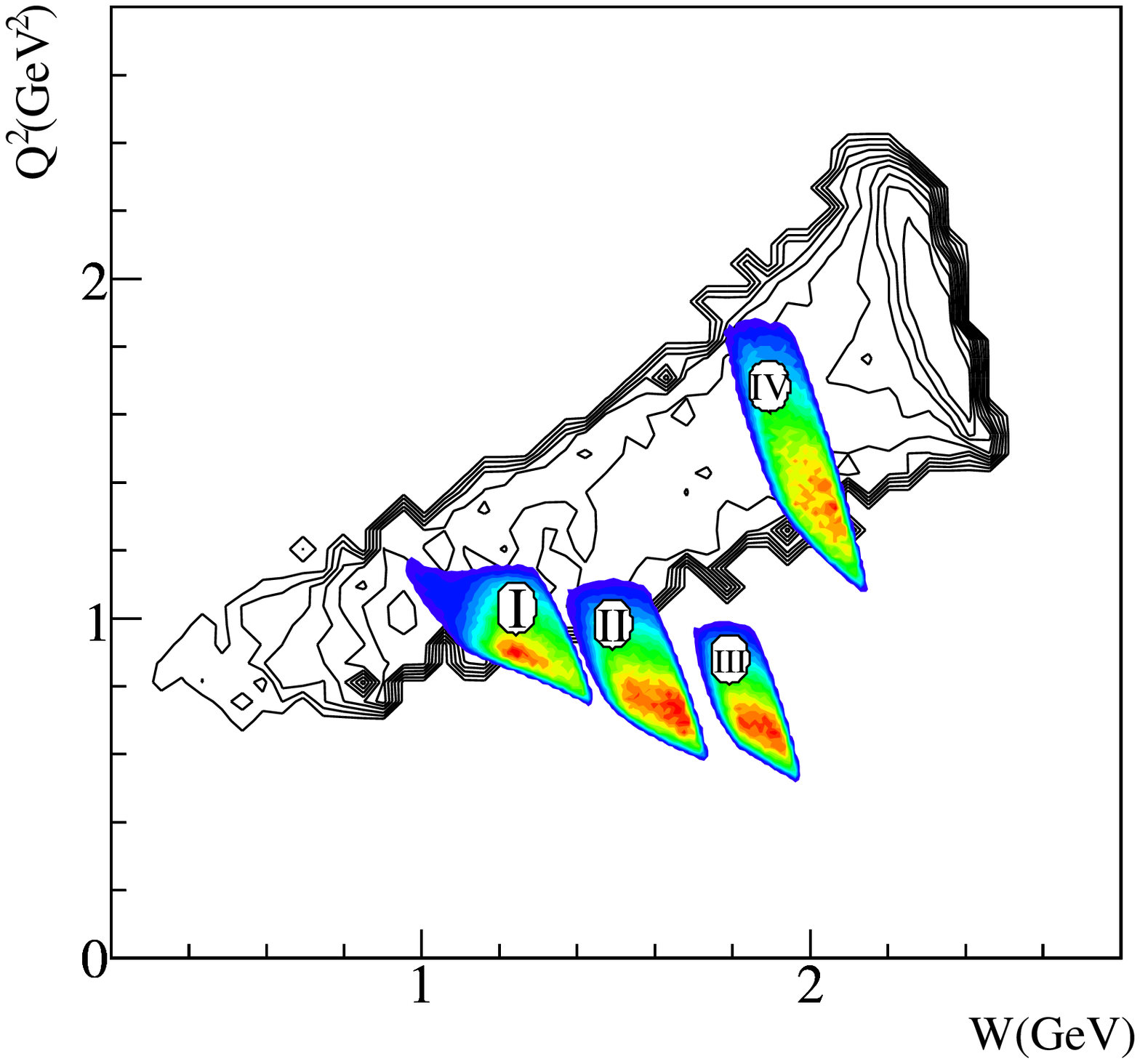}
  \caption{Kinematics coverage of the four resonance measurements (colored
contours), compared with the DIS vertex kinematics (black contours).}
\label{fig:reskine}
 \end{center}
\end{figure}

The physics inputs to HAMC for $e-^2$H elastic, quasi-elastic, DIS, as well as
the cross sections were all based on existing data and the uncertainties are 
small. The uncertainty of the correction was thus dominated by that from the resonance
asymmetry models. 
The validity of these models were evaluated by comparing the measured 
asymmetries from the resonance kinematics, RES I through IV, 
with calculations from these models. 
The kinematic coverage of resonance measurements is shown in 
Fig.~\ref{fig:reskine}.
These resonance asymmetries were reported in Ref.~\cite{Wang:2013kkc}, 
and it was
found that the data agree well with both resonance 
models~\cite{Matsui:2005ns,Gorchtein:2011mz} except RES I. Results at RES I
agreed with the two models at the two standard deviation level. 
The uncertainty from the resonance models was taken to be either the 
observed difference between resonance data and model, or the statistical 
uncertainty of the resonance asymmetry measurement, 
whichever is larger. This gives different model uncertainties as follows:
\begin{itemize}
 \item  For $W^2 < 1.96$~(GeV)$^2$ or the $\Delta(1232)$ region: 
RES I locates primarily in this region. 
The observed $25\%$ relative discrepancy between RES I data and the calculation was 
used as the model uncertainty in this region;
 \item  For $1.96<W^2 < 3.0$~(GeV)$^2$: RES II locates primarily in this region.
Since the RES II asymmetry result agreed well with both models,  
the 10.0\% relative statistical uncertainty of the RES II asymmetry 
was used as the model uncertainty in this region.
\item  For $3.0<W^2 <4.0$~(GeV)$^2$: Both RES III and IV locate in this region. Since the 
agreement with the calculations was well within the statistical uncertainties, 
the relative uncertainties for RES III and IV (8.9\% and 15.4\% respectively)
were combined, and the resulting value of $7.7\%$ was used as the model uncertainty 
in this region.
\end{itemize}
For radiative corrections at DIS kinematics, the resonance models affect the 
denominator, but not the numerator of Eq.~(\ref{eq:frc}). 
Therefore the above model uncertainty affects directly the DIS
corrections. These uncertainties were combined with the fractional events whose 
vertexes fell within the corresponding $W$ region to estimate the uncertainty 
on $\langle A(Q_\q{vtx}^2, x_\q{vtx})\rangle$ and $\bar f_\q{rc}$. 
For radiative corrections at resonance kinematics, the resonance models
affect both the denominator and the numerator of Eq.~(\ref{eq:frc}). The uncertainty
of the model itself therefore cancels out in principle in the correction factor
$\bar f_\q{rc}$.  For resonance kinematics, a conservative $20\%$ relative uncertainty
was used for $\bar f_\q{rc}$. 

The radiative correction factor $1+\bar f_{rc}$ obtained from the above procedure is shown in 
Table~\ref{tab:radcorfactor} for the two models separately. The 
average value of the two models were applied to the measured asymmetries
of this experiment. 

\begin{widetext}
\begin{table}[!htp]
\begin{center}
\begin{tabular}{c|c|c|c|c|c}\hline\hline
Kinematics     &  Resonance Models  
               & $A(\langle Q_\q{det}^2\rangle,\langle x_\q{det}^2\rangle)$ &  $\langle A(Q_\q{vtx}^2,x_\q{vtx}^2)\rangle$ & $1+\bar f_{rc}$ & $1+\bar f_{rc}$ \\ 
     & used    &     ppm     &   ppm   &     &  average  \\ \hline
\multirow{2}{*}{DIS \#1} & Ref.~\cite{Matsui:2005ns}   &  $-88.6$  &  $-86.8$  & $1.021\pm 0.020$   & $1.015\pm 0.021$   \\ 
                         & Ref.\cite{Gorchtein:2011mz} &  $-88.6$  &  $-87.8$  & $1.009\pm 0.020$   & \\ \hline
\multirow{2}{*}{DIS \#2} & Ref.~\cite{Matsui:2005ns}   & $-159.6$  & $-156.6$  & $1.019\pm 0.004$ & $1.019\pm 0.0043$  \\
                         & Ref.\cite{Gorchtein:2011mz} & $-159.6$  & $-156.7$  & $1.019\pm 0.004$ & \\ \hline
\multirow{2}{*}{RES I}   & Ref.~\cite{Matsui:2005ns}   &  $-93.4$ &  $-82.2$ & $1.137\pm 0.027$  & $1.1095\pm 0.0352$ \\ 
                         & Ref.\cite{Gorchtein:2011mz} &  $-89.0$ &  $-82.2$ & $1.082\pm 0.016$  & \\  \hline
\multirow{2}{*}{RES II}  & Ref.~\cite{Matsui:2005ns}   &  $-65.5$ &  $-65.5$ & $1.0002\pm 0.0000$& $1.0205\pm 0.0207$ \\ 
                         & Ref.\cite{Gorchtein:2011mz} &  $-71.1$ &  $-68.3$ & $1.0408\pm 0.0082$& \\  \hline
\multirow{2}{*}{RES III} & Ref.~\cite{Matsui:2005ns}   &  $-58.6$ &  $-59.1$ & $0.9930\pm 0.0014$& $1.0005\pm 0.0076$ \\ 
                         & Ref.\cite{Gorchtein:2011mz} &  $-62.5$ &  $-62.0$ & $1.0079\pm 0.0016$& \\  \hline
\multirow{2}{*}{RES IV}  & Ref.~\cite{Matsui:2005ns}   & $-117.5$ & $-116.7$ & $1.0063\pm 0.0013$& $1.0170\pm 0.0112$ \\ 
                         & Ref.\cite{Gorchtein:2011mz} & $-123.7$ & $-120.4$ & $1.0276\pm 0.0055$& \\  \hline
\multirow{2}{*}{RES V}   & Ref.~\cite{Matsui:2005ns}   & $-103.9$ & $-101.4$ & $1.0241\pm 0.0048$& $1.0134\pm 0.0110$ \\ 
                         & Ref.\cite{Gorchtein:2011mz} & $-103.9$ & $-103.6$ & $1.0027\pm 0.0005$& \\  \hline
\hline
\end{tabular}
\end{center}
\caption{Radiative correction factors. For each kinematics, the simulated 
asymmetries using two resonance models are shown. 
In kinematic regions where the resonance
models are not available, the cross-section-scaling model was used. 
These asymmetries were input to 
Eq.~(\ref{eq:frc}) to obtain the radiative correction factors. 
Results from the two models were 
averaged to provide the final correction $1+\bar f_{rc}$, and the
difference between the two was combined with uncertainties of resonance models 
themselves to provide the total uncertainty on $\bar f_{rc}$.}\label{tab:radcorfactor}
\end{table}
\end{widetext}

\subsection{Box Diagram Corrections}\label{sec:ana_box}

Box diagram corrections refer to effects that arise 
when the electron simultaneously exchanges two bosons 
($\gamma\gamma$, $\gamma Z$, or $ZZ$ box) with the target, and 
are dominated by the $\gamma\gamma$ and the $\gamma Z$ box diagrams. 
For PVES asymmetries, the box diagram effects include  those from the interference between 
$Z$-exchange 
and the $\gamma\gamma$ box, the interference between $\gamma$-exchange and 
the $\gamma Z$ box, and the effect of the $\gamma\gamma$ box on the 
electromagnetic cross sections.
It is expected that there is at least partial cancellation among these
three terms. 
The box-diagram corrections were applied as 
\begin{eqnarray}
  A^\q{box-corrected} &=& (1+\bar f_\q{box})A_e^{\q{meas}}~.
\end{eqnarray}
Corrections for the $\gamma\gamma$ box effect to the measured 
electron asymmetry were estimated to be 
$\bar f_{\gamma\gamma \q{box}}=-0.002$ and $-0.003$
for DIS \#1 and \#2, respectively. For these DIS kinematics, the effects of the 
$\gamma Z$ and $ZZ$ boxes were treated as part of 
the electroweak radiative corrections and will be described in 
Sec.~\ref{sec:result_pdf} [Eqs.~(\ref{eq:c1uth}-\ref{eq:c2dth})].  
For resonance kinematics, the combined corrections for 
$\gamma\gamma$ and $\gamma Z$ boxes (i.e. the full box correction) 
were estimated to be $\bar f_{\gamma\gamma,\gamma Z \q{boxes}}=+0.005$. 
A relative 100\% uncertainty was used for 
all box-diagram corrections.

\section{Results}\label{sec:results}


\subsection{Asymmetry results for both DIS and resonance settings}\label{sec:results_asym}

Table~\ref{tab:allAsym} presents the measured asymmetries along with their kinematics, 
all corrections, and the final physics asymmetry results. 
The $x$ and $Q^2$ values were obtained from the data and therefore were weighted by the scattering 
cross section. 
The dithering-corrected asymmetries were used as $A^\q{bc,raw}$ and the difference 
between dithering and regression methods were used as the systematic uncertainty of
$A^\q{bc,raw}$ (see Table~\ref{tab:beamcorr}). 
In addition to the corrections and uncertainties 
presented in Sections~\ref{sec:optics} through~\ref{sec:ana_box}, 
deadtime corrections from Ref.~\cite{Subedi:2013jha} were also applied to
the asymmetries. 
We chose asymmetries measured by the narrow triggers of the DAQ as $A^\q{bc,raw}$ 
because of the smaller counting deadtime and the associated uncertainty. 
All corrections were applied using Eq.~(\ref{eq:Aphys_cor}).
The largest corrections are due to beam polarization, DAQ deadtime, 
and electromagneic radiative corrections. The largest uncertainties come from 
the beam normal asymmetry and determination of the $Q^2$ values. 
We also note that the pair-production background, though very small for the
present experiment, causes an uncertainty typically one order of magnitude larger than
that from the charged pion background because one cannot reject pair-production 
background with PID detectors. 

\begin{widetext}
\begin{table}[!htp]
  \begin{center}
    \begin{tabular}{c|c|c|c|c|c|c|c|c}
      \hline\hline

\multicolumn{9}{c}{Kinematics}\\\hline
               &   DIS\#1  & Left DIS\#2 & Right DIS\#2     & RES I    & RES II    & RES III  & RES IV    & RES V\\\hline
$E_b$ (GeV)    & $6.067$   & \multicolumn{2}{c|}{$6.067$}   & $4.867$  & $4.867$   & $4.867$   & $6.067$   & $6.067$ \\
$\theta_0$     &$12.9^\circ$&\multicolumn{2}{c|}{$20.0^\circ$}&$12.9^\circ$&$12.9^\circ$&$12.9^\circ$&$15.0^\circ$&$14.0^\circ$\\
$E_0^\prime$ (GeV)& $3.66$    &\multicolumn{2}{c|}{ $2.63$}   & $4.00$    & $3.66$    & $3.10$    & $3.66$    & $3.66$ \\
$\langle Q^2\rangle_\q{data}$ [(GeV/$c$)$^2$]
               & $1.085$   & \multicolumn{2}{c|}{$1.901$}  & $0.950$   & $0.831$   & $0.757$   & $1.472$  & $1.278$ \\
$\langle x\rangle_\q{data}$ 
               & $0.241$   &\multicolumn{2}{c|}{ $0.295$}  & $0.571$   & $0.335$   & $0.228$   & $0.326$   & $0.283$ \\
$\langle W\rangle_\q{data}$ (GeV)
               & $2.073$   & \multicolumn{2}{c|}{$2.330$}  & $1.263$   & $1.591$   & $1.857$   & $1.981$   & $2.030$ \\\hline
$Y_3$          & $0.434$   & \multicolumn{2}{c|}{$0.661$}  & $0.340$   & $0.353$   & $0.411$   & $0.467$   & $0.451$ \\
$R_V$          & $0.808$   & \multicolumn{2}{c|}{$0.876$}  & $-$   & $-$   & $-$   & $-$   & $-$ \\
$Y_3R_V$       & $0.351$   & \multicolumn{2}{c|}{$0.579$}  & $-$   & $-$   & $-$   & $-$   & $-$ \\\hline
$A^\q{bc,raw}$ (ppm) 
               & $-78.45$  & $-140.30$    & $-139.84$     & $-55.11$  & $-63.75$  & $-54.38$  & $-104.04$ & $-67.87$\\
    (stat.)    &$\pm 2.68$ & $\pm 10.43$  & $\pm 6.58$    &$\pm 6.77$ &$\pm 5.91$ & $\pm 4.47$&$\pm 15.26$& $\pm 21.25$\\ 
    (syst.)    &$\pm 0.07$ & $\pm 0.16$   & $\pm 0.46$    &$\pm 0.10$ &$\pm 0.15$ & $\pm 0.24$&$\pm 0.26$ & $\pm 0.72$ \\
\hline 
\multicolumn{9}{c}{Corrections with systematic uncertainties}\\
\hline 
$P_b$          & $88.18\%$ &$89.29\%$     & $88.73\%$     & $90.40\%$ &  $90.40\%$& $90.40\%$ & $89.65\%$ & $89.65\%$ \\
$\Delta P_b$   &$\pm 1.76\%$&$\pm 1.19\%$ &$\pm 1.50\%$   &$\pm 1.54\%$&$\pm 1.54\%$&$\pm 1.54\%$&$\pm 1.24\%$&$\pm 1.24\%$ \\\hline
$1+\bar f_\q{depol}$ & $1.0010$&\multicolumn{2}{c|}{$1.0021$}&$1.0005$&$1.0003$ & $1.0009$& $1.0006$&$1.0008$ \\
 (syst.)       & $<10^{-4}$   & \multicolumn{2}{c|}{$<10^{-4}$}  & $<10^{-4}$  & $<10^{-4}$  & $<10^{-4}$  & $<10^{-4}$ \\\hline
$1+\bar f_{\q{Al}}$ & $0.9999$  & $0.9999$  & $0.9999$ & $0.9997$    & $0.9998$   & $0.9999$   & $0.9998$ & $0.9999$ \\
 (syst.)       &$\pm 0.0024$  &$\pm 0.0024$  &$\pm 0.0024$ &$\pm 0.0043$ &$\pm 0.0043$&$\pm 0.0043$& $\pm 0.0043$& $\pm 0.0043$ \\\hline
$1+\bar f_\q{dt}$ & $1.0147$  & $1.0049$ & $1.0093$     & $1.0148$    & $1.0247$   & $1.0209$   & $1.0076$ & $1.0095$ \\
 (syst.)       &$\pm 0.0009$& $\pm 0.0004$ & $\pm 0.0013$ &$\pm 0.0006$ &$\pm 0.0023$&$\pm 0.0041$&$\pm 0.0004$ & $0.0007$\\\hline
$1+\bar f_\q{rc}$& $1.015$ & \multicolumn{2}{c|}{$1.019$} & $1.1095$    &$1.0205$    &  $1.0005$  & $1.0170$ & $1.0134$ \\
 (syst.)     &$\pm 0.020$ & \multicolumn{2}{c|}{$\pm 0.004$} & $\pm 0.0352$ &$\pm 0.0207$&$\pm 0.0076$& $\pm 0.0112$ & $0.0110$\\\hline
$1+\bar f_\q{\gamma\gamma box}$       
                & $0.998$     & $0.997$    & $-$    & $-$ & $-$     & $-$   & $-$    & $-$ \\
$1+\bar f_\q{\gamma\gamma,\gamma Z boxes}$       
                & $-$         & $-$        & $1.005$    & $1.005$ & $1.005$     & $1.005$    & $1.005$    & $1.005$ \\
  (syst.)       & $\pm 0.002$ & $\pm 0.003$&$\pm 0.005$ & $\pm 0.005$ & $\pm 0.005$ & $\pm 0.005$&$\pm 0.005$ & $\pm 0.005$ \\
\hline
\multicolumn{9}{c}{Systematic uncertainties $\Delta A^\q{phys}/A^\q{phys}$ with no correction}\\
\hline 
charged pion
                &$\pm 9\times 10^{-5}$ &$\pm 6\times 10^{-5}$ &$\pm 3\times 10^{-5}$ & $\pm 1.8\times 10^{-4}$&$\pm 4.6\times 10^{-4}$&$\pm 1.9\times 10^{-4}$& $\pm 3\times 10^{-5}$  & $\pm 1.0\times 10^{-4}$ \\
pair production
                & $\pm 0.0004$ & $\pm 0.004$  & $\pm 0.002$  & $\pm 0.003$  & $\pm 0.003$ & $\pm 0.003$ & $\pm 0.003$& $\pm 0.003$\\
beam $A_n$
                & $\pm 0.025$  & $\pm 0.025$  & $\pm 0.025$  & $\pm 0.025$  & $\pm 0.025$  & $\pm 0.025$  & $\pm 0.025$  & $\pm 0.025$\\
$Q^2$           &$\pm 0.0085$  & $\pm 0.0064$ &$\pm 0.0065$  & $\pm 0.0081$ & $\pm 0.0073$ &$\pm 0.008$  & $\pm 0.035$  & $\pm 0.037$ \\
rescattering & $\ll 0.002$ & $\ll 0.002$ & $\ll 0.002$ & $\ll 0.002$ & $\ll 0.002$ & $\ll 0.002$ & $\ll 0.002$ & $\ll 0.002$ \\
target impurity   & $\pm 0.0006$ & $\pm 0.0006$ & $\pm 0.0006$ & $\pm 0.0006$ & $\pm 0.0006$ & $\pm 0.0006$ & $\pm 0.0006$ & $\pm 0.0006$ \\
\hline
\multicolumn{9}{c}{Asymmetry Results}\\
\hline
$A^\q{phys}$ (ppm) 
               & $-91.10$    & \multicolumn{2}{c|}{$-160.80$} & $-68.62$   & $-73.75$   & $-61.49$  & $-118.97$     & $-77.50$ \\
(stat.)        & $\pm 3.11$  & \multicolumn{2}{c|}{$\pm 6.39$}    & $\pm 8.43$  & $\pm 6.84$ & $\pm 5.05$ & $\pm 17.45$ & $\pm 24.27$\\
(syst.)        & $\pm 2.97$  & \multicolumn{2}{c|}{$\pm 3.12$}    & $\pm 3.26$  & $\pm 2.78$ & $\pm 2.06$ & $\pm 5.54$  & $\pm 3.84$\\ 
(total)        & $\pm 4.30$  & \multicolumn{2}{c|}{$\pm 7.12$}    & $\pm 9.04$  & $\pm 7.38$ & $\pm 5.46$ & $\pm 18.31$ & $\pm 24.57$ \\
\hline
\hline\hline
      \end{tabular}
  \end{center}
  \caption{Asymmetry results on $\vec e-^2$H 
parity-violating scattering from the PVDIS experiment at JLab. The DIS results were 
previously published in Ref.~\cite{Wang:2014bba}.
The kinematics shown include the beam energy $E_b$, central 
angle and momentum settings of the spectrometer $\theta_0, E_0^\prime$,
the actual kinematics averaged from the data (cross-section-weighted) $\langle Q^2\rangle$ and 
$\langle x\rangle$, the kinematics factor $Y_3$ [calculated using $\langle Q^2\rangle$, 
$\langle x\rangle$, $E_b$ and Eq.~(\ref{eq:Apvdis1})], the PDF valence quark distribution
function ratio $R_V$ calculated from MSTW2008~\cite{Martin:2009iq} Leading-Order parameterization 
and Eq.~\ref{eq:Rpdf}, 
and the product $Y_3R_V$ that provides the lever arm for isolating the $C_{2q}$
contribution to the asymmetry. 
The electron asymmetries obtained from the narrow trigger
of the DAQ with beam dithering corrections, $A^\q{bc,raw}$,
were corrected for the effects from the beam polarization $P_b$ 
and many systematic effects including:
the beam depolarization effect $\bar f_\q{depol}$, 
the target aluminum endcap $\bar f_{\q{Al}}$,
the DAQ deadtime $\bar f_\q{dt}$~\cite{Subedi:2013jha},
the radiative correction $\bar f_\q{rc}$ that includes effects
from energy losses of incoming and scattered electrons as well as
the spectrometer acceptance and detector efficiencies, 
and the box-diagram correction $\bar f_\q{\gamma\gamma box}$ (for DIS)
and $\bar f_{\gamma\gamma,\gamma Z \q{boxes}}$ (for resonances).
Systematic effects that do not require a correction to the asymmetry 
include: 
the charged pion and the pair production background , 
the beam normal asymmetry, 
the uncertainty in the determination of $Q^2$, 
the re-scattering background, 
and the target impurity.
Final results on the physics asymmetries $A^\q{phys}$ are shown
with their statistical, systematic, and total uncertainties.
}
\label{tab:allAsym}
\end{table}
\end{widetext}

\subsection{Group trigger asymmetry results for resonance kinematics}\label{sec:results_group}

\begin{figure}[!htp]
 \hspace*{0.1\textwidth}
 \includegraphics[width=0.4\textwidth]{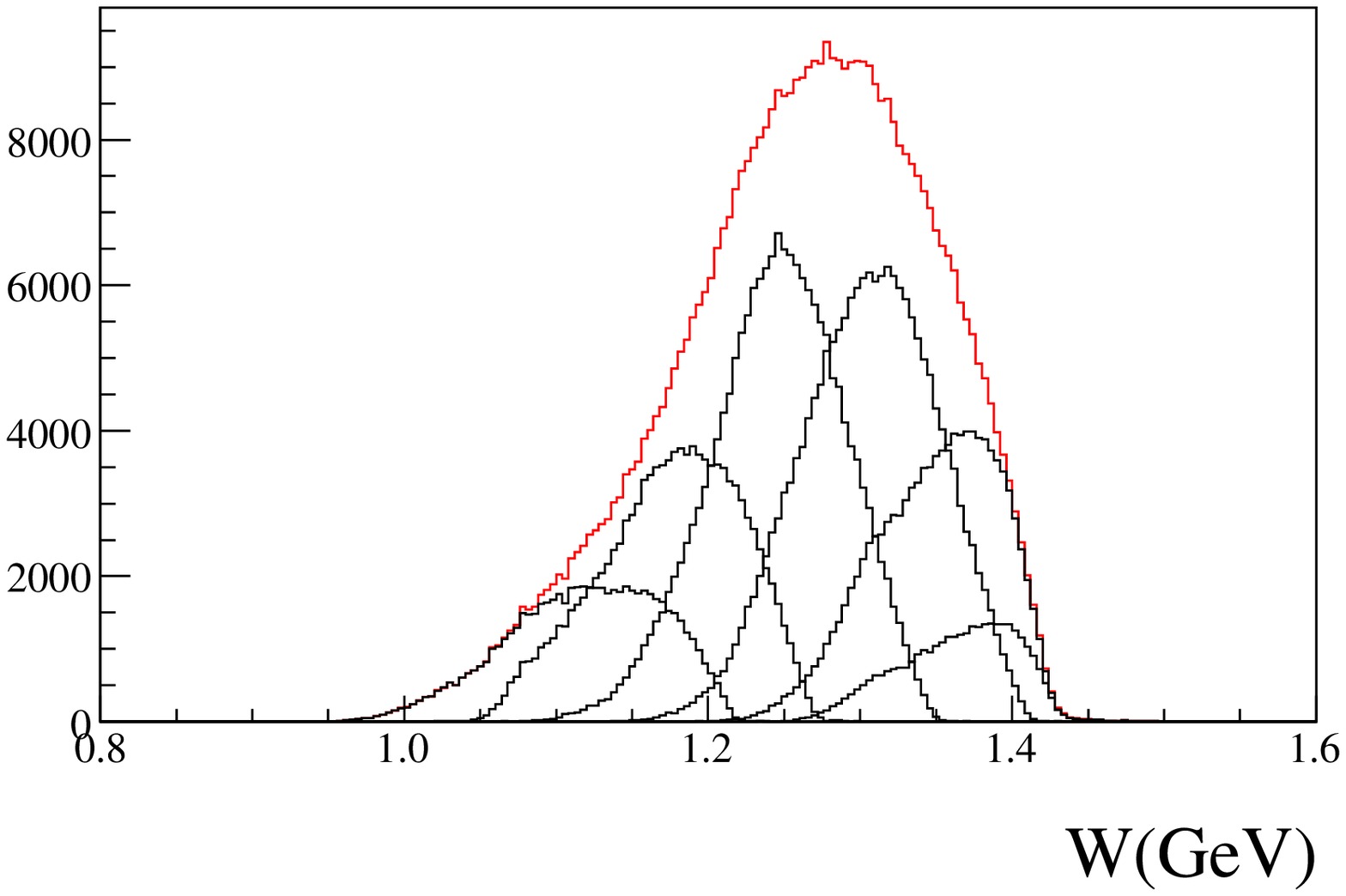}
 \includegraphics[width=0.4\textwidth]{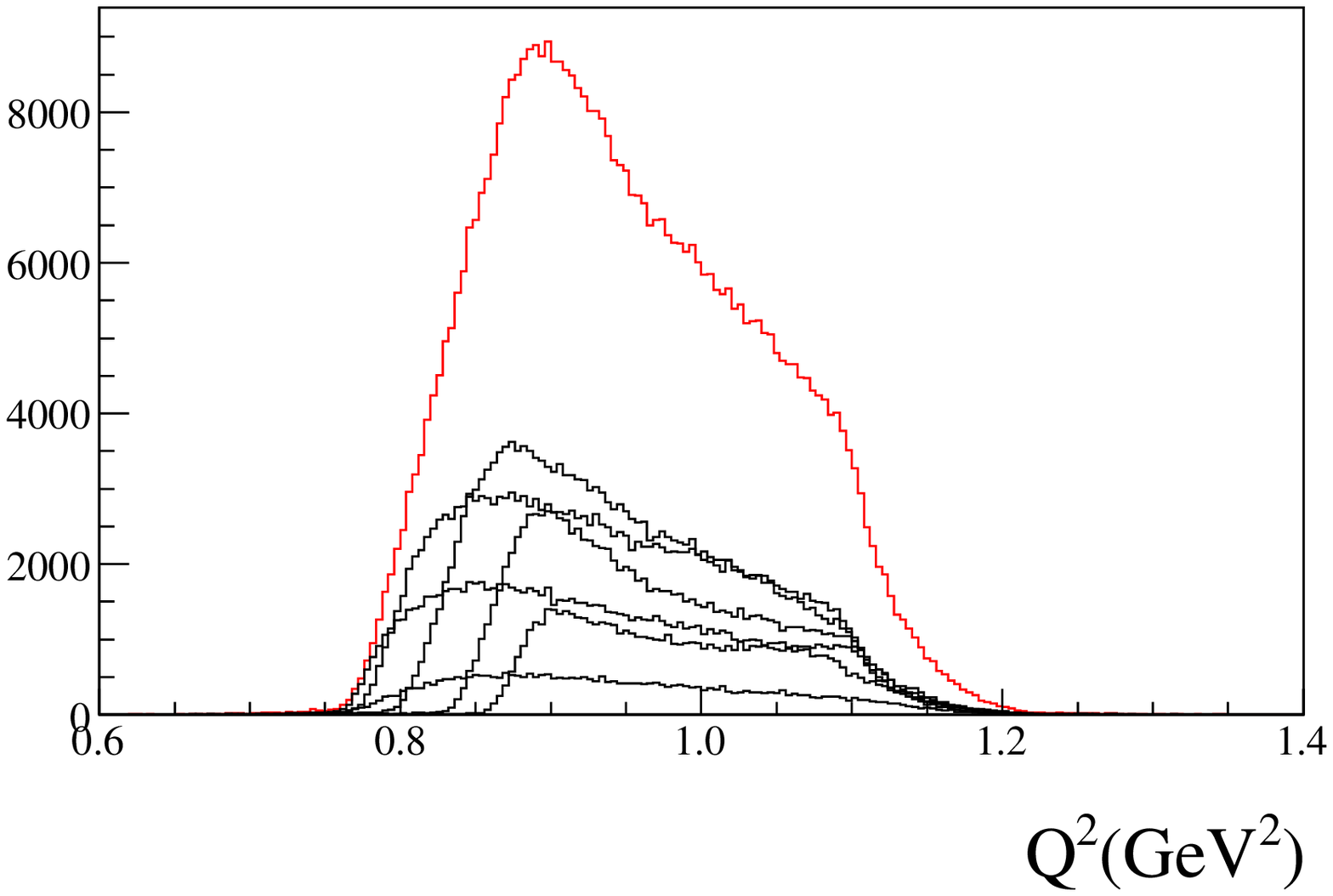}
 \caption{Event distributions in $W$ (left) and $Q^2$ (right) for the six
DAQ groups taken at setting RES I. The coverage in $W$ increases monotonously 
from group 1 to 6. The red (the highest) histogram shows the global trigger events.}
\label{fig:q2wresgroups}
\end{figure}

The asymmetry data taken in the resonance region are of particular value:
they provided the first PVES asymmetries over the complete nucleon 
resonance region, and the first
test of quark-hadron duality for electroweak observables.  
For nucleon resonance studies, fine-binning in $W$ is often desired to reveal 
detailed resonance structure. 
As described in Ref.~\cite{Subedi:2013jha}, in addition to the so-called 
global electron triggers that lead to the main results presented in the 
previous section, the detector package was divided into groups, for which
group electron triggers were constructed, and data recorded in the same
way as global triggers. Settings RES I, II, IV and V on the left HRS 
had six groups, while setting RES III on the right HRS 
had eight groups.  The kinematics coverage varies between group 
triggers, providing different coverage in $W$. Figure~\ref{fig:q2wresgroups}
shows the $Q^2$ and $W$ coverage of the six groups for setting RES I.
As one can see, the $Q^2$ range is similar but the $W$ coverages of the six
groups are different. 

Because there were overlaps in the detector grouping of the DAQ (that is, some 
lead glass blocks were used as inputs to two group triggers), approximately 
(10-30)\% events were recorded simultaneously by two adjacent groups and 
the group trigger events were not completely uncorrelated. Nevertheless, asymmetries extracted 
for individual groups allowed a study of the $W$-dependence of the asymmetry.  
Corrections to the raw asymmetry from group triggers were applied in the same
manner presented in the previous section. Among all 
corrections, two corrections were expected to vary among groups to an 
observable level, and must be evaluated carefully
for individual groups: deadtime (rate-dependent) 
and electromagnetic radiative corrections (kinematic-dependent). 
All other corrections either do not depend on groups, or their kinematic
variation is expected to be well below the statistical uncertainty of
the measurement.

Tables~\ref{tab:grp_rawl} and~\ref{tab:grp_rawr} show respectively for 
the left and the right HRS: the average kinematics $\langle W\rangle$ 
and $\langle Q^2\rangle$, the raw measured asymmetries, the two
group-dependent corrections for individual groups, and the physics
asymmetry results.
Corrections and uncertainties that do not
depend on groups are the same as in Table~\ref{tab:allAsym}. 
Similar to DIS results, we used the dithering-corrected asymmetries
measured from the narrow path triggers of the DAQ as raw-asymmetry 
inputs to the analysis because the narrow path had smaller counting
deadtime and associated uncertainties.

\begin{table}[!htp]
\begin{center}
\begin{tabular}{c|cccccc}
\hline\hline
Group              & 1         & 2         & 3         & 4         & 5         & 6     \\\hline
\multicolumn{7}{c}{RES I} \\ \hline
$\langle Q^2 \rangle_\mathrm{data} \q{[(GeV/c)^2]}$ 
                   & $0.992$   &$0.966$    &$0.948$    &$0.940$    &$0.931$    &$0.940$ \\
$\langle W\rangle_\mathrm{data} \q{(GeV)}$          
                   & $1.119$   &$1.175$    &$1.245$    &$1.305$    &$1.350$    &$1.364$ \\\hline
$A_\q{dit}^\q{bc,raw}\q{(ppm)}$       
                   & $-30.84$  &$-57.65$   &$-54.01$   &$-46.12$   &$-60.24$   &$-95.49$ \\
(stat.)            & $18.31$   &$14.34$    &$11.51$    &$11.33$    &$14.41$    &$23.85$ \\ \hline
$1+\bar f_\q{dt}$  & $1.0077$  &$1.0089$   &$1.0105$   &$1.0106$   &$1.0088$   &$1.0069$ \\
(syst.)            & $0.0004$  &$0.0009$   &$0.0004$   &$0.0010$   &$0.0008$   &$0.0009$ \\ \hline
$1+\bar{f_{rc}}$   & $1.359$   &$1.150$    &$1.045$    &$1.024$    &$1.011$    &$1.010$ \\
(syst.)            & $0.155$   &$0.031$    &$0.014$    &$0.005$    &$0.004$    &$0.004$ \\
\hline
$A^\q{phys}$ (ppm) &$-46.95$   &$-74.35$   &$-63.37$   &$-53.05$   &$-68.26$   &$-107.89$ \\
(stat.)            &$\pm 27.87$&$\pm 18.49$&$\pm 13.50$&$\pm 13.03$&$\pm 16.33$&$\pm 26.95$ \\
(syst.)            &$\pm 7.42$ &$\pm 3.36$ &$\pm 2.26$ &$\pm 1.77$ &$\pm 2.26$ &$\pm 3.58$ \\
(total)            &$\pm 28.84$&$\pm 18.80$&$\pm 13.69$&$\pm 13.15$&$\pm 16.48$&$\pm 27.18$ \\

\hline
\multicolumn{7}{c}{RES II} \\ \hline
$\langle Q^2 \rangle_\mathrm{data} \q{[(GeV/c)^2]}$ 
                   &$0.856$    &$0.849$    & $0.834$   & $0.820$   & $0.808$   & $0.819$ \\
$\langle W\rangle_\mathrm{data} \q{(GeV)}$ 
                   &$1.503$    &$1.533$    & $1.583$   & $1.629$   & $1.662$   & $1.672$ \\\hline
$A_\q{dit}^\q{bc,raw}\q{(ppm)}$ 
                   &$-60.67$   &$-55.15$   &$-77.16$   & $-65.46$  & $-65.92$  & $-61.73$ \\
(stat.)            &$13.24$    &$11.18$    & $10.55$   & $10.57$   & $12.95$   & $20.71$ \\ \hline
$1+\bar f_\q{dt}$  &$1.0134$   &$1.0152$   & $1.0160$  & $1.0158$  & $1.0135$  & $1.0107$ \\
(syst.)            &$0.0008$   &$0.0017$   & $0.0006$  & $0.0014$  & $0.0012$  & $0.0015$ \\ \hline
$1+\bar{f_{rc}}$   &$1.032$    &$1.017$    & $1.012$   & $1.000$   & $0.995$   & $0.995$ \\
(syst.)            &$0.006$    &$0.003$    & $0.002$   & $<0.001$  & $0.001$   & $0.001$ \\
\hline
$A^\q{phys}$ (ppm) &$-70.56$   &$-63.31$   &$-88.21$   &$-73.94$   &$-73.91$   &$-69.02$ \\
(stat.)            &$\pm 15.40$&$\pm 12.83$&$\pm 12.06$&$\pm 11.94$&$\pm 14.52$&$\pm 23.16$ \\
(syst.)            &$\pm 2.35$ &$\pm 2.09$ &$\pm 2.89$ &$\pm 2.42$ &$\pm 2.42$ &$\pm 2.26$ \\
(total)            &$\pm 15.58$&$\pm 13.00$&$\pm 12.40$&$\pm 12.18$&$\pm 14.72$&$\pm 23.27$ \\
\hline
\multicolumn{7}{c}{RES IV} \\ \hline
$\langle Q^2 \rangle_\mathrm{data} \q{[(GeV/c)^2]}$ 
                   & $1.531$   & $1.533$   & $1.473$   & $1.442$   & $1.427$   & $1.378$\\
$\langle W\rangle_\mathrm{data} \q{(GeV)}$ 
                   & $1.901$   & $1.922$   & $1.978$   & $2.020$   & $2.049$   & $2.071$\\\hline
$A_\q{dit}^\q{bc,raw}\q{(ppm)}$ 
                   & $-103.29$ & $-91.13$  & $-82.82$  & $-117.19$ & $-142.95$ & $87.30$\\
(stat.)            & $32.87$   & $32.21$   & $27.24$   & $27.00$   & $37.52$   & $96.85$\\ \hline
$1+\bar f_\q{dt}$  & $1.0057$  & $1.0057$  & $1.0061$  & $1.0061$  & $1.0055$  & $1.0049$ \\
(syst.)            & $0.0003$  & $0.0004$  & $0.0003$  & $0.0004$  & $0.0004$  & $0.0003$ \\ \hline
$1+\bar{f_{rc}}$   & $1.013$   & $1.013$   & $1.020$   & $1.027$   & $1.031$   & $1.032$ \\
(syst.)            & $0.003$   & $0.003$   & $0.004$   & $0.005$   & $0.006$   & $0.006$ \\
\hline
$A^\q{phys}$ (ppm) &$-118.02$  &$-104.13$  &$-95.32$   &$-135.81$  &$-166.21$  &$101.54$ \\
(stat.)            &$\pm 37.56$&$\pm 36.80$&$\pm 31.35$&$\pm 31.29$&$\pm 43.62$&$\pm 112.65$ \\
(syst.)            &$\pm 5.43$ &$\pm 4.79$ &$\pm 4.39$ &$\pm 6.28$ &$\pm 7.70$ &$\pm 4.71$ \\
(total)            &$\pm 37.95$&$\pm 37.11$&$\pm 31.66$&$\pm 31.91$&$\pm 44.30$&$\pm 112.75$ \\
\hline\hline
\end{tabular}
\caption{From left HRS group triggers: $\langle W\rangle$ 
and $\langle Q^2\rangle$ from data (cross-section weighted), 
beam-(dithering-)corrected raw asymmetries from narrow triggers, and group-dependent
corrections. Corrections and uncertainties that do not
depend on groups are the same as in Table~\ref{tab:allAsym} and are not
shown here.
After all corrections are applied, the final asymmetries are shown in the last row for each setting.
}\label{tab:grp_rawl}
\end{center}
\end{table}

\begin{table}
\begin{center}
\begin{tabular}{c|cccccccc}
\hline\hline
Group              &1 &2 &3 &4 &5 &6 &7 &8 \\\hline
\multicolumn{9}{c}{RES III} \\\hline
$\langle Q^2 \rangle_\mathrm{data} \q{[(GeV/c)^2]}$ 
                   & $0.731$   & $0.719$   & $0.730$   & $0.744$   & $0.761$   & $0.777$   & $0.796$  & $0.799$ \\
$\langle W\rangle_\mathrm{data} \q{(GeV)}$ 
                   & $1.928$   & $1.923$   & $1.905$   & $1.880$   & $1.851$   &$1.820$    & $1.790$  & $1.771$ \\ \hline
$A_\q{dit}^\q{bc,raw}\q{(ppm)}$ 
                   & $-58.62$  & $-38.74$  &$-56.02$   & $-56.74$  & $-56.67$  & $-57.15$  & $-52.57$ & $-35.99$ \\
(stat.)            & $26.82$   & $13.05$   & $9.95$    & $9.57$    & $9.58$    & $9.97$    & $11.13$  & $24.24$ \\\hline
$1+\bar f_\q{dt}$  & $1.0127$  & $1.0148$  & $1.0169$  & $1.0174$  & $1.0173$  & $1.0170$  & $1.0161$ & $1.0127$ \\
(syst.)            & $0.0011$  & $0.0010$  & $0.0011$  & $0.0010$  & $0.0010$  & $0.0010$  & 0$.0011$ & $0.0012$ \\\hline
$1+\bar{f_{rc}}$   & $1.022$   & $1.021$   & $1.024$   & $1.026$   & $1.025$   & $1.024$   & $1.020$  & $1.010$\\
(syst.)            & $0.004$   & $0.004$   & $0.005$   & $0.005$   & $0.005$   & $0.005$   & $0.004$  & $0.002$\\
\hline
$A^\q{phys}$ (ppm) &$-67.50$   &$-44.66$   &$-64.90$   &$-65.90$   &$-65.75$   &$-66.22$   &$-60.62$   &$-40.96$ \\
(stat.)            &$\pm 30.88$&$\pm 15.05$&$\pm 11.53$&$\pm 11.12$&$\pm 11.12$&$\pm 11.55$&$\pm 12.83$&$\pm 27.59$ \\
(syst.)            &$\pm 2.25$ &$\pm 1.49$ &$\pm 2.17$ &$\pm 2.21$ &$\pm 2.20$ &$\pm 2.21$ &$\pm 2.02$ &$\pm 1.36$ \\
(total)            &$\pm 30.97$&$\pm 15.12$&$\pm 11.73$&$\pm 11.33$&$\pm 11.33$&$\pm 11.76$&$\pm 12.99$&$\pm 27.62$ \\
\hline\hline
\end{tabular}
\end{center}
\caption{From right HRS group triggers: $\langle W\rangle$ 
and $\langle Q^2\rangle$ from data (cross-section-weighted), 
beam-(dithering-)corrected raw asymmetries from narrow triggers, and group-dependent
corrections. Corrections and uncertainties that do not
depend on groups are the same as in Table~\ref{tab:allAsym} and are not
shown here.
After all corrections are applied, the final asymmetries are shown in the last row for each setting.  
We did not perform a group analysis for setting RES V because of the very-low statistics. }\label{tab:grp_rawr}
\end{table}

\subsection{Test of quark-hadron duality using resonance PV asymmetries}

\begin{figure}
 \begin{center}
 \includegraphics[width=0.7\textwidth]{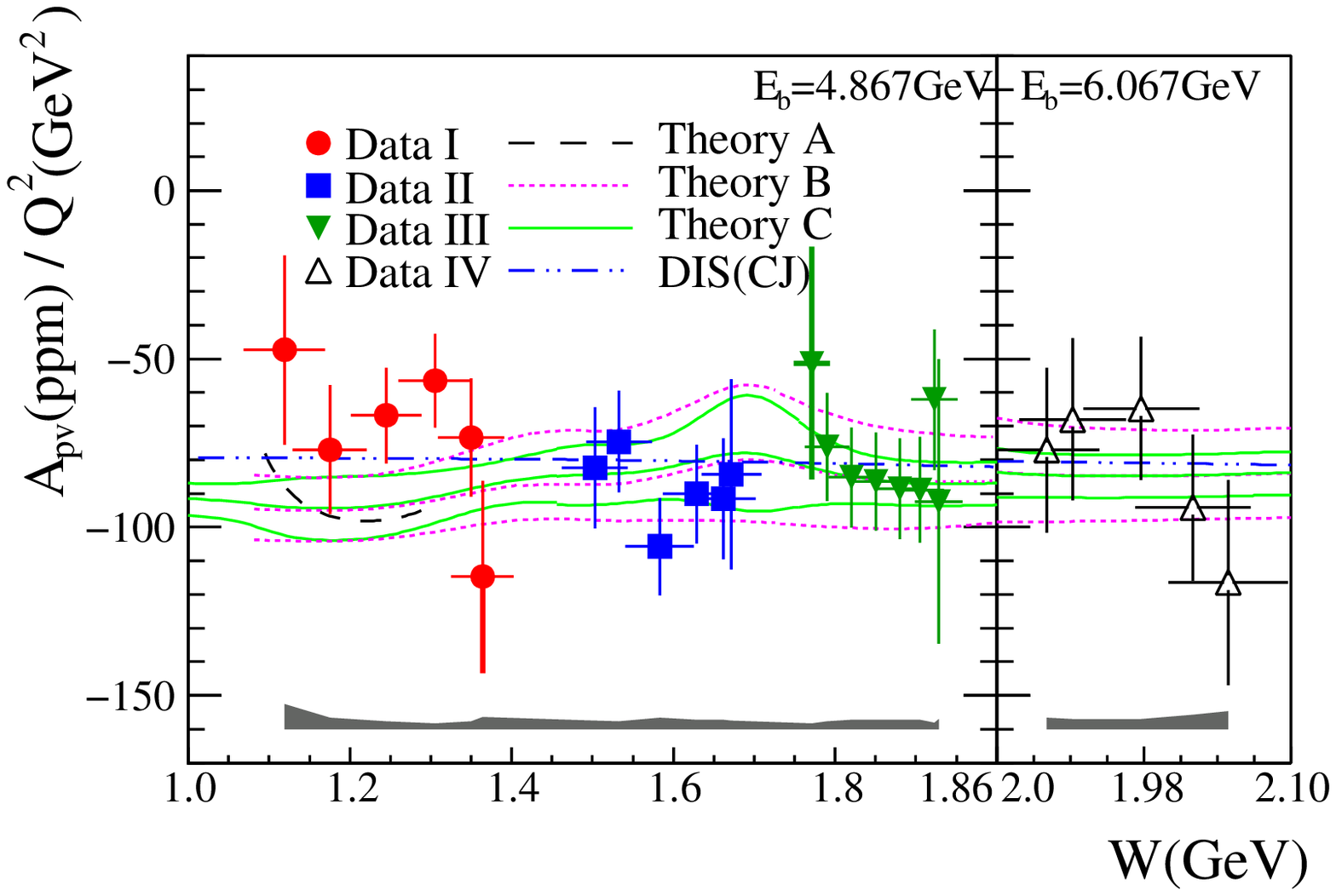}
 \caption{({\it Color online}) From Ref.~\cite{Wang:2013kkc}: 
$W$-dependence of the parity-violating asymmetries in  
$\vec e-^2$H scattering in the nucleon resonance region. The 
physics asymmetry results $A_{PV}^\q{phys}$ for the four kinematics 
RES I, II, III and IV (solid circles, solid squares, solid triangles, and
open triangles, respectively), in parts per million (ppm), 
are scaled by $1/Q^2$ and compared with calculations from 
Ref.~\cite{Matsui:2005ns} (Theory A, dashed), Ref.~\cite{Gorchtein:2011mz} 
(Theory B, dotted), 
Ref.~\cite{Hall:2013hta} (Theory C, solid) and the 
DIS estimation (dash-double-dotted) using Eq.~(\ref{eq:Apvdis_R}) with 
the extrapolated CJ PDF~\cite{Owens:2012bv}. 
The vertical error bars for the data are statistical uncertainties, while
the horizontal error bars indicate the root-mean-square values of the
$W$ coverage of each bin. The experimental
systematic uncertainties are shown as the shaded bands at the bottom. 
For each of the four
kinematics, calculations were
performed at the fixed $E_b$ and $Q^2$ values of each of the RES I, II, 
III and IV settings  
and with a variation in $W$ to match the coverage of the data.
Theories B and C each have three curves showing the central values
and the upper and the lower bounds of the calculation. Uncertainties
of the DIS calculation were below 1~ppm and are not visible.
}\label{fig:Apv_gr}
\end{center}
\end{figure}

Figure~\ref{fig:Apv_gr} shows the $W$-dependence of the group-trigger resonance 
asymmetry results $A_{PV}^\q{phys}$ of Tables~\ref{tab:grp_rawl} 
and~\ref{tab:grp_rawr}, scaled by $1/Q^2$. 
The data of adjacent bins in each kinematics 
typically have a 20-30\% overlap and are 
thus correlated, while the lowest and the highest bins of each kinematics have
larger overlaps with 
their adjacent bins. 
Figure~\ref{fig:Apv_gr} illustrates that all asymmetry data
are consistent with the three resonance
models and with the DIS estimation. 
No significant resonance structure is observed in the $W$-dependence
of the asymmetries.

The agreement with DIS-based calculations indicates
that quark-hadron duality holds for 
PVES asymmetries on the deuteron at the 
$(10-15)\%$ level throughout the resonance region, 
for $Q^2$ values just below 1~(GeV/$c$)$^2$. 
These results are comparable to 
the unpolarized electromagnetic structure
function data which verified duality at the $(5-10)\%$ level for the 
proton and $(15-20)\%$ for the neutron at similar $Q^2$ values, 
although the unpolarized 
measurements provided better resolution in $W$ and covered a broader
kinematic range~\cite{Malace:2009dg,Malace:2009kw,Malace:2011ad}.

\subsection{Extraction of electron-quark effective coupling $C_{2q}$ from DIS asymmetries}

\subsubsection{Calculation of PVDIS asymmetry sensitivity to $C_{2q}$}\label{sec:result_pdf}
In order to extract the electron-quark VA couplings $C_{2q}$, one must first study
the sensitivity of the measured PVDIS asymmetry to $C_{2q}$. Equation~\ref{eq:Apvdis1} was used
for this purpose. In this section, inputs to Eq.~\ref{eq:Apvdis1} will be 
explained in detail, including all physical constants and couplings and the structure
function evaluation. Uncertainties due to higher twist effects will be discussed at 
the end.

Electroweak radiative corrections were applied to all couplings used in the
calculation of the asymmetry. The electromagnetic fine structure constant
 $\alpha$ was evolved to the measured $Q^2$-values from 
$\alpha_{EM}\vert_{Q^2=0}=1/137.036$~\cite{Beringer:2012}. 
The evaluation takes into account purely electromagnetic vacuum polarization.
The Fermi constant is $G_F=1.1663787(6)\times 10^{-5}$ GeV$^{-2}$~\cite{Beringer:2012}. 
The $C_{1q,2q}$ were evaluated using Table 7 and Eq.~(114-115) of 
Ref.~\cite{Erler:2013xha} 
at our measured $Q^2$-values in the modified minimal subtraction 
($\overline{\q{MS}}$) scheme using a fixed
Higgs mass $M_H=125.5$~GeV:
\begin{eqnarray}
 C_{1u}^\q{SM} &=&-0.1887-0.0011\times {2 \over 3} \ln(\langle Q^2\rangle/0.14 \q{GeV}^2)\label{eq:c1uth}\\
 C_{1d}^\q{SM}&=&0.3419-0.0011\times {-1 \over 3} \ln(\langle Q^2\rangle/0.14 \q{GeV}^2)\label{eq:c1dth}\\
 C_{2u}^\q{SM}&=&-0.0351-0.0009\ln(\langle Q^2\rangle/0.078~\q{GeV}^2)\label{eq:c2uth}\\
 C_{2d}^\q{SM}&=&0.0248+0.0007\ln(\langle Q^2\rangle/0.021~\q{GeV}^2)\label{eq:c2dth}
\end{eqnarray}
and it is expected that the uncertainty is negligible. 
Equations~(\ref{eq:c1uth}-\ref{eq:c2dth}) include the ``charge radius effect'' and 
an estimate of the 
interference between $\gamma$-exchange and the $\gamma Z$ box, but not the 
effect from the $\gamma\gamma$ box. The effect from the $\gamma\gamma$ box was 
applied as a correction to the measured asymmetry as 
described in previous sections.

To express the measured asymmetries in terms of $2C_{1u}-C_{1d}$ and $2C_{2u}-C_{2d}$, 
we calculated the $F_{1,3}^{\gamma,\gamma Z}$ structure functions in 
Eqs.~(\ref{eq:Apvdis1}, \ref{eq:a11}, \ref{eq:a31}) and the resulting $a_{1,3}$ contribution to 
the asymmetry, see Table~\ref{tab:Apvdis_calc}. Here the approximation $Y_1=1$ was used, which
is valid if $R^\gamma=R^{\gamma Z}$. 
Also shown in Table~\ref{tab:Apvdis_calc} are values of  $2C_{1u}-C_{1d}$ and $2C_{2u}-C_{2d}$
evaluated at the $Q^2$-values of the measurement.
Three different parton 
distribution functions (PDFs) were used: the CTEQ/JLab (``CJ'') fit~\cite{Owens:2012bv} 
which provides structure functions at the next-to-leading order (NLO), 
the CT10~\cite{Lai:2010vv} (NLO only), and the leading-order (LO) MSTW2008~\cite{Martin:2009iq} fits. 
The CT10 and the MSTW2008 fits provide only PDF values but not the structure functions. 
For these two fits the quark-parton model (QPM) [Eqs.~(\ref{eq:F1gqpm}-\ref{eq:F3gzqpm})] 
was used to calculate structure functions from PDFs.  
The parametrization most suitable for our kinematics is the CJ fit, and it provides
three different sets: the medium (mid), minimum, and maximum. However the CJ fit 
is not applicable for $Q^2$-values below $1.7$~(GeV/$c$)$^2$.  
From the $Q^2=1.901$~(GeV/$c$)$^2$ comparison we found that the result of the
LO MSTW2008 fit is closest to CJ, therefore it was used to interpret the 
$Q^2=1.085$~(GeV/$c$)$^2$ result. 
Results in Table~\ref{tab:Apvdis_calc} were also used for uncertainty estimation: 
the variation between various fits (three fits for $Q^2=1.901$~(GeV/$c$)$^2$ and 
two fits for $Q^2=1.085$~(GeV/$c$)$^2$) are at the level of relative 0.5\% for the $a_1$ term 
and relative 5\% for the $a_3$ term of the asymmetry. 
The ``valence quark only'' values [Eq.~(\ref{eq:a1a3nopdf})] 
are also shown in Table~\ref{tab:Apvdis_calc}. These values differ from the 
PDF-based calculation by not more than 2\% and 20\% for the $a_1$ and the $a_3$ terms
respectively, which explains in part why the calculations are in-sensitive to 
the choice of the PDF fits.

\begin{table}[!htp]
\begin{center}
\begin{tabular}{p{5cm}|c|c}\hline\hline
   & $\langle Q^2\rangle =1.085$, & $\langle Q^2\rangle=1.901$, \\
   & $\langle x\rangle=0.241$ & $\langle x\rangle=0.295$\\\hline
 \multicolumn{3}{c}{Physical couplings used in the Calculation} \\\hline
$\alpha_{EM}(Q^2)$           & $1/134.45$        & $1/134.20$  \\
 $C_{1u}^\q{SM}$              & $-0.1902$         & $-0.1906$ \\
 $C_{1d}^\q{SM}$              & $0.3427$          & $0.3429$ \\
$2C_{1u}^\q{SM}-C_{1d}^\q{SM}$  & $-0.7231$ & $-0.7241$  \\
 $C_{2u}^\q{SM}$            & $-0.0375$           & $-0.0380$ \\
 $C_{2d}^\q{SM}$            & $0.0276$           & $0.0280$ \\
$2C_{2u}^\q{SM}-C_{2d}^\q{SM}$ & $-0.1025$  & $-0.1039$  \\\hline\hline
 \multicolumn{3}{c}{ $a_1$, $a_3$ terms in $A_\q{SM}$, in ppm}\\\hline
 ``valence quark only''            & $-83.07, -5.11$     & $-145.49, -14.28$\\
  CTEQ/JLab (CJ) full fit, mid  & NA               & $-147.37, -12.12$\\
              \hspace*{3cm}min &  NA               & $-147.41, -12.99$\\
              \hspace*{3cm}max &  NA                & $-147.40, -13.07$\\
  ``PDF+QPM'' MSTW2008 LO    & $-83.61, -4.13$     & $-146.43,-12.48$ \\
  ``PDF+QPM'' CT10 (NLO)     & $-84.06, -4.35$     & $-146.64, -12.89$\\ \hline\hline
 \multicolumn{3}{c}{ coefficients for $2C_{1u}-C_{1d}$, $2C_{2u}-C_{2d}$ in $A_\q{SM}$, in ppm}\\\hline
 ``valence quark only''            & $114.88,49.82$     & $200.92,137.51$\\
  CTEQ/JLab (CJ) full fit, mid  & NA               & $203.52,116.68$\\
               \hspace*{3cm}min &   NA                 & $203.58,125.01$\\
               \hspace*{3cm}max &   NA                 & $203.56, 125.78$\\
  ``PDF+QPM'' MSTW2008 LO    & $115.63,40.26$     & $202.22, 120.08$ \\
  ``PDF+QPM'' CT10 (NLO)     & $116.25,42.41$     & $202.51, 124.08$\\ \hline\hline
\end{tabular}
 \caption{From Supplemental Tables of Ref.~\cite{Wang:2014bba}: 
Comparison of Standard-Model (SM) prediction for the asymmetry, 
$A_\q{SM}$, using different structure functions: 
LO MSTW2008~\cite{Martin:2009iq}, (NLO) CT10~\cite{Lai:2010vv}, 
and the CTEQ/JLab 
(CJ)~\cite{Owens:2012bv} fits. The CJ fits include 3 sets -- middle, 
minimal, and maximal -- to provide the nominal value of the PDF  
and the uncertainties.
Values for $\alpha_{EM}(Q^2)$ were calculated using $\alpha_{EM}(Q^2=0)=1/137.036$. 
The weak couplings at the measured $Q^2$-values, $C_{1,2}^\q{SM}(Q^2)$, 
were based on Table 7 and Eq.~(114-115) of Ref.~\cite{Erler:2013xha}.
}\label{tab:Apvdis_calc}
\end{center}
\end{table}

As can be seen from Eq.~(\ref{eq:a1}, \ref{eq:a3}), the $a_{1,3}$ 
terms of the asymmetry are proportional to the $C_{1,2}$ couplings,  
respectively. This proportionality, i.e. the coefficient for 
 $2C_{1u}-C_{1d}$ or $2C_{2u}-C_{2d}$ in the asymmetry, describes 
quantitatively the sensitivity to these couplings. To interpret the asymmetry results for
both $Q^2$ values consistently, we used the MSTW2008 LO 
values in Table~\ref{tab:Apvdis_calc} as the nominal values and found 
for DIS setting \#1, $A_\q{SM} = -87.7\pm 0.7$~ppm where the uncertainty is dominated by that
from the PDFs. The sensitivity to the effective couplings is
\begin{eqnarray}
 A_\q{SM}&=&(115.63~\mathrm{ppm})(2C_{1u}-C_{1d})+(40.26~\mathrm{ppm})(2C_{2u}-C_{2d})\\
         &=&(1.156\times 10^{-4})\left[(2C_{1u}-C_{1d})+0.348(2C_{2u}-C_{2d})\right]\label{eq:adis_int_loq}
\end{eqnarray} 
For DIS setting \#2, $A_\q{SM} =(-158.9\pm 1.0)$ ppm and
\begin{eqnarray}
 A_\q{SM}&=&(202.22~\mathrm{ppm})(2C_{1u}-C_{1d})+(120.08~\mathrm{ppm})(2C_{2u}-C_{2d})\\
         &=&(2.022\times 10^{-4})\left[(2C_{1u}-C_{1d})+0.594(2C_{2u}-C_{2d})\right].\label{eq:adis_int_hiq}
\end{eqnarray}
The uncertainties in the sensitivity to $2C_{1u}-C_{1d}$ and $2C_{2u}-C_{2d}$ are
0.5\% and 5\%, respectively, as described in the previous paragraph. 
The resulting uncertainty in the $2C_{2u}-C_{2d}$ extraction 
due to the PDF fits is $\Delta(2C_{2u}-C_{2d})(\mathrm{PDF})=\pm 0.011$.

The above calculation used the approximation that $Y_1=1$ which is valid if 
$R^\gamma=R^{\gamma Z}$. 
The effect of possible differences between $R^{\gamma Z}$ and $R^\gamma$ was 
studied in Ref.~\cite{Hobbs:2008mm}:
to account for a shift of 1 ppm in the asymmetry, 7.7\% and 4.5\% differences
between $R^{\gamma Z}$ and $R^\gamma$ are needed, for DIS settings \#1 and \#2, respectively. 
Such large differences were considered highly 
unlikely and the uncertainty in the asymmetry due to the possible difference 
between $R^{\gamma Z}$ and $R^\gamma$ was considered to be negligible compared to 
the statistical uncertainties of the measurement. 

The higher-twist (HT) effects refer to the interaction between quarks inside 
the nucleon at low $Q^2$, where QCD perturbation theory breaks down. 
At a relatively low $Q^2$, but not low enough for the effective QCD coupling 
to diverge, the HT effects introduce a $1/Q^2$-dependence 
to the structure functions in addition to the $\ln Q^2$ perturbative 
QCD evolution. 
The HT effects modify the PVDIS asymmetry through a change in the
absorption cross-section ratio $R^\gamma$ in 
Eqs.~(\ref{eq:y11},\ref{eq:y13}), or through changes in the
structure function ratios $a_1$ and $a_3$ of Eq.~(\ref{eq:a31}). 
The effect on $R^\gamma$ was estimated in Ref.~\cite{Alekhin:2007fh} 
and was found to be negligible.
Studies of the HT effects on the PVDIS asymmetry through changes in the 
structure functions can be dated back to the SLAC E122
experiment~\cite{Bjorken:1978ry,Wolfenstein:1978rr}, 
where it was argued that the HT effects on the $a_1$ term of the asymmetry 
are very small. The most recent discussions on HT effects of 
the PVDIS asymmetry, represented by work in 
Refs.~\cite{Mantry:2010ki,Belitsky:2011gz,Seng:2013fia}, 
indicated that the HT contribution to the $a_1$ term is at or below
the order of $0.5\%/Q^2$ for the $x$ range of this experiment, 
where $Q^2$ is in units of (GeV/$c$)$^2$. 

There is no theoretical estimation of the HT effects on the $a_3$ term 
of the asymmetry. However, this term is bounded by data on the neutrino
structure function $H_3^\nu$~\cite{Alekhin:2007fh}, which has the same 
quark content as $F_3^{\gamma Z}$. If applying the observed $H_3^\nu$ higher-twist
$Q^2$-dependence to $F_3^{\gamma Z}$ alone, one expects the asymmetry to shift 
by $+0.7$~ppm and $+1.2$~ppm for DIS\#1 and \#2, respectively. 
We used these values as the uncertainty in the $a_3$ term due to HT effects.

Overall, a combination of theoretical and experimental 
bounds on the HT effects indicate that they do not exceed 1\% of our
measured asymmetry. The uncertainties in the $a_1$ and the $a_3$
terms due to HT were evaluated separately, and the corresponding uncertainty 
in $2C_{2u}-C_{2d}$ is $\pm 0.012$, and is quite small 
compared to the experimental uncertainties. 


\subsubsection{Global fit to effective couplings $C_{1q}$ and $C_{2q}$}\label{sec:globalfit}
Including the two DIS points obtained by our experiment, there are enough data
to perform a simultaneous fit to the three linear combinations of effective couplings,
$C_{1n} \equiv C_{1u} + 2C_{1d}$, $2C_{1u} - C_{1d}$, and $2C_{2u} - C_{2d}$.
To do this, we used the constraint extracted from atomic parity violation in Cs~\cite{Wood:1997zq}
as quoted in Ref.~\cite{Erler:2013xha},
\begin{eqnarray}
188\, C_{1u} + 211\, C_{1d} = 36.35 \pm 0.21\ ,
\end{eqnarray}
where we relied on the most recent atomic structure calculation in Ref.~\cite{Dzuba:2012kx}.
We also employed the latest $C_{1q}$ result from Ref.~\cite{Androic:2013rhu}: 
\begin{eqnarray}
2\, C_{1u} + C_{1d} - 0.0004 = - 0.032 \pm 0.006\ ,
\end{eqnarray}
where the small adjustment on the left-hand side is from the electron charge radius~\cite{Erler:2013xha}.
Finally, we included the 11 data points of the SLAC--E122 experiment~\cite{Prescott:1979dh}. 
For the E122 asymmetries, we employed Eq.~(\ref{eq:Apvdis_R}) with $\alpha = \alpha(Q^2)$ and $R_C = 0$,
while the values of $R_S$ and $R_V$ are shown in Table~\ref{tab:e122_kine}.
To account for the different $Q^2$ values of these measurements, we adjusted
the effective couplings using Eq.~(\ref{eq:c1uth}-\ref{eq:c2dth}). 
Note that these corrections were applied to our DIS points as well, 
see Table~\ref{tab:Apvdis_calc}.

There are various E122 point-to-point errors which we added in quadrature
(following the original publication~\cite{Prescott:1979dh}),
and then we added the result again quadratically to the statistical errors
(rather than linearly as in Ref.~\cite{Prescott:1979dh}).
In addition, the polarization uncertainty was common to all data points.
This resulted in a 5\% correlated uncertainty in the scale of the asymmetries.
We constructed the corresponding covariance matrix and included it in our fits.

As for the two DIS points of the present experiment, we erred on the conservative side
and approximated their systematic (see Table~\ref{tab:allAsym}) and theory uncertainties as fully correlated.
The latter are composed of PDF uncertainties of 0.76\% and errors originating from
higher twist (quark-quark correlation) effects.
The higher twist uncertainties enter separately and uncorrelated for the $a_1$ 
and the $a_3$ terms. As explained in the previous section, the HT uncertainty 
on $a_1$ term was taken to be $0.5\%/Q^2$ with $Q^2$ in (GeV/$c$)$^2$, or 0.39~ppm 
averaged over DIS\#1 and \#2, and that for the $a_3$ term
was estimated from $H_3^\nu$ data to be 0.7~ppm and 1.2~ppm, respectively, 
for DIS\#1 and DIS\#2.

\begin{figure}[!htp]
 \begin{center}
 \includegraphics[width=0.9\textwidth]{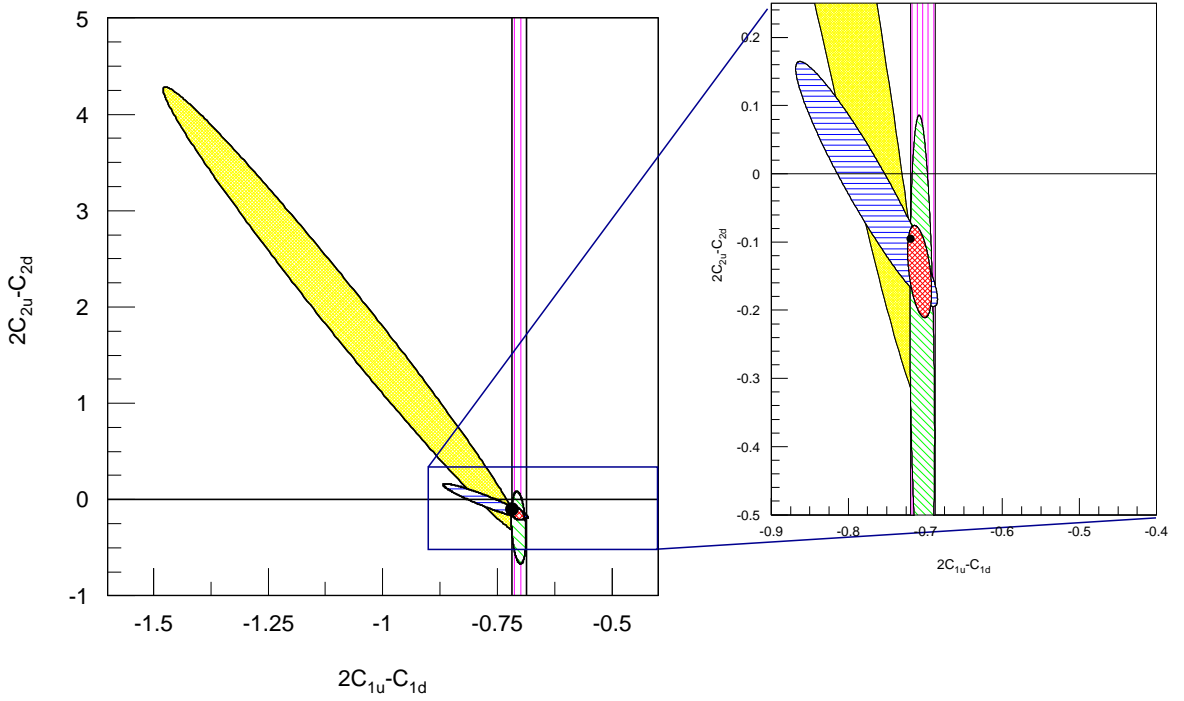}
 \caption{From Ref.~\cite{Wang:2014bba}: results on $(2C_{1u}-C_{1d})\vert_{Q^2=0}$ and 
$(2C_{2u}-C_{2d})\vert_{Q^2=0}$ from the present experiment. 
The right panel shows an enlarged view with the vertical and the 
horizontal axis at the same scale. The new results  
(blue horizontal-line-hatched ellipse) are compared
with SLAC E122 (yellow ellipse)~\cite{Prescott:1978tm,Prescott:1979dh}. 
The latest data on 
$C_{1q}$~\cite{Androic:2013rhu} (from PVES and Atomic 
Cs~\cite{Wood:1997zq,Bennett:1999pd,Ginges:2003qt,Dzuba:2012kx}) 
are shown as the magenta vertical-line-hatched band.
The green slanted-line-hatched ellipse shows the combined result of SLAC E122 and 
the latest 
$C_{1q}$, while the red line-cross-hatched ellipse shows the combined result of 
SLAC E122, the present 
experiment, and the latest $C_{1q}$. The Standard Model value 
$2C_{2u}-C_{2d}\vert _{Q^2=0}=-0.0950\pm 0.0004$ is shown as
the black dot, where the size of the dot is for visibility.
}
 \label{fig:c2q}
 \end{center}
\end{figure}

We then obtain the best fit result and correlation matrix,
\begin{eqnarray}
\label{fitresult}
\boxed{\begin{array}{l|ccc}
C_{1u} + 2\, C_{1d} = \phantom{-}0.489 \pm 0.005 & \phantom{-}1.00 & -0.94 & \phantom{-}0.42 \\
2\, C_{1u} - C_{1d} = - 0.708 \pm 0.016 & -0.94 & \phantom{-}1.00 & -0.45 \\
2\, C_{2u} - C_{2d} = - 0.145 \pm 0.068 & \phantom{-}0.42 & - 0.45 & \phantom{-}1.00
\end{array}}
\end{eqnarray}
where the $\chi^2$ per degree of freedom is 17.3/12, corresponding to a 14\% probability. 
These results are shown in Fig.~\ref{fig:c2q}.
Figure~\ref{fig:c2q} shows our results have greatly improved the uncertainty on the effective
coupling $C_{2u,2d}$ and are in good agreement with the Standard Model prediction. 
The result on $C_{2q}$ alone is~\cite{Wang:2014bba}
\begin{eqnarray}
 \left(2C_{2u}-C_{2d}\right)\vert_{Q^2=0}&=&-0.145\pm 0.066~\mathrm{(exp.)}\pm 0.011~\mathrm{(PDF)}\pm 0.012~\mathrm{(HT)}\\
    &=& -0.145\pm 0.068~\mathrm{(total)}.\label{eq:c2q_result}
\end{eqnarray}
We note that this is the first time we observe the combination $(2C_{2u}-C_{2d})$ to 
be non-zero at the two standard deviation level. Because the $C_{2q}$ is axial-vector in nature 
at the quark vertex, the result of Eq.~(\ref{eq:c2q_result}) can be interpreted as
the first direct evidence that quarks do exhibit a chirality preference when interacting 
with electrons through the neutral weak force~\cite{Marciano:2014hva}.

\subsubsection{Extracting mass limits}\label{sec:masslimit}
A comparison of the present result on $C_{1q,2q}$ with the Standard Model prediction
can be used to set mass limits $\Lambda$ below which new interactions are unlikely to 
occur. For the cases of electron and quark compositeness, we used 
the conventions from Ref.~\cite{Eichten:1983hw} and the procedure followed by the LEP~2 Collaborations,
described in Ref.~\cite{Schael:2013ita}.
The new-physics effective Lagrangian for $eq$ interactions is given by~\cite{Eichten:1983hw}
\begin{eqnarray}
\label{ELPoperator}
\mathcal{L}_{eq} = \frac{g^2}{\Lambda^2} \sum_{i,j = L,R} \eta_{ij}\, \bar{e}_i \gamma_\mu e_i\, \bar{q}_j \gamma^\mu q_j\ ,
\end{eqnarray}
where $\Lambda$ is defined~\cite{Eichten:1983hw} for strong coupling, {\em i.e.} relative to $g^2 = 4\pi$.
For $\eta_{LL} = \eta_{RL} = - \eta_{LR} = - \eta_{RR} = 1$, and adding the SM contribution, one then obtains
\begin{eqnarray}
\label{lagrangian}
\mathcal{L}_{eq}
&=& \left[ \frac{G_F}{\sqrt{2}} C_{2q}({\rm SM}) + \frac{g^2}{\Lambda^2} \right] \bar{e} \gamma_\mu e\, \bar{q} \gamma^\mu \gamma^5 q\\
&\equiv& \frac{C_{2q}({\rm SM}) + \delta C_{2q}({\rm new})}{2 v^2}\, \bar{e} \gamma_\mu e \bar{q}\, \gamma^\mu \gamma^5 q
\equiv \frac{C_{2q}}{2 v^2}\, \bar{e} \gamma_\mu e\, \bar{q} \gamma^\mu \gamma^5 q\ ,
\end{eqnarray}
where $\delta C_{2q}$(new) is the deviation in $C_{2q}$ from the SM value
that may be related to beyond-the-SM physics, 
and the quantity $v = (\sqrt{2}\, G_F)^{-1/2} = 246.22$~GeV is the Higgs vacuum 
expectation value which sets the electroweak scale.

If a measurement of the effective coupling, $C_{2q}$,
or a fit to some data set, finds a central value $\bar C_{2q}$,
then the best estimate of the new physics contribution would be given by
\begin{eqnarray}
\label{scale}
\frac{g^2}{\Lambda^2} =
\frac{4\pi}{\Lambda^2} =
\frac{\bar C_{2q} - C_{2q}({\rm SM})}{2 v^2}\ .
\end{eqnarray}
For the expected (projected) limits, one assume $\bar C_{2q} = C_{2q}{\rm (SM)}$,
in which case the 90\% confidence-level (CL) central range for $C_{2q}$ is given by
\begin{eqnarray}
- 1.645\, \Delta C_{2q}  < \delta C_{2q}({\rm new}) <  1.645\, \Delta C_{2q}\ ,
\end{eqnarray}
where $\Delta C_{2q}$ is the total (statistical + systematic + theoretical) 1~$\sigma$ uncertainty from the extraction.
The endpoints of this range can be interpreted as the 95\% CL upper and lower limits of $C_{2q}$.
However, it is conventional to consider the two possible sign choices of $g^2/\Lambda^2$ as
two different ``models", quoting two separate limits, $\Lambda_\pm$.
Half of the probability distribution is then excluded by construction and one has to renormalize
the remaining part.  This amounts to the 95\% CL: 
\begin{eqnarray}
\label{expected}
|\delta C_{2q}({\rm new})| < 1.96\, \Delta C_{2q}\ .
\end{eqnarray}
In the general case, $\bar C_{2q} \neq C_{2q}{\rm (SM)}$, we find instead the 95\% CL limits,
$$
|C_{2q}|^\pm =
\pm \left[ \bar C_{2q} - C_{2q}{\rm (SM)} \right] + \sqrt{2}\, \Delta C_{2q}\,
{\rm erf}^{-1}\left[ 0.95 \mp 0.05\, {\rm erf} \left( \frac{\bar C_{2q} - C_{2q}{\rm (SM)}}{\sqrt{2}\, \Delta C_{2q}} \right) \right]\ ,
$$
where
\begin{eqnarray}
{\rm erf}(x) \equiv \frac{2}{\sqrt{\pi}} \int_0^x dt \,e^{-t^2}
\end{eqnarray}
is the Gauss error function and ${\rm erf}^{-1}(x)$ its inverse.

A complication arises if a given observable or data set (such as the case at hand)
is not sensitive to a specific flavor operator.
In the case where $u$ and $d$ quarks are involved, we can rewrite,
\begin{eqnarray}
\label{udoperator1}
\mathcal{L}_{eu} + \mathcal{L}_{ed}
= \frac{\bar{e} \gamma_\mu e}{2 v^2}
\left[ C_{2u}\, \bar{u} \gamma^\mu \gamma^5 u + C_{2d}\, \bar{d} \gamma^\mu \gamma^5 d \right],
\end{eqnarray}
in terms of two rotated operators,
\begin{eqnarray}
\mathcal{L}_{eu} + \mathcal{L}_{ed}
&=& \frac{\bar{e} \gamma_\mu e}{2 v^2}
\left( \cos\xi\, C_{2u} + \sin\xi\, C_{2d} \right)
\left( \cos\xi\, \bar{u} \gamma^\mu \gamma^5 u + \sin\xi\, \bar{d} \gamma^\mu \gamma^5 d \right)\nonumber\\
&& + \frac{\bar{e} \gamma_\mu e}{2 v^2}
\left( - \sin\xi\, C_{2u} + \cos\xi\, C_{2d} \right)
\left( - \sin\xi\, \bar{u} \gamma^\mu \gamma^5 u + \cos\xi\, \bar{d} \gamma^\mu \gamma^5 d \right).\label{eq:udoperator2}
\end{eqnarray}
For example, in the operator basis in which
$$
\tan\xi = - \frac{1}{2}\ ,
$$
Eq.~(\ref{eq:udoperator2}) becomes
\begin{eqnarray}
\mathcal{L}_{eu} + \mathcal{L}_{ed}
&=& \frac{\bar{e} \gamma_\mu e}{2 v^2}
\frac{\left(2C_{2u} - C_{2d} \right)}{\sqrt{5}}
\frac{\left( 2\bar{u} \gamma^\mu \gamma^5 u - \bar{d} \gamma^\mu \gamma^5 d \right)}{\sqrt{5}}\nonumber\\
&& + \frac{\bar{e} \gamma_\mu e}{2 v^2}
\frac{\left( C_{2u} + 2 C_{2d} \right)}{\sqrt{5}}
\frac{\left( \bar{u} \gamma^\mu \gamma^5 u + 2\bar{d} \gamma^\mu \gamma^5 d \right)}{\sqrt{5}}.\label{udoperator3}
\end{eqnarray}
Experiments in PVDIS on isoscalar targets
are only sensitive to the operator in the first line of Eq.~(\ref{udoperator3}).
The same applies to the analogously defined rotation angle
between the couplings $C_{1u}$ and $C_{1d}$.  
In this case, the second line turns out to be proportional to the weak charge of the neutron.
In other words, the weak charge of the neutron (but not that of the proton)
contains exactly orthogonal information to that provided by our experiment.

\begin{figure}[!htp]
 \begin{center}
 \includegraphics[width=0.5\textwidth]{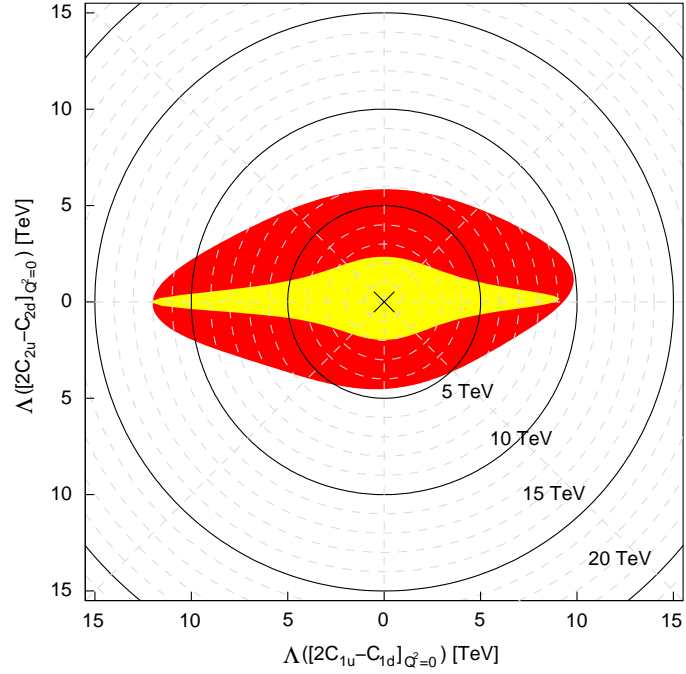}
 \caption{From Ref.~\cite{Wang:2014bba}: 
Mass exclusion limits on the electron and quark compositeness 
and contact interactions obtained from the zero-$Q^2$ values of 
$2C_{1u}-C_{1d}$ and $2C_{2u}-C_{2d}$ 
at the 95\% confidence level. The yellow contour shows the limit
obtained from SLAC E122 asymmetry 
results~\cite{Prescott:1978tm,Prescott:1979dh} combined with the best $C_{1q}$ 
values~\cite{Androic:2013rhu}. The red contour shows the limit with
our new results added. 
}
 \label{fig:masslimits}
 \end{center}
\end{figure}

We determined the combination, $2\, \bar C_{2u} - \bar C_{2d}$, in the last line of the fit result in~(\ref{fitresult}).
Currently, the SM prediction is $[2\, C_{2u} - C_{2d}]({\rm SM})= -0.0949$, and so
the new physics scale corresponding to this operator is bounded (at the 95\% CL) by,
\begin{eqnarray}
\Lambda_+ &>& v\sqrt{\frac{\sqrt{5}\, 8\pi}{|2C_{2u}-C_{2d}|^+}}
       = v \sqrt{\frac{\sqrt{5}\, 8\pi}{0.104}} = 5.7~{\rm TeV},\label{eq:masslimit1}\\
\Lambda_- &>&  v\sqrt{\frac{\sqrt{5}\, 8\pi}{|2C_{2u}-C_{2d}|^-}}
       = v \sqrt{\frac{\sqrt{5}\, 8\pi}{0.170}} = 4.5~{\rm TeV}.\label{eq:masslimit2}
\end{eqnarray}
Results on the new mass limits are shown in Fig.~\ref{fig:masslimits}. 
The improvement on the $C_{2q}$ mass limit is approximately a factor of $\sqrt{5}$. 
We note that while collider experiments have set higher limits on new compositeness that are
vector-electron and axial-vector-quark in nature, their observables are sensitive to 
a combination of different chiral structures, and such limits can only be derived by 
assuming all other chiral terms are zero. 
Such an assumption is not necessary for the present experiment since we measured
$C_{2q}$ directly. Equations~(\ref{eq:masslimit1}-\ref{eq:masslimit2}) provide 
model-independent mass limits on the electron-quark VA contact interactions 
and should be satisfied by any model of new physics.

\section{Summary}\label{sec:summary}

In this paper we document the PVDIS experiment performed at Jefferson Lab using the 6 GeV
longitudinally-polarized electron beam. We archive the experimental setup, the data
analysis procedure, all corrections applied to the asymmetry, and all asymmetry results. 
Asymmetry results from DIS settings (Table~\ref{tab:allAsym}) were used to 
extract the electron-quark effective couplings $C_{1q,2q}$ and the associate mass limits
on new contact interactions. These DIS results have been published in Ref.\cite{Wang:2014bba}. 
Our results on $C_{2q}$ improved 
over existing data by a factor of five and agreed well with the Standard Model prediction. 
They also showed for the first time that $2C_{2u}-C_{2d}$ is non-zero at the two 
standard-deviation level, indicating that the parity-violating asymmetry measured in 
electron deep inelastic scattering does receive a contribution from the quarks' chiral 
preference in neutral weak interaction. Mass limits on new electron-quark VA contact 
interactions were extracted from our $2C_{2u}-C_{2d}$ result, and have improved over
existing limits from PVES by a factor $\sqrt{5}$. Our mass limits are valid for all 
new electron-quark contact interactions that have the VA chiral structure, and are complementary
to limits obtained from collider experiments.

Asymmetries in the nuclear resonance region are reported in Table~\ref{tab:allAsym} 
and their $W$-dependence in Tables~\ref{tab:grp_rawl} and \ref{tab:grp_rawr}. These
results were published previously in Ref.~\cite{Wang:2013kkc}. Our resonance asymmetry
results are in good agreement with theoretical predictions. They also agree well with 
DIS calculations extended to our kinematics, and do not show distinct resonance
structure. This indicates that quark-hadron duality works for PVES asymmetries at 
the 10-15\% level. 

We also report on parity-violating asymmetries of inclusive pion production 
(Tables~\ref{tab:results_Api_dis} and~\ref{tab:results_Api_res}), 
pair production (Table~\ref{tab:Apositron}),
and beam-normal asymmetries (Table~\ref{tab:ATbg}). The results are useful for
background evaluation for other PVES experiments, including those planned for the 
JLab 12 GeV program.

\appendix
\section{Re-analysis of E122 asymmetry results}\label{sec:app_e122}

To study the 
sensitivity of the E122 asymmetry results to $C_{2q}$ couplings, we show these 
kinematics in Table~\ref{tab:e122_kine} including the values for $Y_3$ and $R_V$. 
Calculations of $R_V$ were based on the MSTW2008 
parameterization~\cite{Martin:2009iq} 
of the parton distribution functions. 
Also shown are the simplified value of $Y_3$ which were used in the original
analysis~\cite{Prescott:1979dh}:
\begin{eqnarray}
 Y_3^\mathrm{simplified} &=& \frac{1-(1-y)^2}{1+(1-y)^2}~,\label{eq:y3simp}
\end{eqnarray}
and which we continued to use in this re-analysis. Note, however, that the use 
of Eq.~(\ref{eq:y3simp}) tends to overestimate the already small sensitivity to the $C_{2q}$.
Equation~(\ref{eq:Apvdis_R}) illustrates that the product $Y_3R_V$ 
provides the lever arm to isolate the $C_{2q}$ contribution to the asymmetry. 
The relatively small values and coverage of $Y_3R_V$ 
in E122 were largely due to the small and fixed scattering angle (4$^\circ$), 
and were not ideal for isolating the $C_{2q}$ term. 

\begin{table}[!htp]
\begin{center}
 \begin{tabular}{c|c|c|c|c|c|c|c|c}\hline
  $E_b$ (GeV) & $Q^2$ (GeV/$c$)$^2$ & $x$ & $y$ & $Y_3$ & $Y_3^\mathrm{simplified}$ & $R_S$ & $R_V$ & $Y_3^\mathrm{simplified}R_V$\\ \hline
  16.2 & 0.92 & 0.14 & 0.22 & 0.19 & 0.24 & $0.071 \pm 0.014$ & $0.623 \pm 0.014$ & 0.152 \\
  19.4 & 1.53 & 0.28 & 0.15 & 0.15 & 0.16 & $0.022 \pm 0.005$ & $0.859 \pm 0.012$ & 0.138 \\
  19.4 & 1.52 & 0.26 & 0.16 & 0.16 & 0.17 & $0.027 \pm 0.006$ & $0.836 \pm 0.012$ & 0.144 \\
  19.4 & 1.33 & 0.16 & 0.23 & 0.21 & 0.26 & $0.068 \pm 0.012$ & $0.671 \pm 0.014$ & 0.171 \\
  19.4 & 1.28 & 0.14 & 0.25 & 0.23 & 0.28 & $0.082 \pm 0.013$ & $0.630 \pm 0.014$ & 0.176 \\
  19.4 & 1.25 & 0.13 & 0.26 & 0.24 & 0.29 & $0.090 \pm 0.013$ & $0.608 \pm 0.013$ & 0.178 \\
  19.4 & 1.16 & 0.11 & 0.29 & 0.26 & 0.33 & $0.107 \pm 0.013$ & $0.563 \pm 0.013$ & 0.186 \\
  19.4 & 1.07 & 0.09 & 0.32 & 0.29 & 0.37 & $0.127 \pm 0.014$ & $0.518 \pm 0.012$ & 0.190 \\
  19.4 & 0.93 & 0.07 & 0.36 & 0.33 & 0.42 & $0.148 \pm 0.017$ & $0.471 \pm 0.011$ & 0.197 \\
  22.2 & 1.96 & 0.28 & 0.17 & 0.17 & 0.18 & $0.027 \pm 0.005$ & $0.860 \pm 0.011$ & 0.158 \\
  22.2 & 1.66 & 0.15 & 0.26 & 0.24 & 0.29 & $0.081 \pm 0.012$ & $0.654 \pm 0.014$ & 0.191 \\\hline

 \end{tabular}\caption{Kinematics for the SLAC E122 experiment. Values 
for $E_b$, $Q^2$, $x$ and $y$ are from Ref.~\cite{Prescott:1979dh}.
Values for $R_S$ and $R_V$ are calculated
using the MSTW2008~\cite{Martin:2009iq} leading-order parameterization. 
The product $Y_3R_V$ provides the lever arm for isolating the $C_{2q}$
contribution to the asymmetry. We used $Y_3^\mathrm{simplified}$ in line with the 
original publication~\cite{Prescott:1979dh}. }\label{tab:e122_kine}
\end{center}
\end{table}

\section{Formalism for beam depolarization calculation}\label{sec:app_depol}
The beam depolarization was calculated using Eq.(9.11) of Ref.~\cite{Olsen:1959zz}:
\begin{eqnarray}
  D(\vec p_1, \vec \zeta_1) &=& 
    \frac{k^2\left[\psi_1-\zeta^2_{1z}(\psi_1-{2\over 3}\psi_2)\right]}
         {(\epsilon_1^2+\epsilon_2^2)\psi_1-{2\over 3}\epsilon_1\epsilon_2\psi_2}
\end{eqnarray}
where $\epsilon_{1,2}$ are the energy of the 
electron before and after bremsstrahlung in unit of the electron mass $m_e c^2$, 
$k$ is the bremsstrahlung photon energy in units of $m_e c^2$, 
$\vec\zeta$ is the polarization vector of the electron with $\zeta_{1z}=1$ for 
longitudinally polarized electrons,
and $\psi_{1,2}$ are given in the ``complete screening'' limit by 
\begin{eqnarray}
 \psi_1 &=& 4\ln(111 Z^{-1/3})+2-4f(Z)=4[\ln(183 Z^{-1/3})-f(Z)],\\
 \psi_2 &=& 4[\ln(183 Z^{-1/3})-f(Z)] - {2\over 3}.
\end{eqnarray}
The function $f(Z)$ is 
\begin{eqnarray}
 f(Z)&=& a^2\sum_{n=1}^{\infty} \frac{1}{n(n^2+a^2)}~,
\end{eqnarray}
with $a=(Ze^2/\hbar/c)$.

The ``complete screening'' limit is defined as $\beta_i\xi/\delta \gg 1$ where 
$\beta_i=(Z^{1/3}/121)b_i$ with $b_1=6$, $b_2=1.2$ and $b_3=0.3$; 
$\xi\equiv 1/(1+u^2)$ 
with $u=p_1\theta_1$; and $\delta\equiv 
k/(2\epsilon_1\epsilon_2)$. 
Here $\vec p_1, \vec p_2$ are the momenta of the electron
before and after bremsstrahlung in units of $m_ec$, and $\theta_1,\theta_2$ are 
the angles between $\vec p_1, \vec p_2$ and the photon $\vec k$, respectively. 
Because for high energy electrons $\theta_1$ is very small, $u\approx 0$ and 
$\xi\approx 1$. Putting all notations together, the complete screening limit is 
\begin{eqnarray}
 \frac{\beta_i\xi}{\delta} &=& 
   \frac{\frac{Z^{1/3}}{121} b_i}
        {\left(1+\epsilon_1^2\theta_1^2\right)\frac{k}{2\epsilon_1\epsilon_2}}
   \approx \frac{\frac{Z^{1/3}}{121} b_i}
        {\frac{k}{2\epsilon_1\epsilon_2}+{1\over 2}k\theta_1^2} \gg 1
\end{eqnarray}
where the approximation is valid if $k\ll\epsilon_1$ (which implies $\epsilon_1\approx \epsilon_2$
and $k\ll\epsilon_2$) and the complete screening
condition is satisfied if $\epsilon_{1,2}\gg 1$. For the 6-GeV beam used in this experiment, 
$\epsilon_1\approx 12000$ and $k\ll \epsilon_1$, therefore the complete screening limit can be used.

\end{document}